\documentclass[sn-nature]{sn-jnl}

\usepackage{graphicx}%
\usepackage{multirow}%
\usepackage{amsmath,amssymb,amsfonts}%
\usepackage{amsthm}%
\usepackage{mathrsfs}%
\usepackage[title]{appendix}%
\usepackage[table]{xcolor}%
\usepackage{textcomp}%
\usepackage{manyfoot}%
\usepackage{booktabs}%
\usepackage{algorithm}%
\usepackage{algorithmicx}%
\usepackage{algpseudocode}%
\usepackage{listings}%

\theoremstyle{thmstyleone}%
\theoremstyle{thmstyletwo}%

\theoremstyle{thmstylethree}%

\usepackage{multirow}
\usepackage{subcaption}
\usepackage{rotating}
\usepackage{bm}
\usepackage{ulem}
\usepackage{lineno}

\newcommand{\rev}[1]{{#1}}
\newcommand{\revv}[1]{{#1}}
\newcommand{\revvv}[1]{{#1}}

\begin{document}

\title[Volatile depletion in rocky planets as a chemical fingerprint of hybrid accretion]{Volatile depletion in rocky planets as a chemical fingerprint of hybrid accretion}



\author*[1]{\fnm{Haiyang S.} \sur{Wang}}\email{haiyang.wang@sund.ku.dk}
\author*[1]{\fnm{Anders} \sur{Johansen}}\email{anders.johansen@sund.ku.dk}

\author[1]{\fnm{Ziyan} \sur{Xu}} 
\author[2]{\fnm{Marie-Luise} \sur{Steinmeyer}} 
\author[1]{\fnm{Michiel} \sur{Lambrechts}}
\author[1]{\fnm{Elishevah} \sur{van Kooten}}
\author[3]{\fnm{Chao-Chin} \sur{Yang}}
\author[4]{\fnm{Zhaohuan} \sur{Zhu}}
\author[1,5]{\fnm{Dante S.} \sur{Lauretta}}
\author[1]{\fnm{Martin} \sur{Bizzarro}}

\affil*[1]{\orgdiv{Center for Star and Planet Formation, Globe Institute}, \orgname{University of Copenhagen}, \orgaddress{\city{1350 Copenhagen}, \country{Denmark}}} 

\affil[2]{\orgdiv{Institute for Particle Physics and Astrophysics}, \orgname{ETH Z\"urich}, \orgaddress{\city{8093 Z\"urich}, \country{Switzerland}}} 

\affil[3]{\orgdiv{Department of Physics and Astronomy}, \orgname{University of Alabama}, \orgaddress{\city{Tuscaloosa}, \postcode{AL 35487-0324}, \country{USA}}} 

\affil[4]{\orgdiv{Department of Physics and Astronomy and Nevada Center for Astrophysics}, \orgname{University of Nevada}, \orgaddress{\city{Las Vegas}, \postcode{NV 89154-4002}, \country{USA}}} 

\affil[5]{\orgdiv{Lunar and Planetary Laboratory}, \orgname{University of Arizona}, \orgaddress{\city{Tucson}, \postcode{AZ85721}, \country{USA}}} 


\abstract{
Volatile depletion in rocky planets relative to their host stars is commonplace in both the Solar System and exoplanetary systems, yet the connections between planet formation and composition remain elusive. Here we model devolatilization during pebble accretion in combination with collisional growth from volatile-depleted planetesimals to explore the formation pathways of Earth and Mars. Using Bayesian inference, we find that bulk silicate Earth is best reproduced by ${\gtrsim}$75\% contribution from two protoplanets formed via pebble accretion, supplemented by up to $\sim$25\% material from planetesimals that are compositionally akin to the asteroid Vesta. Using instead a planetesimal volatile-depletion curve that is not observed among known meteorite parent bodies would allow the planetesimal contribution to reach 40$^{+15}_{-14}$\%. In comparison, bulk silicate Mars reflects 27$\pm$5\% pebble-accreted material and 73${\pm}$5\% Vesta-like planetesimals.
We identify volatile depletion as a chemical fingerprint of hybrid accretion, in which both pebble accretion and collisional assembly contribute to terrestrial planet growth. By quantitatively linking formation pathways to volatile budgets, our findings demonstrate how planetary accretion histories can be inferred from elemental signatures, with broad implications for interpreting the chemical diversity of rocky exoplanets.
}

\maketitle


The compositions of the inner Solar System planets Earth and Mars show as a common feature that their mantle abundances of moderately volatile lithophile elements (MVLEs) are depleted relative to the proto-Sun (\cite{Albarede2009, Wang2018, Yoshizaki2020}; Fig.~1). MVLEs are defined as elements with condensation/sublimation temperatures lower than those of the major rock-forming elements Mg and Si but higher than those of atmophile species such as H$_2$O, and that preferentially remain in the silicate mantle during core-mantle differentiation. This common ``devolatilization" feature of MVLEs \cite{Wang2019, Braukmuller2019} has been attributed to several processes. These include (i) incomplete condensation during an early, ultra-hot stage of the solar protoplanetary disk \cite{Bond2010,Sossi2022,Lodders2025}, (ii) melting and partial evaporation of planetesimals driven by radiogenic heating and/or collisional kinetic energy \cite{Eatson2024, Hin2017, Halliday2001}, and (iii) partial vaporization of protoplanets by giant impacts \cite{Norris2017, Saurety2025}.  Importantly, devolatilization in rocky planet formation may be a universal process \cite{Wang2019}. Support for this hypothesis comes from the inferred compositions of rocky exoplanetary material, based on spectroscopic measurements of polluted white dwarf atmospheres \cite{Harrison2021} and binary/co-moving stars with signatures of planet ingestion \cite{Spina2021, Liu2024}.

Anchored by MVLE depletions in both Earth and Mars (Fig.~1; Methods Section~1; Supplementary Table~1), we probe here potential formation pathways of terrestrial planets, which remain highly debated \cite{Morbidelli2025, Johansen2024}, particularly between rapid pebble accretion \cite{Johansen2021, Bizzarro2025} and prolonged mutual collisions of planetesimals and protoplanets \cite{Burkhardt+etal2021, Morbidelli2022}. Our goal is not to contrast these two competing theories, but to assess the viability of a potential \textit{hybrid scenario}. This scenario is motivated by simulations showing that both pebble accretion and mutual collisions have the potential to drive planetary growth in the inner regions of protoplanetary disks where rocky planets form \cite{Lambrechts2019}. If several processes indeed contribute to the assembly of rocky planets, then those processes must imprint various signatures in the volatile depletion of terrestrial planets, based on which the relative contributions of these processes can be retrieved.

A high ambient temperature during the formation of planetary building blocks in the protoplanetary disk is often assumed to explain the depletion of MVLEs in Earth \cite{Bond2010, Sossi2022, Lodders2025}. However, in the emerging disk wind paradigm for protoplanetary disks, mass accretion through the disk is mainly facilitated through winds launched by large-scale magnetic fields \cite{Bethune2017}. The temperature increase of the disk caused by wind-driven accretion is far lower than in the traditional model of turbulence-driven accretion, with a mid-plane temperature remaining $\lesssim$ 300 K at 1 au \cite{Mori2021}. 
Indeed, the presence of moderately volatile siderophile elements (MVSEs), along with carbon, nitrogen and water, in the parent bodies of iron meteorites also point to the formation of even the earliest planetesimal generations under such cool conditions \cite{Grewal2021, Grewal2022, Grewal2024, Grewal2025}.
Further, sample returns from the asteroids Ryugu and Bennu (Fig.~1) demonstrate that dust similar in composition to CI chondrites -- pristine and solar-like, except for the most volatile elements (H, C, N and noble gases) -- was widespread in the Solar System \cite{Nakamura2022, Lauretta2024}.

Alternatively, melting and evaporation of early-formed planetesimals has been suggested as a viable mechanism of volatile loss, caused by the decay of short-lived radionuclides (e.g., $^{26}$Al and $^{60}$Fe, \cite{Lichtenberg2016, Eatson2024}) and/or energetic planetesimal collisions \cite{Halliday2001, Hin2017, Grewal2025}.  Meteorites from the differentiated asteroid Vesta, a body sometimes categorized as a surviving planetary embryo \cite{Russell2012}, show strong depletion of elements with sublimation temperatures below $\sim$$1200\,{\rm K}$ (\cite{Mittlefehldt2015}; Fig.~1; Methods Section~2). It is not understood whether Vesta's depletion of volatile elements is due to high ambient disk temperatures \cite{Sossi2022} or evaporation from a magma ocean \cite{Fang2024}, and some volatile elements may even have become stored in the core during differentiation of Vesta \cite{Steenstra2018}. Here we adopt the view that Vesta represents a population of early-formed planetesimals (Methods Section~3) that degassed their lithophile volatiles before contributing to the accretion of Earth and Mars (we will assess the effect of potentially volatile-bearing core material \cite{Grewal2025} separately). We note that the angrite meteorite group shows an even stronger elemental depletion than meteorites from Vesta, being depleted even in Si and Mg \cite{Hu2022}; this provides additional support for melting as a widespread mechanism of volatile loss of early-formed planetesimals.

{On the other hand, thermal processing of pebbles during their accretion, which is typically not considered in previous work \cite{Hin2017, Sossi2022, Morbidelli2025}, offers a new mechanism for impoverishing volatiles \cite{Johansen2021}.} Here we use a 1-D model of the gas envelope of a protoplanet undergoing pebble accretion \cite{Johansen2021, Steinmeyer2023} to calculate the loss of MVLEs through thermal processing of pebbles as they descend through the H$_2$/He-dominated gaseous envelope (Methods Section~4). Figure~2a shows the cumulative sublimation fraction of both refractory and volatile lithophile elements over the growth of a protoplanet. When the sublimation temperature of a volatile lithophile ($T_{\rm sub}$; Extended Data Table~1) is lower than the temperature at the bottom of the envelope ($T_{\rm env}$; Extended Data Fig.~1), the element is assumed to sublimate and subsequently escape into the disk through envelope convection. Importantly, this mass loss does not apply to refractory species (Si and Mg and more refractory); these elements sublimate instead into a deep silicate vapor atmosphere in equilibrium with the underlying magma ocean, protected from diffusion to the upper envelope by an inner radiative zone where silicate cloud formation strongly inhibits the convection \cite{Steinmeyer2023}.

To validate our assumption of complete loss of a volatile element once it has been sublimated into the envelope, we perform 3-D hydrodynamical simulations of the convective flow pattern in the envelope (Fig.~2b; Methods Section~5). Within the convective zone (distance from the planet surface $R \lesssim R_{\rm B}$, where $R_{\rm B}$ is the Bondi radius), the released volatiles are transported upward by convective mixing. Upon reaching the recycling region ($R \gtrsim R_{\rm cyc}$), they are carried away by large-scale recycling flows from the protoplanetary disk that penetrate the Hill sphere, and they are thus lost from the envelope \cite{KurokawaTanigawa2018, Kuwahara2026}. Figure~2b presents the temporal evolution of volatiles in the envelope (assuming $R_{\rm cyc}=2 R_{\rm B}$). Within 1 year, only $\sim$$30\%$ of the volatiles remain in the envelope, and only $\lesssim$$3\%$ remain by 5 years. This is consistent with a diffusion timescale of $\sim$$1$ year measured from the velocity dispersion of the particles ({Extended Data} Fig.~2a). Our assumption of $R_{\rm cyc}=2 R_{\rm B}$ is conservative, as previous simulations indicate that $R_{\rm cyc}$ is likely close to $R_{\rm B}$ \cite{KurokawaTanigawa2018}, reducing the diffusion timescale to $\sim$0.1--1 years ({Extended Data} Fig.~2b). Recent simulations of envelope convection coupled realistically to the gas flow from the protoplanetary disc \cite{Kuwahara2026} show that the tracer particles in a fully convective envelope will indeed be depleted within a timescale of a couple of years. Notably, efficient escape applies to volatiles both in the gas phase and recondensed into (or onto) grains of sizes $\lesssim$$10\,\mu{\rm m}$, which are well-coupled to the gas and do not sediment appreciably towards the protoplanet. Even under the extreme assumption that volatiles would recondense onto sedimenting pebble-sized particles (mm-cm), these volatiles would simply go on to sublimate again deep in the envelope and be more likely to condense onto $\mu{\rm m}$-sized grains (due to the higher total surface of such small grains) at their next upwards passage across their sublimation temperature.

\section*{Results}\label{sec2}

Here we show the results of our evaluation of the viability of a potential {\it hybrid scenario} -- in which rocky planets grow by a combination of pebble accretion, planetesimal accretion, and giant impacts -- against the bulk silicate (lithophile) compositions of Earth and Mars as a function of elemental sublimation temperature (Extended Data Table~1 and Supplementary Table~1) by applying both a reduced-$\chi^2$ minimization and a Bayesian inference framework. 

\subsection*{A hybrid pathway of Earth formation}

For Earth (Fig.~3), we consider formation primarily from a three-component mixture: the proto-Earth (\textbf{PE}) and one (giant) impactor (\textbf{Imp}) -- both of which grew via pebble accretion (and subject to devolatilization during the process) -- along with early-formed planetesimals (\textbf{Plm}) that are assumed to have a Vesta-like composition (Methods Section~3). We have also examined a series of alternative scenarios: in particular, including \textit{or not} a population of volatile-depleted (Vesta-like) and/or volatile-rich (CI-like) planetesimals, accretion of pebbles that are volatile-poor (Methods Section~6), as well as varying the number of impactors onto our proto-Earth (Table~1, column 1). We note that EH chondrites from the inner Solar System have CI-like levels of MVLEs (Fig.~1), but we will nevertheless  for simplicity of notation refer to our pristine, non-depleted planetesimal component as `CI-like'.

We start with a reduced-$\chi^2$ minimization approach (Methods Section~7) to explore the plausible parameter space for the primary three-component scenario ({\bf PE+Plm+Imp}) by setting up a grid of the relative mass fractions of those three components (in terms of planetesimal mass fraction, $f_{\rm Plm}$, and the mass ratio of impactor to proto-Earth, $r_{M}$, shown in the 2D space as in Fig.~3b). The minimum of the reduced-$\chi^2$ ($\chi^2/N$, where $N=13$ is the number of lithophile elements considered in the fit: Al, Ti, Ca, Cr, Mg, Si, Li, Mn, Na, K, F, Zn, and In) is 3.3. Within an increase of $\chi^2/N$ of unity from the minimum value, a large range of solutions for both $r_M$ (0.1--1.0) and $f_{\rm Plm}$ (0--0.4) can be achieved.

We then employ Bayesian inference with dynamical nested sampling (Methods Section~7), along with a set of reasonable priors (Extended Data Table~2), to place probabilistic constraints on the parameter estimation of a series of proposed models. The test results based on (logarithmic) Bayesian evidence ($\ln Z$) and Bayes factor ($\Delta\ln Z$) suggest that the models \textbf{PE+Plm+Imp} and \textbf{PE+Imp} perform equally well in explaining the data (Table~1; Fig.~4a; Supplementary Fig.~3a).
By comparing the volatility trends produced with these two models (Fig.~3a and Extended Data Fig.~3a) and the resulting reduced-$\chi^2$ values (3.3 vs. 3.6; Table~1, column 3), we find that the {\bf Plm} component particularly contributes to a better explanation of the lithium level, which is elevated in Earth relative to {\bf PE}+{\bf Imp}. The {\bf PE}+{\bf Imp}+{\bf Plm} and {\bf PE}+{\bf Imp} models together inform a hybrid accretion pathway of Earth's formation that requires at least ${\sim}75\%$ mass contribution from proto-Earth and an impactor formed via pebble accretion, with up to ${\sim}25\%$ from the Vesta-like planetesimal component (Table~1, column 2).

We go on to test the essentiality of including the pebble-grown {\bf PE} and {\bf Imp} components and we also introduce a volatile-rich CI-like planetesimal population ({\bf CI}) (Extended Data Fig.~3b--e). Without an {\bf Imp}, the {\bf PE+Plm} model is disfavored (relative to the reference {\bf PE+Plm+Imp} model) at the `substantial' level according to Jeffreys' scale (Methods Section~7) that measures the strength of Bayesian evidence ($\Delta\ln Z = -1.31\pm0.07$, Table~1, column 5). Without including any pebble-grown body, the {\bf CI+Plm} model is disfavored `decisively' ($\Delta\ln Z = -11.31\pm0.07$). By re-adding one or both of the pebble-grown bodies, the {\bf PE+CI+Plm} and {\bf PE+CI+Plm+Imp} models return higher $\Delta\ln Z$ values of $-2.07\pm0.07$ and $-1.78\pm0.08$, respectively, while remaining disfavored `substantially' relative to the reference {\bf PE+Plm+Imp} model. The contribution of a CI-like population of planetesimals is nevertheless no more than 10\% $M_{\oplus}$ (within $1\sigma$ uncertainty; Table~1, column~2) in these models.

We have also quantified the range of the mass fraction of an impactor in the two favored {\bf PE+Plm+Imp} and {\bf PE+Imp} models: 8--43\% $M_{\oplus}$ (within $1\sigma$ uncertainty; Table~1, column 2). This relatively large range is in line with the canonical Moon-forming giant impact (Theia) mass ({$\sim$}10--15\% $M_{\oplus}$; \cite{Canup2001, Gabriel2023}) and also the range required for Theia ({$\sim$}28--46\% $M_{\oplus}$) to disrupt a primordial resonant chain evolving into the present-day configuration of the inner Solar System \cite{Huang2025}. The possibility of multiple impactors has also been tested in the {\bf PE+Plm+DualImp} (with dual impactors) and {\bf PE+Plm+TriImp} (with triple impactors) models (Extended Data Fig.~4). The dual-impactor scenario is plausible (given its $|\Delta\ln Z| < 1$), as long as the mass of the secondary impactor is low ($0.04^{+0.08}_{-0.03} M_{\oplus}$), with its posterior distribution skewing towards the lower end of the mass range (Supplementary Fig.~4). The triple-impactor scenario is evidently disfavored (with $\Delta\ln Z = -1.77\pm0.07$), and the mass of the third impactor (even if it exists) would be negligible ($0.01^{+0.02}_{-0.01} M_{\oplus}$). The number of giant impacts preferred in our model is thus somewhat below the mean number of giant impacts seen in $N$-body simulations ($\sim$3, \cite{Quintana2016}); the inclusion of pebble accretion in terrestrial planet formation naturally implies that fewer planetesimal seeds are required to form terrestrial-mass planets, resulting in fewer protoplanets and hence fewer giant impacts.

As an alternative to accreting pebbles of solar-like composition, our tests for volatile-poor pebbles are constructed by scaling the levels of pre-depletion of  pebbles relative to that of H chondrites (Methods Section~6). We find that the reduced-$\chi^2$ of the best fit increases as a function of the increasing pre-depletion factor (Extended Data Fig.~5a) while the relative contributions from {\bf PE}, {\bf Plm} and {\bf Imp} remain similar. This demonstrates that pre-depletion of accreted pebbles gives a poorer fit to the volatile depletion of Earth (Extended Data Fig.~5b--d), even if pre-depleted pebbles cannot be excluded based on Bayesian evidence (with $|\Delta\ln Z| \lesssim 1$). Such volatile-depleted pebbles could nevertheless have contributed to the later stages of proto-Earth's growth (beyond ${\sim}0.15 M_\oplus$; Fig.~2a) where all moderately volatile elements would be anyway lost during the accretion.

\subsection*{A hybrid pathway of Mars formation}

For Mars, we consider its growth through a combination of pebbles and planetesimals accreted within the protoplanetary disk lifetime, motivated by its inferred rapid core formation \cite{Dauphas2011}. Our reference model, based on Bayesian inference, consists of $27{\pm}5\,\%$ by mass via pebble accretion (\textbf{PM}) and $73{\pm}5\,\%$ by mass from planetesimals with a Vesta-like composition (\textbf{Plm}) (Figs.~4b and 5; Table~1). This finding is in line with silicon isotope constraints on Mars' accretion that suggest significant contribution from material akin to Vesta-like planetesimals \cite{Onyett2023}.

For alternative scenarios of Mars' accretion, we have primarily focused on varying the volatile-depletion levels of the accreted planetesimals. As done for Earth, we first introduce a population of volatile-rich CI-like planetesimals as a replacement of the pebble-grown component {\bf PM} (forming the planetesimal-mixture model of {\bf CI+Plm}) and as an addition to the existing components (constituting the {\bf PM+CI+Plm} model). The {\bf CI+Plm} model is disfavored `strongly' (with $\Delta\ln Z=-4.11\pm0.06$), with a worsening fit towards Zn and In in particular (Extended Data Fig.~6a). The addition of a pebble-grown component in the {\bf PM+CI+Plm} model elevates the Bayes factor to $-2.37\pm0.09$, but it remains disfavored `substantially' due to the (unnecessary) model complexity (Extended Data Fig.~6b). The contribution of a CI-like planetesimal population to Mars' final mass remains well below 10\% ($0.03^{+0.04}_{-0.02} M_{\rm Mars}$). Similar as for Earth, this lands somewhat below a previously proposed accretion of 10-15\% CI-like material contribution to Earth \cite{Braukmuller2019}. Considering pre-depleted pebbles for Mars yields a {\bf PM} contribution that increases by about 10\%, but also leads to poorer overall fits to the volatile depletion of Mars (Supplementary Fig.~2).

We then go on to test hypothetical planetesimal populations with varied volatile-depletion scales by modulating the two variables of the logistic model for volatile-depleted planetesimals (Eq.~\ref{eq:vesta}): the inflection point ($T_0$) and the steepness parameter ($\sigma_0$). 
We first apply the reduced-$\chi^2$ minimization in a parameter space of varied $T_0$ ([900, 1600] K) and $f_{\rm Plm}$ ([0.1, 1.0]) (Fig.~5b). The result indicates a confined region -- $T_0 \sim$ 1100--1400 K and $f_{\rm Plm} \sim$ 60--80\% -- where $\chi^2/N$ values are comparably small. Our further Bayesian inference ({Extended Data} Fig.~7a) returns a quantitative value of $T_0=1363^{+43}_{-52}$ K and $f_{\rm Plm}=68{\pm}5\,\%$ (corresponding to the star sign with error bars in orange in Fig.~5b; refer to Table~1 for model details). This implies that an unidentified type of planetesimals with moderately volatile elements even more depleted than Vesta could have contributed to forming Mars. 

We further test the steepness parameter $\sigma_0$ (together with $T_0$) of the planetesimal logistic model in a wide range of 10--500 K (indicating an extended mixture of various volatile-depleted planetesimals) in the case of Mars ({Extended Data} Fig.~7b). The result with the highest Bayesian evidence shows $T_0=1273^{+70}_{-65}$ K and $\sigma_0=227^{+78}_{-75}$, representing the respective, \textit{average} properties of various populations of hypothetical planetesimals. Correspondingly, the total fraction of this mixture is $79^{+11}_{-9}\%$, leaving the remaining $21^{+9}_{-11}\%$ to be via pebble accretion (Table~1, column 2). In all of these tested cases for Mars, the mass contribution from early-formed planetesimals is unanimously dominant, over ${\sim}65\%$, while the contribution from pebble accretion is never less than 10\% (within 1$\sigma$ uncertainty), reinforcing the hybrid-accretion nature of Mars formation.

\section*{Discussion}\label{sec12}
Freeing the assumption of a Vesta-like composition of the planetesimals accreted to Earth (Supplementary Figs.~7--9) further solidify the hybrid-accretion nature of Earth formation. The contribution of such a hypothetical population of planetesimals could reach $40^{+15}_{-14}$\%, when we optimize the composition of the early-formed planetesimals without considering any potential devolatilization of Mg, Si, and Cr of the pebbles before or during pebble accretion (Methods Section~4). Taken together, these results suggest that Earth experienced a relatively high pebble accretion rate while Mars's pebble accretion remained comparably sluggish -- possibly due to gravitational stirring of its orbit by other, more massive protoplanets~\cite{Levison2015}.

Our accretion model for Earth assumes that the giant impact did not lead to any loss of MVLEs. The efficiency of impact-driven devolatilization of planetesimals and protoplanets depends on both impact speed and the mass of the target body \cite{Saurety2025, Calogero2025}.
For the specific case of K loss, ref. \cite{Calogero2025} found collisional devolatilization to be only relevant for planetary bodies up to about Moon mass. Beyond that, insufficient energy is available for vapor to escape from the deep gravitational potential. This upper limit around Moon mass is interesting, since protoplanets of higher mass reach the MVLE devolatilization domain driven by pebble sublimation (Fig.~2a), while radiogenic heating and/or collisional devolatilization may have been responsible for volatile loss on planetesimals and smaller protoplanets \cite{Eatson2024, Grewal2025}. We nevertheless emphasize that collisional devolatilization from impacts during the disk phase is still poorly understood \cite{Kominami2002}, as the  disk gas may facilitate vapor loss.

Our hybrid-accretion models based on MVLEs also carry important implications for the abundances of MVSEs. Unlike MVLEs, those MVSEs are potentially preserved in planetesimal cores. Indeed, non-carbonaceous (NC) iron meteorites, which sample the cores of early-formed planetesimals in the inner solar system, record MVSE-rich compositions \cite{Grewal2025}. Here, we focus on sulfur (S) given its MVSE nature and that it is a major light element candidate in the cores of both Earth and Mars \cite{Wang2018, Yoshizaki2020}. We have adopted an approximate range of the CI-normalized abundance of S in iron meteorite parent bodies (${\sim}$0.1--1; \cite{Grewal2025}) to represent the S abundance in the Vesta-like planetesimals ({\bf Plm}). Based on the best solutions of our reference models ({\bf PE+Plm+Imp} and {\bf PM+Plm}; Table~1), we estimate the protosolar-normalized ranges of S abundance in bulk Earth and Mars: 0.10--0.23 and 0.19--0.85, respectively (Extended Data Fig.~8). These ranges are  consistent with the recent (and higher) estimate of S content in Earth's core \cite{Fischer2025} and with constraints on Mars' core based on seismic data \cite{Stahler2021}, instead of the traditional (and lower) estimates of S content in the terrestrial planets' cores based on BSE/BSM's volatility trends \cite{Wang2018, Yoshizaki2020, Khan2022}.

Our reference model ({\bf PE+Plm+Imp}) of Earth's hybrid accretion aligns well with a recent three-component model \cite{OlsonSharp2023} developed to explain the relatively low excess of $^{182}$W in Earth's mantle. W is siderophile, but $^{182}$W is continuously created in the mantle by decay of $^{182}$Hf with a half-life of 8.9 million years. {The model of \cite{OlsonSharp2023}} invokes 20\% contribution from planetesimals and 80\% contribution from two protoplanets of mass $0.7 \,M_{\rm E}$ and $0.1 \,M_{\rm E}$, respectively, both formed via pebble accretion. If this view is correct, then the low $^{182}$W excess measured for our planet reflects the ongoing delivery of metal to Earth's core {-- segregating in the process W from the mantle --} by planetesimal accretion as well as in the Moon-forming giant impact \cite{Yu2011}.

Based on isotopic compositions, there is an ongoing debate about whether Earth and Mars formed from purely inner Solar System materials or had significant isotopic contribution from pebbles drifting in from the outer Solar System \citep{Morbidelli2025, Johansen2024, Bizzarro2025}. While the composition of Earth lies on a mixing line between inner-solar-system ureilites and outer-solar-system CI-chondrites in the isotopic distribution of many major elements such as Si and Ca \cite{Bizzarro2025}, Earth appears as an endmember in other elements such as Mo and Zr \cite{Burkhardt+etal2021}. The latter has been interpreted as evidence that Earth formed from a lost reservoir of planetesimals \cite{Burkhardt+etal2021} or otherwise that Earth lost an isotopically distinct component embedded in FeS/CaS/MgS grains \cite{Onyett2023} and in water ice \cite{Johansen2021, Bizzarro2025b} when these components were subject to sublimation during pebble accretion. Our thermal processing model for pebbles, which underpins the hybrid accretion of Earth, aligns well with the latter view. Our composition-based results, complementary to isotopic analyses that remain debated \cite{Burkhardt+etal2021, Onyett2023, Johansen2024, Morbidelli2025, Bizzarro2025, Bizzarro2025b, Sossi2026}, help refine the range of viable formation pathways for Earth and Mars.

The loss of volatile elements during the accretion of Earth and Mars carries significant implications for the source of life-essential volatile elements (e.g., C, H, N, O, P, and S) on rocky planets. Those elements are expected to be substantially depleted during the pebble accretion phase, even if the protoplanet grows in the region just exterior {to} the water ice line \cite{Johansen2021, Houge2025}. Volatile delivery to rocky planets thus occurs both through the direct accretion of volatile-rich pebbles -- before the protoplanet reaches the sublimation temperature of the relevant volatile -- and through impacts with smaller protoplanets. Additionally, accretion of primitive planetesimals, which formed late enough to avoid substantial devolatilization, could further contribute to the volatile budget \cite{Raymond2009, Braukmuller2019}. Extending these results to rocky planets around other stars, the observed depletion of moderately volatile elements in the spectra of both Sun-like and evolved stars -- suspected to have recently accreted rocky planetary materials \cite{Harrison2021, Spina2021, Liu2024} -- suggests that volatile loss is a universal process. Understanding the volatile loss process thus provides an important context for evaluating the compositional diversity and potential habitability of rocky exoplanets.

\section*{Methods}\label{sec11}

\subsection*{1. Bulk silicate compositions of Earth and Mars}
We have selected the major rock-forming lithophile elements (Mg, Si, Cr, Al, Ca, Ti, Mn, Na, K, and Zn), supplemented by a few of less abundant but thermally relevant lithophiles (Li, F, In, Cl, Br, and I) to cover a wide range of elemental volatility. We adopt the composition of bulk silicate Earth (i.e., Earth excluding its core) for those lithophiles from \cite{Wang2018}, which is calibrated from various geophysical, geochemical, and cosmochemical models \cite{Lyubetskaya2007, McDonough2008, Palme2014}. We note that the more recent estimates of bulk silicate Earth composition for these selected lithophiles \cite{McDonough2025} are consistent with our adopted composition (within uncertainties). For the composition of bulk silicate Mars (i.e., Mars excluding its core), we adopt the estimates from ref.~\cite{Yoshizaki2020} for the same list of lithophiles as above. The more recent estimates (but for fewer lithophiles) {for Mars} from ref.~\cite{Khan2022} are found to be consistent with the adopted estimates \cite{Yoshizaki2020}, except for Mn, Na, and Zn (these newer estimates are also indicated in Fig.~5). We have checked that if we had instead adopted these estimates of Mn, K, and Zn, then that does not change our best-{inferred} models. We refer to Supplementary Table~1 and \cite{Wang2026a} for the protosolar- and Al-normalized composition of those selected major rock-forming lithophile elements.

The abundances of heavy halogens (Cl, Br, and I) in CI chondrites were revised downward in \cite{Clay2017}, resulting in significantly higher values in the CI-normalized compositions of bulk silicate Earth and Mars (shown in light blue in Figs.~3 and 5). However, this downward revision has been put into question due to the use of a limited sample for the measurement as well as large inconsistency with previous measurements of various CI chondritic samples \cite{Palme2021b, Lodders2023}. It has also been suggested that since heavy halogens are highly incompatible during partial melting, their strong depletion on Earth may result from preferential loss through collisional erosion \cite{Campbell2012}. For these reasons, we have chosen to exclude Cl, Br and I from our quantification of the best models, but we show both sets of estimates for comparison in Figs.~3 and 5.

The abundances of some lithophiles suspected of slight siderophile or chalcophile tendencies, specifically Cr, Mn, and Zn, may not represent volatile depletions in bulk planets. We refer to Supplementary Information for our abundance correction for these three elements and the sensitivity test by excluding them (Supplementary Fig.~1).

\subsection*{2. Pressure-dependent elemental sublimation temperature}
Elemental volatility is essential to understanding volatile depletion, and it is pressure-dependent. We therefore need to compute the ambient pressure that pebbles experience within the envelope during pebble accretion. Although envelope pressures vary from the top to bottom of the envelope, it is the bottom pressure that is the highest and also determines whether materials falling into the envelope would be eventually accreted onto the growing planet or experience sublimation before accretion. Hence, we employ the envelope bottom pressure $P_{\rm env}$ (as well as temperature $T_{\rm env}$) that are calculated over the growth of an Earth-mass planet with a 1-D envelope model of pebble accretion described below.

Based on the pressure-dependent calculations of 50\% condensation temperatures of elements ($T_{\rm c}^{50}$; \cite{Lodders2025}) for a solar-composition gas that is equivalent to our H$_2$/He-dominated envelope gas, we interpolate the $T_{\rm c}^{50}$, as a proxy for $T_{\rm sub}$ in chemical equilibrium (see further discussion in Supplementary Information), for the specific pressure at which an element starts to sublimate and which is found where the pressure-dependent $T_{\rm c}^{50}$ crosses the $P_{\rm env}$--$T_{\rm env}$ profiles for different disk opacity conditions (as described below and Extended Data Fig.~1). We find that the $T_{\rm sub}$ values for an element as adopted here between the upper and lower limits of the opacity conditions are nearly identical (with the largest difference of $\sim$20 K for In; Extended Data Table~1; \cite{Wang2026a}). It is therefore reliable to adopt the mean of the $T_{\rm sub}$ values (with negligible error bars) within the range of opacity conditions as the proxy for elemental volatility for further model exploration. There is a small, systematic difference of $\sim$5 K in $T_{\rm sub}$ between values calculated for the plausible disk conditions at the locations of Earth and Mars. This small systematic shift of elemental volatility does not influence our model parameter exploration and thus is discarded for simplicity.

\subsection*{3. Volatile-depleted planetesimal model}
In our assumption of a volatile-depleted planetesimal model, we have adopted the lithophile composition of Vesta \cite{Mittlefehldt2015} and a logistic curve of volatile depletion fit for it with $T_{\mathrm c}^{50}$ of those lithophiles referred at the canonical pressure of $10^{-4}$ bar \cite{Sossi2022}. However, to include the composition trend of Vesta (i.e. retaining the similar volatile depletion levels) in our model parameter exploration self-consistently, we have re-fit a logistic curve to the protosolar-normalized Vesta lithophile composition ($f$) as a function of $T_{\rm sub}$ of those lithophiles (Extended Data Table~1). This yields a set of $T_{\rm sub}$-adapted inflection point ($T_0=1135\pm35$ K) and steepness parameter ($\sigma_0=62\pm24$ K) of the Vesta logistic model used in this work,
\begin{equation}
f = \frac{1}{2}\left[1 + {\rm erf}\left(\frac{T_{\rm sub}-T_0}{\sigma_0\sqrt{2}}\right)\right] \, ,
\label{eq:vesta}
\end{equation}
\noindent where ${\rm erf}$ is the Gauss error function, related to the cumulative distribution of a normal distribution. These values are nevertheless consistent with those ($T_0=1081\pm29$ K and $\sigma_0=57\pm17$ K) of ref.~\cite{Sossi2022}, in which they interpret these two parameters respectively as the nominal mid-plane temperature and width/extension of the feeding zone where the objects formed/accreted. The equivalent effect on volatile depletion of early-formed planetesimals has also been attributed to short-lived radiogenic heating (e.g., by $^{26}$Al and $^{60}$Fe; \cite{Eatson2024}) or to planetesimal collisions \cite{Halliday2001, Hin2017, Grewal2025}. Here, we are agnostic of the exact pathway(s) through which early-formed planetesimals are thermally processed. We choose to adopt the $T_{\rm sub}$-adapted Vesta-like logistic curve above as the baseline model for volatile-depleted planetesimals, while we also adjust $T_0$ and $\sigma_0$ broadly to extend this Vesta-specific logistic model to a hypothetical, collective model of volatile-depleted planetesimals and thus examine their effects on the accretion of both Earth and Mars. At an extreme case where both $T_0$ and $\sigma_0$ are set to near zero, the logistic function would mimic compositionally CI-like (non-volatile-depleted) planetesimals. In addition, we also test further in Supplementary {Information} (i) an alternative, exponential model for Vesta-like planetesimals ({Supplementary Fig.~9a}): $\log_{10}(f) = b(\frac{T_0^{\prime} - T_{\rm sub}}{T_0^{\prime}})$, where constants $b=-5.78\pm0.18$ and $T_0^{\prime}=1284\pm4$ -- the latter representing the kink point between the slope (depletion) and the horizontal line (i.e., non-depletion); (ii) a scenario without a model fit to the Vesta (HED meteoritic) data {(Supplementary Fig.~9b)}.

\subsection*{4. 1-D envelope model of pebble accretion}
The 1-D model used to calculate the $P_{\rm env}$--$T_{\rm env}$ profile at the bottom of the envelope (Fig.~2a and Extended Data Fig.~1) is based on analytical derivations described in detail in ref.~\cite{PisoYoudin2014}. The outer boundary $P$--$T$ conditions are given at the Hill radius $R_{\rm H} = [M_{\rm p}/(3 M_\odot)]^{1/3} r$ -- where $M_{\rm p}$ is the planetary mass, $M_\odot$ is the stellar mass and $r$ is the distance  of the planet from the star -- with the relevant conditions in the protoplanetary disk at 1 au ($T_{\oplus} \approx 120\,{\rm K} $ and $P_{\oplus} \approx 0.2\,{\rm Pa}$). The depth of the radiative-convective boundary (RCB), below which energy is transported purely by convection, depends on the luminosity of the planet and the opacity of the gas. We calculate the planet luminosity, assuming planetary growth with an e-folding time-scale of $\tau_{\rm acc}= 0.7\,{\rm Myr}$, as $L_{\rm p} = G M_{\rm p} \dot{M}_{\rm p}/R_{\rm p}$, where $\dot{M}_{\rm p}$ is the mass accretion rate ($M_{\rm p}/\tau_{\rm acc}$) and $R_{\rm p}$ is the radius of the protoplanet. The opacity in the envelope is given by $\kappa = \kappa_{\rm d} (T_{\rm env}/T_{\rm disk})^2$, where $T_{\rm disk}=T_{\oplus}(r/{\rm au})^{-3/7}$ \cite{Ida2016} and $\kappa_{\rm d}=[0.1, \,1]\,{\rm m^2\,kg^{-1}}$ are the temperature and opacity conditions in the surrounding protoplanetary disk, with the opacity arising mainly from sub-micron-sized dust grains as in ref.~\cite{PisoYoudin2014}. Knowing the temperature and pressure at the RCB, we then construct an adiabatic solution for the convective envelope -- with mean molecular weight and adiabatic index from a solar mixture of H$_2$ and He -- down to its bottom (that is in touch with the surface of the planet). The metallicity and opacity of the envelope gas could be increased from dust and recondensed MLVEs released from the infalling pebbles \cite{Brouwers2020}. We have experimented with our model by increasing the opacity parameter $\kappa_{\rm d}$ by a factor of 3 (and even higher) over the unity value. The resultant $P_{\rm env}$--$T_{\rm env}$ profile does not change significantly (Extended Data Fig.~1), due to the nearly-adiabatic temperature profile. The $T_{\rm sub}$ values of elements only weakly depend on pressure (Extended Data Fig.~1), so a small increase in metallicity will only increase those values monotonically and limited to a few of tens of K \cite{Timmermann2023, Zaveri2026}. Therefore, we do not expect it will affect our results and conclusions.
We have also experimented with higher values of the ambient disk temperature, to represent conditions interior to the water ice line, and found similar results. 

Following previous work \cite{Steinmeyer2023, HabibPierrehumbert2024}, we assume that an inner radiative zone (IRZ) emerges following the rapid building-up of a SiO-dominated vapor at the innermost envelope, where $T_{\rm env}$ reaches the sublimation temperature ($\gtrsim 1450$ K; Extended Data Table~1) of silicates in accreted pebbles at the conditions of the innermost envelope. This IRZ (i.e., the saturated SiO-dominated cloud layer) prevents sublimated refractory materials (Si, Mg, as well as more refractory elements) from diffusing to the upper envelope and hence from loss by the recycling flow \cite{KurokawaTanigawa2018, Popovas2018, Wang2023, Kuwahara2026}. Before pebbles reach the IRZ, they will have already sublimated their MVLEs including Li and Mn (with $T_{\rm sub}\lesssim 1320$ K; Extended Data Table~1). Given this physical separation between the sublimation of the moderately volatile elements and the IRZ, we adopt a characteristic temperature $\sim$$1400$ K to represent the transition between MVLEs (thermally processed and diffused in the envelope) and refractory elements (protected by the IRZ). This results in the jump in the volatile-depletion curve between moderately volatile Li and Mn and refractory Mg and Si in the Earth (Fig.~3a). 

We nevertheless caution that loss of moderately refractory elements (Si, Mg and Cr) is physically plausible if early-accreted pebbles experience some degree of sublimative mass loss before a protective IRZ builds up \cite{Steinmeyer2023, HabibPierrehumbert2024}, since 10--20\% depletion of this group of elements has been found in the terrestrial planet \cite{Wang2019, Wang2019b} as well as in potentially rocky exoplanetary materials inferred from polluted white dwarfs \cite{Aguilera-Gomez2024}. Our pebble-devolatilization model is not yet precise enough to capture this transient regime. Alternatively, such a depletion may also be explained by accretion of more volatile-depleted planetesimals (like angrite parent bodies \cite{Hu2022}). We further discuss the effect of inclusion/exclusion of this group of elements on our model-fitting outcome in Supplementary Information.

\subsection*{5. 3-D convection simulations in the envelope}
We perform 3-D hydrodynamical simulations of a convective gas envelope (Fig.~2b) using the \texttt{Athena++} magnetohydrodynamic code \cite{Stone2020}, which uses a high-order Godunov scheme for solving the fluid dynamics. Volatiles released in the envelope are treated as passive particles that move with the local velocity of the gas. These particles are implemented using the Lagrangian tracer particles module based on the particle-mesh method \cite{HuLi2022, Baronett2024}. When studying the volatile loss process, particles that cross the convection-recycling boundary $R_{\rm cyc}$ are manually removed to mimic the recycling flow. This approach allows us to explore volatile loss across different values of $R_{\rm cyc}$ (and thus varied diffusion timescales; {Extended Data} Fig.~2). 

Our setup generally follows the ``isolated spherical envelope'' case of ref.~\cite{Zhu2021}, where convection is driven by accretion heating, which is distributed throughout the envelope to follow a prescribed pebble accretion luminosity profile $(1-R_{\rm in}/R)L_{\rm p}$, where $R_{\rm in}$ is the inner boundary radius of the simulation. $L_{\rm p}$ is again the planet luminosity, which can be calculated with the mass accretion timescale $\tau_{\rm acc}$.
Instead of radiative cooling assumed in ref. \cite{Zhu2021}, we adopt Newtonian cooling, which damps the gas temperature beyond $R_{\rm B}$ to the ambient disk temperature $T_{\rm disk}$ on a cooling timescale of $0.16$ years, which is on the order of the disk dynamical timescale. Since the convective envelope is typically optically thick around 1 au region \cite{Kuwahara2026}, radiative cooling is negligible, making this approach a reasonable and computationally efficient approximation {of} the cooling process. We confirm its robustness by verifying that the thermal structure and flow velocity patterns in the envelope remain consistent with simulations using full radiative transfer. 

Our simulations consider the envelope around a planet of mass $M_{\rm p}=M_{\oplus}$, located at 1 au from a solar-mass star. We note that the convection speed is not expected to vary much as a function of planet mass (from $0.1M_{\oplus}$ to $3M_{\oplus}$; \cite{Johansen2020}). The ambient disk temperature at 1 au is set to be $T_{\rm disk} \approx 250\,{\rm K}$ and the midplane gas density is set as $\approx$$10^{-9}\,{\rm g}\,{\rm cm}^{-3}$. The pebble accretion luminosity is determined by an accretion timescale of $\tau_{\rm acc}=10^6\, {\rm yr}$. We note that this disk temperature is higher than that assumed in the 1-D model, and the accretion timescale is slightly longer than in the 1-D model. However, these descrepancies are not expected to significantly impact the results. If any effect occurs, our adopted higher disk temperature and longer accretion timescale likely result in weaker convection, thus lowering volatile loss efficiency, which makes our results a conservative estimate. Our simulation domain covers from $R_{\rm in}=7R_{\oplus}(=0.1R_{\rm B})$ to $R_{\rm out}=2111R_{\oplus}(=30R_{\rm B})$, with 140 radial grid cells uniformly distributed in logarithmic space. The domain is discretized with 160 cells on the azimuthal direction and 40 cells on the poloidal direction. This ensures that the convective flow, with a characteristic lengthscale of $R_{\rm B}$, is well-resolved. Our simulations incorporate a total number of $\sim$$9\times 10^5$ tracer particles, providing robust statistic sampling of the volatile dynamics.

\subsection*{6. Pre-depletion model of accreted pebbles}
We test plausible levels of pre-depletion in the accreted pebbles by scaling the composition of the pebbles in MVLEs between proto-solar composition and H-type ordinary chondrite composition (H-chondrites are one of the most volatile-depleted chondrite groups in the inner protoplanetary disk \cite{Lodders2021}). To do so, we first fit a common exponential model to the H-chondrite composition $f$ relative to Al and proto-Sun,
\begin{equation}
  \log_{10} f = b^{\rm H}\cdot\frac{T_0^{\rm H} - T_{\rm sub}}{T_0^{\rm H}} \, ,
\label{eq:pebble_H_model}
\end{equation}
where $T_0^{\rm H}$ describes the temperature below which the H-chondrites are volatile-depleted (we assume solar composition above this threshold temperature) and $b^{\rm H}$ describes the steepness of the depletion. We find $b^{\rm H} = -3.12$ and $T_0^{\rm H}=997$ K for the H-chondrite depletion model. Different levels of pre-depletion ($b$) of accreted pebbles are then scaled to $b^{\rm H}$ -- namely, $b=[0, 1]b^{\rm H}$, while fixing at $T_0^{\rm H}$. We note that alternative models such as polynomial and power-law forms could also be fit to H chondrite data, but they are all consistent with one another within the uncertainties of the fits \cite{Fischer2025}. Since here we only use the H-chondrite model as baseline for scaling various levels of volatile pre-depletion in pebbles, the exact form and its uncertainty is not important. For a broader discussion of pre-depleted pebbles, we refer to the Supplementary Information.

\subsection*{7. Model parameter estimation}
{\textit{Reduced-$\chi^2$ minimization}}
\\ \noindent
To assess the fit between the modeled and observed lithophile element abundances, we first employ a reduced chi-squared ($\chi^2$) minimization approach \cite{Wang2026b}. For each element, we compare the observed value to the model prediction at the corresponding temperature or composition by identifying the closest match in the model grid of relative fractions of different components -- pebble-accreting bodies (protoplanet and impactor) and Vesta-like planetesimals. The $\chi^2$ contribution for each element is computed in linear space. Asymmetric data uncertainties are conservatively handled by adopting the larger of the upper and lower bounds. The total $\chi^2$ is then summed across all elements for fitting and normalized by the degrees of freedom (set to be the number of fit elements, $N$) to yield the reduced statistic ($\chi^2/N$). We did not consider the number of model parameters for simplicity, and this metric provides only a first-order quantitative measure of the goodness-of-fit. We have also made an alternative computation of $\chi^2/N$ in the logarithmic space and found that the solutions agree with those found in the linear space. \\

\noindent 
{\textit{Probabilistic inference via Bayesian nested sampling}}
\\ \noindent
To quantify the compositional contributions of various planetary building blocks, we also employ a Bayesian inference framework \cite{Wang2026b} using nested sampling via the \texttt{dynasty} Python package, specifically employing its dynamic nested sampling mode. This approach allows for efficient exploration of complex posterior distributions and simultaneous estimation of the Bayesian evidence. Compared to traditional Markov Chain Monte Carlo (MCMC) methods, nested sampling is particularly advantageous for problems involving multi-modal posteriors, strong parameter correlations, or pronounced degeneracies \cite{Skilling2006} -- all of which are relevant to the inference problem addressed here. A set of priors (Extended Data Table~2) are applied to perform the Bayesian inference. The initial values of three parameters controlling the dynamic sampling efficiency and stop criteria are set moderately: \texttt{nlive\_init}=600, \texttt{nlive\_batch}=300, and \texttt{dlogz}=0.05, which are kept the same for all cases.

Based on the posterior distributions of the Bayesian inference, the [16th, 50th, 84th] percentiles of individual model parameters are calculated and taken as the summary of the parameter space (Table~1). To report the strength of Bayesian evidence for model comparison, we adopt the following criteria of Jeffreys' scale \cite{jeffreys1998theory, Trotta2008} on the absolute Bayes factor ($|\Delta\ln Z|$):
\begin{itemize}
    \item[] 0--1 \,\,\, `Not worth mentioning (insignificant)'
    \item[] 1--2.5 `Substantial'
    \item[] 2.5--5 `Strong'
    \item[] $>5\,$ \,\,\,\,`Decisive'
\end{itemize}

For clarity in describing whether a model is disfavored or not relative to the reference models, the true (instead of absolute) value of a Bayes factor is often used, while the strength of Bayesian evidence is measured against the criteria above.

\subsection*{Data availability}
All data supporting the findings of this study are available within the paper and its Supplementary Information. In addition, the source data reproducing the data figure (Figure 1) are also provided in a spreadsheet as Source Data alongside the paper and available via figshare at \url{https://doi.org/10.6084/m9.figshare.33085055} (ref. \cite{Wang2026a}). 

\subsection*{Code availability}
The codes essential to reproduce the work are available via Code Ocean at \url{https://doi.org/10.24433/CO.8499849.v1} (ref. \cite{Wang2026b}). 


\subsection*{Acknowledgements}
We thank B. Fegley and K. Lodders for discussing pressure-dependent condensation temperatures, H. St.C. O'Neill, W. F. McDonough, and P. A. Sossi for Earth and Mars compositions, and Y. Tian for Bayesian inference principles. 

\subsection*{Funding Statement}
H.S.W, A.J., Z.X. were supported by the Carlsberg Foundation’s Semper Ardens grant (`FIRSTATMO’). C.C.Y. acknowledges the support from NASA via the Emerging Worlds program (\#80NSSC23K0653), the Astrophysics Theory Program (\#80NSSC24K0133), and the Theoretical and Computational Astrophysical Networks (\#80NSSC21K0497). Z.Z. acknowledges support from NASA XRP grant (\#80NSSC25K7144). D.S.L. acknowledges support from the Kephalos Research Fund. 

\subsection*{Author contributions}
Conceptualization: H.S.W. and A.J.; Formal analysis: H.S.W., A.J. and Z.X.; Methodology: H.S.W., A.J., M.-L.S., Z.X., C.-C.Y. and Z.Z.; Resources: A.J., M.L., E.v.K., D.S.L. and M.B.; Validation: all authors; Visualization: H.S.W. and Z.X.; Writing – original draft: H.S.W., A.J. and Z.X.; Writing – review \& editing: all authors.

\subsection*{Competing interests}
The authors declare no competing interest.

\subsection*{Additional information}


\noindent\textbf{Correspondence and requests for materials} should be addressed to Haiyang S. Wang and Anders Johansen.


\clearpage
\begin{figure}[htbp]
\centering
\includegraphics[width=1.0\textwidth]{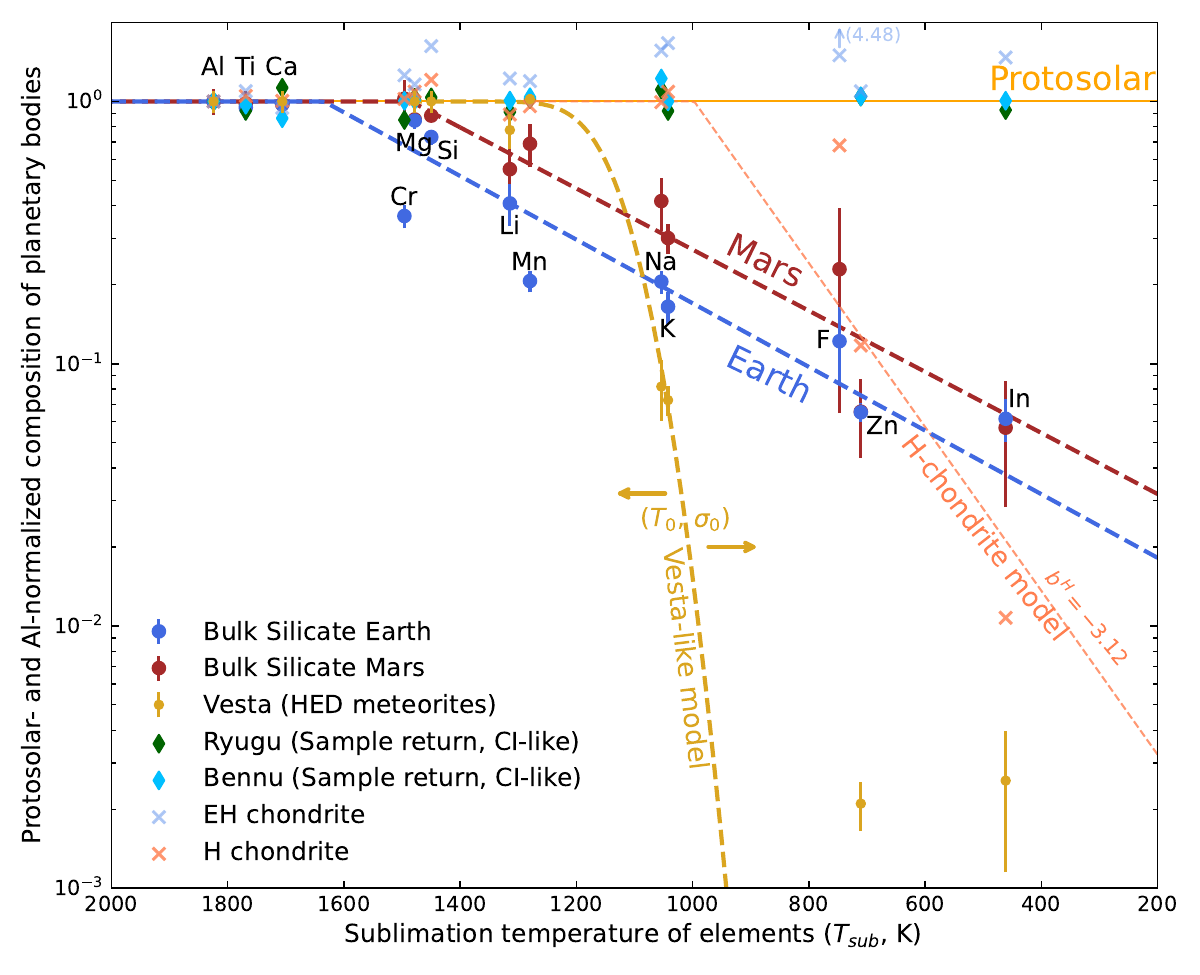} 
\caption{\textbf{Lithophile element compositions of Solar System rocky bodies as a function of elemental volatility.} Protosolar- and Al-normalized composition of major rock-forming lithophile elements in bulk silicate Earth \cite{Wang2018}, bulk silicate Mars \cite{Yoshizaki2020}, the asteroid Vesta (based on howardite–eucrite–diogenite, HED, meteorites \cite{Mittlefehldt2015}), two asteroid sample returns (Ryugu \cite{Nakamura2022} and Bennu \cite{Lauretta2024}), as well as EH and H chondrites \cite{Lodders2021} (Supplementary Table 1) are sorted as a function of elemental sublimation temperature (proxy for elemental volatility; Extended Data Table~1). Compositional data for Earth, Mars, and Vesta are presented as mean values $\pm 1\sigma$ SD, while those for Ryugu, Bennu, and EH and H chondrites are presented without error bars.  The protosolar composition (normalized as $\equiv$1; \cite{Wang2019}) is denoted by the horizontal solid line (orange). The volatile-depletion trends for Earth and Mars (dashed lines in blue and red, respectively) are illustrative only. A logistic function with two variables ($T_0$ and $\sigma_0$; Eq. \ref{eq:vesta}) recommended in ref.~\cite{Sossi2022} is used to fit the composition of Vesta as a function of $T_{\rm sub}$ (yellow dashed line). The apparent deviations of Zn and In from the Vesta curve likely caused by their recondensation \cite{Fang2024}. The red dash line indicates an exponential model of H chondrites (with its slope $b^H=-3.12$; Eq. \ref{eq:pebble_H_model}), based on which we will scale various degrees of volatile pre-depletion in pebbles.}
\label{fig:composition_compare}
\end{figure}


\begin{figure}[htbp!]
\centering
\includegraphics[width=1.0\textwidth]{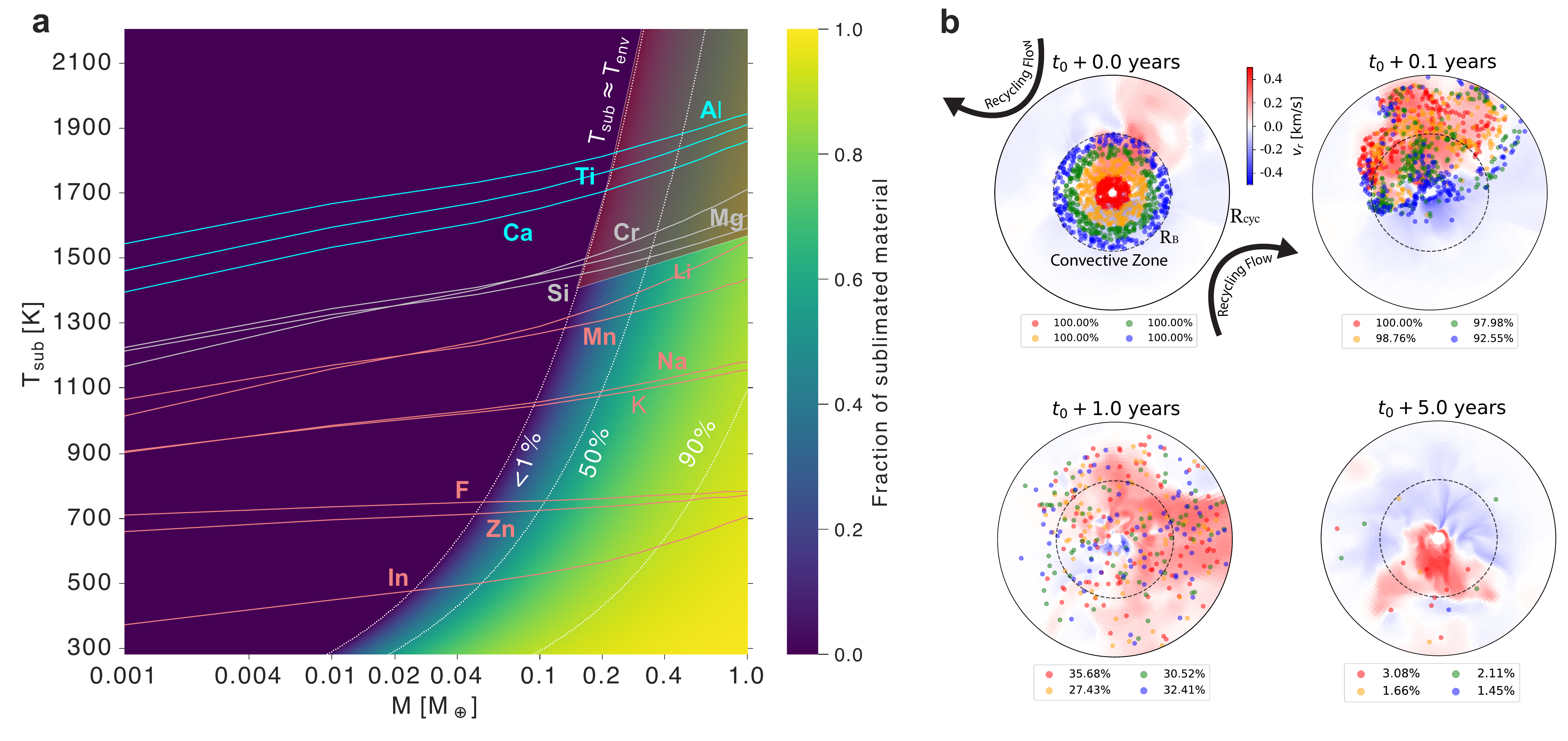}
\caption{\textbf{Devolatilization during pebble accretion}. ({\bf a}) Sublimation of major rock-forming lithophile elements in the envelope of a protoplanet that grows through pebble accretion. The color map represents the cumulative fraction of sublimated material as a function of the mass of the protoplanet ($M$, relative to Earth mass, $M_{\oplus}$) and elemental sublimation temperature ($T_{\rm sub}$), with the `$<$1\%', `50\%', and `90\%' fractions shown in dotted curves for guidance. The colored lines indicate $T_{\rm sub}$ for selected elements. For moderately volatile lithophiles (red: Li, Mn, Na, K, F, Zn, In), their sublimation and loss back to the protoplanetary disk is assumed to be 100\% efficient after their $T_{\rm sub}$ is overtaken by the temperature at the bottom of the envelope ($T_{\rm env}$; along the `$<$1\%' dotted curve). Major refractory lithophiles (light blue: Al, Ti, and Ca) and moderately refractory lithophiles (gray: Mg, Si and Cr), even if sublimated, are assumed to be retained as vapor in equilibrium with the surface magma ocean, protected from loss by a radiative zone (indicated by a shadow in light brown on the top right of the panel).
({\bf b}) 3-D hydrodynamical simulations of the temporal evolution of embedded particles (representing sublimated materials) in the convective envelope of a protoplanet during pebble accretion. Particles are initially placed in the spatial intervals 0.1-0.3 $R_{\rm B}$ (red), 0.3-0.6 $R_{\rm B}$ (yellow), 0.6-0.8 $R_{\rm B}$ (green), and 0.8-1 $R_{\rm B}$ (blue) (top-left panel in {\bf b}), where $R_{\rm B}$ is the Bondi radius (indicated by a dashed circle), beyond which the temperature is damped to that of the ambient disk and convection thus suppressed. The transition to the recycling region $R_{\rm cyc}$, where particles escape back to the protoplanetary disk, is assumed to be at $2 R_{\rm B}$. The gas radial velocity $v_{\rm r}$ in the envelope is illustrated with red and blue indicating upward and downward flows, respectively. The embedded particles have largely vanished by five years after their release (bottom-right panel in {\bf b}).}
\label{fig:sublimation_loss}
\end{figure}


\begin{figure} [htbp]
\centering
\includegraphics[width=1.0\textwidth]{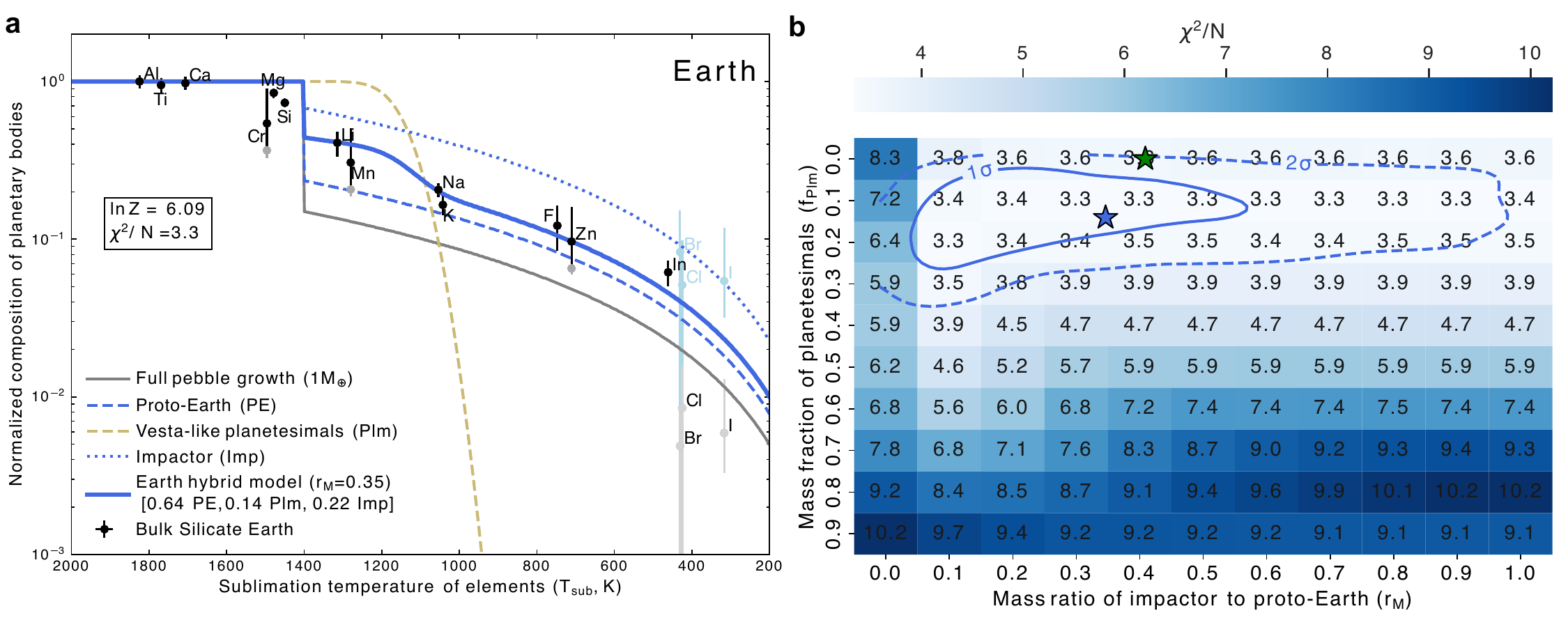} 
\caption{\textbf{Volatile depletion reveals hybrid accretion of Earth.} ({\bf a}) Earth's volatile depletion pattern (solid curve in blue) resulting from the median (central-estimate) solution ([0.64 \textbf{PE}, 0.14 \textbf{Plm}, 0.22 \textbf{Imp}]) of the Bayesian inference of the model combining proto-Earth (\textbf{PE}, dashed curve in blue) and Impactor (\textbf{Imp}, dotted curve in blue) that grew via pebble accretion, together with a population of Vesta-like planetesimals (\textbf{Plm}, dashed curve in yellow). The solution is constrained by the protosolar- and Al-normalized composition of major rock-forming lithophile elements (dots in black) in Earth as a function of elemental sublimation temperature (Extended Data Table~1). The compositional data are presented as mean values $\pm 1\sigma$ SD. The Bayesian evidence ($\ln Z$) and (post-calculated) reduced-$\chi^2$ ($\chi^2/N$, where $N$ is the number of major rock-forming lithophile elements adopted for the inference, 13 in total) are shown in the box for reference. Lithophiles with a suspected siderophile or chalcophile tendency (Cr, Mn, and Zn) have been corrected (with uncorrected values shown in gray; see Supplementary Information and Supplementary Fig.~1 for a sensitivity test). 
Heavy halogens (Cl, Br and I) are not included in the inference, due to their disputed abundances in CI chondrites (with two sets of data shown in gray and light blue, respectively). ({\bf b}) Color-coded grid of the reduced-$\chi^2$ as a function of the mass ratio of impactor to proto-Earth ($r_M$) and the mass fraction of planetesimals ($f_{\mathrm{Plm}}$). It is superimposed by the 1$\sigma$ and 2$\sigma$ contours of the posterior distributions of the Bayesian inference of the {\bf PE+Plm+Imp} model. The central-estimate solutions of the two most favorable scenarios for Earth's hybrid accretion ({\bf PE+Plm+Imp} and {\bf PE+Imp}; Table~1) are denoted by star signs in blue and green, respectively.
} 
\label{fig:earth_bestmodel}
\end{figure}

\begin{figure}[htbp!]
\centering
\includegraphics[width=1.0\textwidth]{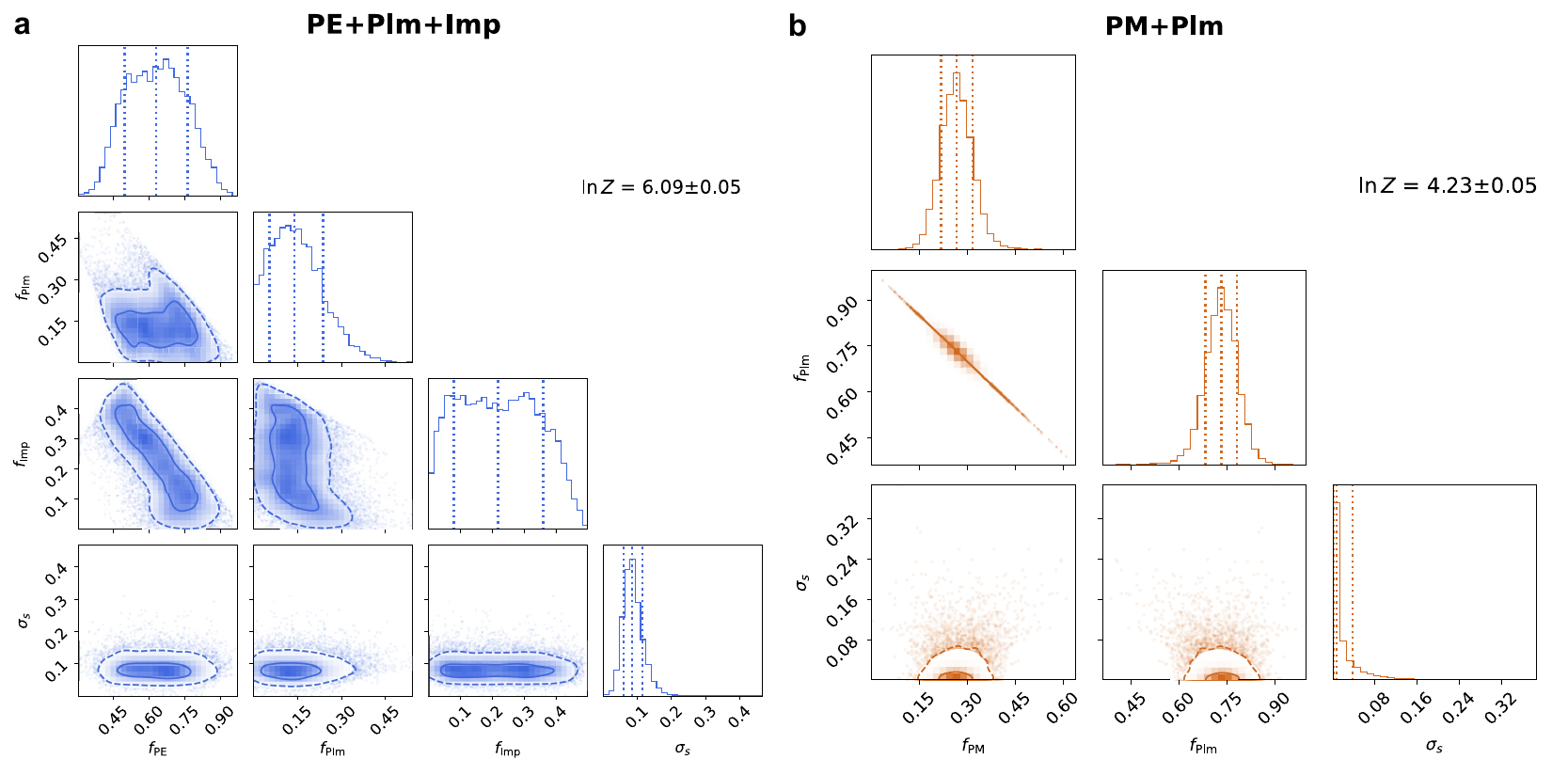}
\caption{\textbf{Bayesian posterior distributions of formation model parameters for Earth and Mars.} Panels a and b respectively represent their reference models: \textbf{PE+Plm+Imp} (Earth) and \textbf{PM+Plm} (Mars). The model parameters consists of individual component fractions ($f$) and an inference intrinsic scatter ($\sigma_s$). The 1$\sigma$- and 2$\sigma$-contours are shown in solid and dashed curves on each corner. On the 1D-histograms of individual parameters along the diagonal panels, the [16th, 50th, 84th] percentiles are shown in dotted lines (with the 50th percentile reported as the median and 16th-84th percentiles as the error bars of those quantities). The Bayesian evidence ($\ln Z$) with 1$\sigma$ uncertainty is also shown for each case.}
\label{fig:earth_mars_bayesian}
\end{figure}


\begin{figure} [htbp]
\centering
\includegraphics[width=1.0\textwidth]{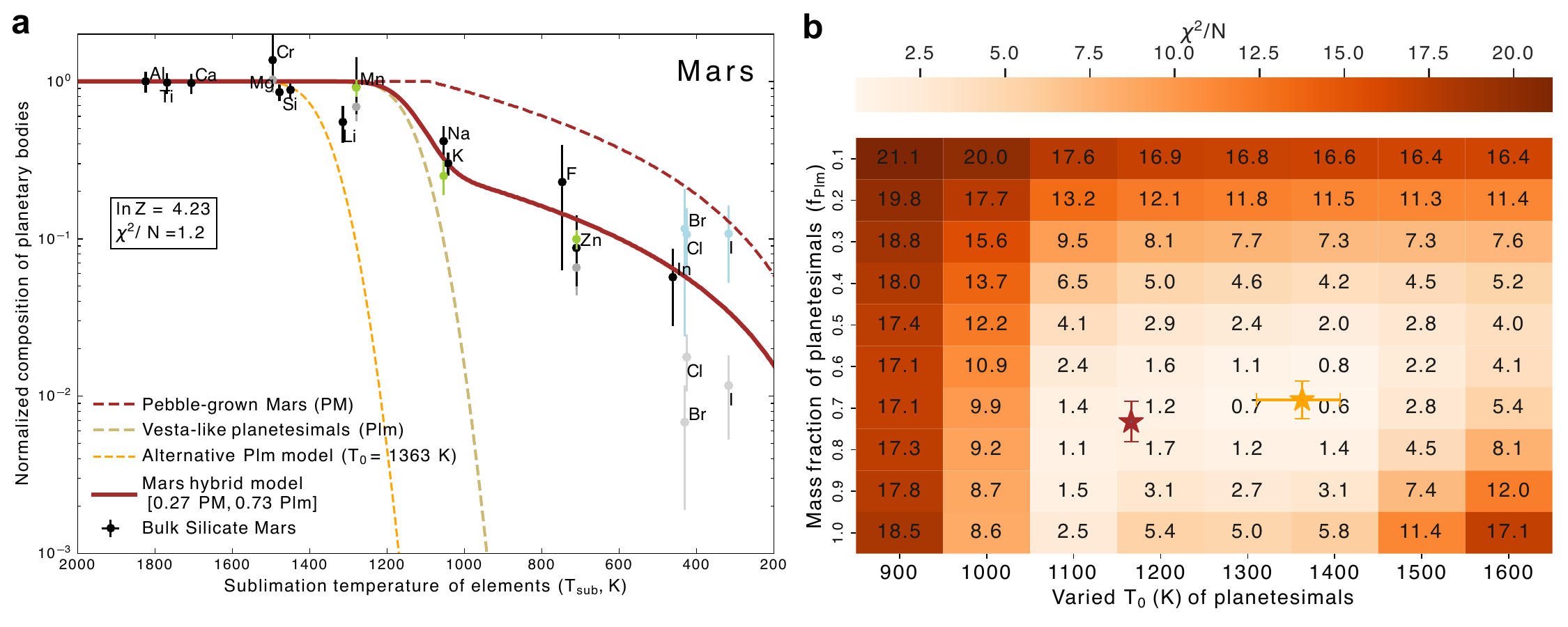} 
\caption{\textbf{Volatile depletion reveals hybrid accretion of Mars.} ({\bf a}) Mars' volatile depletion pattern (solid curve in red) resulting from the median (central-estimate) solution ([0.27 \textbf{PM}, 0.73 \textbf{Plm}]) of the Bayesian inference of the model combining a pebble-grown Mars component (\textbf{PM}, dashed curve in red) and a population of Vesta-like planetesimals (\textbf{Plm}, dashed curve in yellow). An alternative planetesimal model of higher $T_0$ is shown for comparison. The compositional data are presented as mean values $\pm 1\sigma$ SD. The Bayesian evidence ($\ln Z$) and (post-calculated) reduced-$\chi^2$ are shown for reference. Similar to Earth's case, the abundances of Cr, Mn, and Zn are corrected for their potential partition into the core and heavy halogens (Cl, Br and I) are not included in the inference. The recent update of the Martian mantle composition (green dots; \cite{Khan2022}) is consistent within error bars of \cite{Yoshizaki2020}, with major differences mainly for Mn, Na and Zn. ({\bf b}) Color-coded grid of reduced-$\chi^2$ as a function of the inflection point ($T_0$) of a logistic model of volatile-depleted planetesimals ($x$-axis) and of the relative mass contribution of such planetesimals ($f_{\mathrm{Plm}}$; $y$-axis). For guidance, $f_{\mathrm{Plm}}$ (=$0.73{\pm}0.05$) for the reference model (panel A) with a Vesta-like composition (with $T_0=1135\pm35$ K) is shown in the star sign (with 16th-84th percentile error bars) in brown. Given a varied $T_0$ (and thus varied volatile depletions) of early-formed planetesimals, the alternative solution is found at $T_0=1363^{+43}_{-52}$ K and $f_{\mathrm{Plm}}=0.68\pm0.05$ (the orange star sign with 16th-84th percentile error bars; for more details see Table~1 and Extended Data Fig.~7 and Supplementary Fig.~6). 
} 
\label{fig:mars_bestmodel}
\end{figure}

{\renewcommand{\arraystretch}{1.25}
\begin{table}[htbp!]
\centering
\caption{\textbf{Summary of Bayesian inference results of various hybrid-accretion scenarios of both Earth and Mars.}}
\begin{tabular}{llllll}
\hline
\rowcolor{gray!20}
& \textbf{Model} & $\bm{[f_0, ..., f_i]}$ & 
$\bm{\chi^2/N}$  &$\bm{\ln Z}$ & $\bm{\Delta\ln Z}$ \\ 
\hline
\multirow{11}{*}{\rotatebox{90}{\large{\textbf{Earth}}}} &
\multirow{1}{*}{\bf PE+Plm+Imp}  & $[0.64^{+0.12}_{-0.14}, \, 0.14^{+0.10}_{-0.09}, \, 0.22^{+0.14}_{-0.15}]$ & 
\multirow{1}{*}{3.3} &
\multirow{1}{*}{$6.09\pm0.05$} & \multirow{1}{*}{Reference} \\
& \multirow{1}{*}{\bf PE+Imp}  & $[0.71\pm0.14, \, 0.29\pm0.14]$ &
\multirow{1}{*}{3.6}  &
\multirow{1}{*}{$6.16\pm0.04$}               & \multirow{1}{*}{$0.07\pm0.06$} \\
& \multirow{1}{*}{\bf PE+Plm}  & $[0.73\pm0.11, \, 0.27\pm0.11]$ &
\multirow{1}{*}{5.9}  &
\multirow{1}{*}{$4.78\pm0.05$} & \multirow{1}{*}{$-1.31\pm0.07$} \\
& \multirow{1}{*}{\bf CI+Plm}  & $[0.12^{+0.10}_{-0.08}, \, 0.88^{+0.08}_{-0.10}]$ &
\multirow{1}{*}{$11.2$}  &
\multirow{1}{*}{$-5.22\pm0.05$} & \multirow{1}{*}{{$-11.31\pm0.07$}} \\
& \multirow{1}{*}{\bf PE+CI+Plm}  & $[0.74^{+0.09}_{-0.10}, \, 0.06\pm0.04, \, 0.20^{+0.11}_{-0.10}]$ &
\multirow{1}{*}{$3.6$}  &
\multirow{1}{*}{$4.02\pm0.06$} & \multirow{1}{*}{{$-2.07\pm0.07$}} \\ 
& \multirow{1}{*}{\bf {PE+CI+Plm+Imp}}  & {$[0.66^{+0.11}_{-0.16}, \, 0.03^{+0.04}_{-0.02}, \, 0.12^{+0.10}_{-0.08}, \, 0.19^{+0.08}_{-0.15}]$} &
\multirow{1}{*}{{$3.9$}}  &
\multirow{1}{*}{{$4.31\pm0.06$}} & \multirow{1}{*}{{$-1.78\pm0.08$}} \\ 
& \multirow{1}{*}{\bf {PE+Plm+DualImp}}  & {$[0.61^{+0.13}_{-0.15}, \, 0.10^{+0.09}_{-0.07}, \, 0.25^{+0.11}_{-0.15}, \, 0.04^{+0.08}_{-0.03}]$} & 
\multirow{1}{*}{{$3.8$}} &
\multirow{1}{*}{{$5.55\pm0.05$}} & \multirow{1}{*}{{$-0.54\pm0.06$}} \\ 
& \multirow{1}{*}{\bf {PE+Plm+TriImp}}  & \multirow{1}{*}{\parbox[t]{5.5cm}{{$[0.60^{+0.12}_{-0.16}, \, 0.08^{+0.09}_{-0.06}, \, 0.25^{+0.09}_{-0.15}, \, 0.05^{+0.08}_{-0.03},$ $\, 0.01^{+0.02}_{-0.01}]$}} } & 
\multirow{1}{*}{{$5.2$}} &
\multirow{1}{*}{{$4.32\pm0.06$}} & \multirow{1}{*}{$-1.77\pm0.07$} \\ 
& \multirow{1}{*}{}  & \multirow{1}{*}{} & 
\multirow{1}{*}{} &
\multirow{1}{*}{} & \multirow{1}{*}{} \\
& \multirow{1}{*}{\bf {PE+Plm+Imp ($\mathbf{b^H}$)}}  & \multirow{1}{*}{{$[0.64^{+0.13}_{-0.15}, \, 0.14^{+0.11}_{-0.09}, \, 0.22^{+0.14}_{-0.15}]$}} & 
\multirow{1}{*}{$5.9$} &
\multirow{1}{*}{$5.03\pm0.05$} & \multirow{1}{*}{{$-1.06\pm0.06$}} \\
& \multirow{1}{*}{\bf {PE+Plm+Imp ($\mathbf{0.5b^H}$)}}  & \multirow{1}{*}{{$[0.64^{+0.13}_{-0.14}, \, 0.14^{+0.10}_{-0.09}, \, 0.22^{+0.14}_{-0.15}]$}} & 
\multirow{1}{*}{{$5.3$}} &
\multirow{1}{*}{{$5.32\pm0.05$}} & \multirow{1}{*}{{$-0.77\pm0.06$}} \\
& \multirow{1}{*}{\bf {PE+Plm+Imp ($\mathbf{0.1b^H}$)}}  & \multirow{1}{*}{{$[0.64^{+0.12}_{-0.14}, \, 0.14^{+0.10}_{-0.09}, \, 0.22\pm0.14]$}} & 
\multirow{1}{*}{{$3.8$}} &
\multirow{1}{*}{{$6.01\pm0.05$}} & \multirow{1}{*}{{$-0.08\pm0.06$}} \\
\hline
\multirow{5}{*}{\rotatebox{90}{\large{\textbf{Mars}}}} &
\multirow{1}{*}{\bf PM+Plm}  & $[0.27\pm0.05, \, 0.73\pm0.05]$ & 
\multirow{1}{*}{1.2} &
\multirow{1}{*}{$4.23\pm0.05$}               & \multirow{1}{*}{Reference} \\
& \multirow{1}{*}{\bf {CI+Plm}}  & $[0.15^{+0.07}_{-0.05}, \, 0.85^{+0.05}_{-0.07}]$ & 
\multirow{1}{*}{{$2.4$}} &
\multirow{1}{*}{{$0.12\pm0.05$}}               & \multirow{1}{*}{{$-4.11\pm0.06$}} \\
& \multirow{1}{*}{\bf {PM+CI+Plm}}  & {$[0.22^{+0.06}_{-0.07}, \, 0.03^{+0.04}_{-0.02}, \, 0.75^{+0.05}_{-0.06}]$} & 
\multirow{1}{*}{{$1.3$}} &
\multirow{1}{*}{{$1.86\pm0.07$}}               & \multirow{1}{*}{{$-2.37\pm0.09$}} \\
& {\bf PM+Plm (var. $T_0$)}   & \multirow{1}{*}{$[0.32\pm0.05, \, 0.68\pm0.05]$} & 
\multirow{1}{*}{0.6}  &
\multirow{1}{*}{$5.52\pm0.06$}               & \multirow{1}{*}{$1.29\pm0.08$} \\
& {\bf PM+Plm (var. $T_0$\&$\sigma_0$)}  & \multirow{1}{*}{$[0.21^{+0.09}_{-0.11}, \, 0.79^{+0.11}_{-0.09}]$} & 
\multirow{1}{*}{0.2} &
\multirow{1}{*}{$6.81\pm0.07$}               & \multirow{1}{*}{$2.58\pm0.09$} \\
\bottomrule
\multicolumn{6}{p{15cm}}{\footnotesize Notation $\bm{[f_0, ..., f_i]}$ represents the relative mass fractions (with 50th percentile as the median and 16th--84th percentiles as the error bars) of accreted components of a planet, listed in the same order as their appearance in the named models -- e.g., \textbf{PE} (proto-Earth)+\textbf{Plm} (Planetesimals)+\textbf{Imp} (Impactor) as the reference model for Earth and \textbf{PM} (pebble-grown Mars component)+\textbf{Plm} as the reference model for Mars. {\bf CI} represents a population of volatile-rich CI-like planetesimals. \textbf{DualImp} and \textbf{TriImp} represent dual and triple impactors, respectively. $\mathbf{b^H}$, $\mathbf{0.5b^H}$, and $\mathbf{0.1b^H}$ indicate different levels of pre-depletion of accreted pebbles (Methods). Please refer to Methods for calculation details of the reduced-$\chi^2$ ($\bm{\chi^2/N}$) and of Bayesian inference including the logarithmic Bayesian evidence ($\bm{\ln Z}$) and Bayes factor ($\bm{\Delta\ln Z} = \ln Z - \ln Z_{\rm ref}$, where 'ref' denotes the adopted reference model correspondingly), as well as the definitions of $T_0$ and $\sigma_0$ of the adopted volatile-depleted planetesimal model.}
\end{tabular}
\label{tab:models_main_summary}
\end{table}
}

\clearpage

\newpage
\onecolumn
\renewcommand{\figurename}{Extended Data Fig.}
\renewcommand{\tablename}{Extended Data Table}

\begin{figure}[hbt!]
\centering
\includegraphics[width=0.9\textwidth]{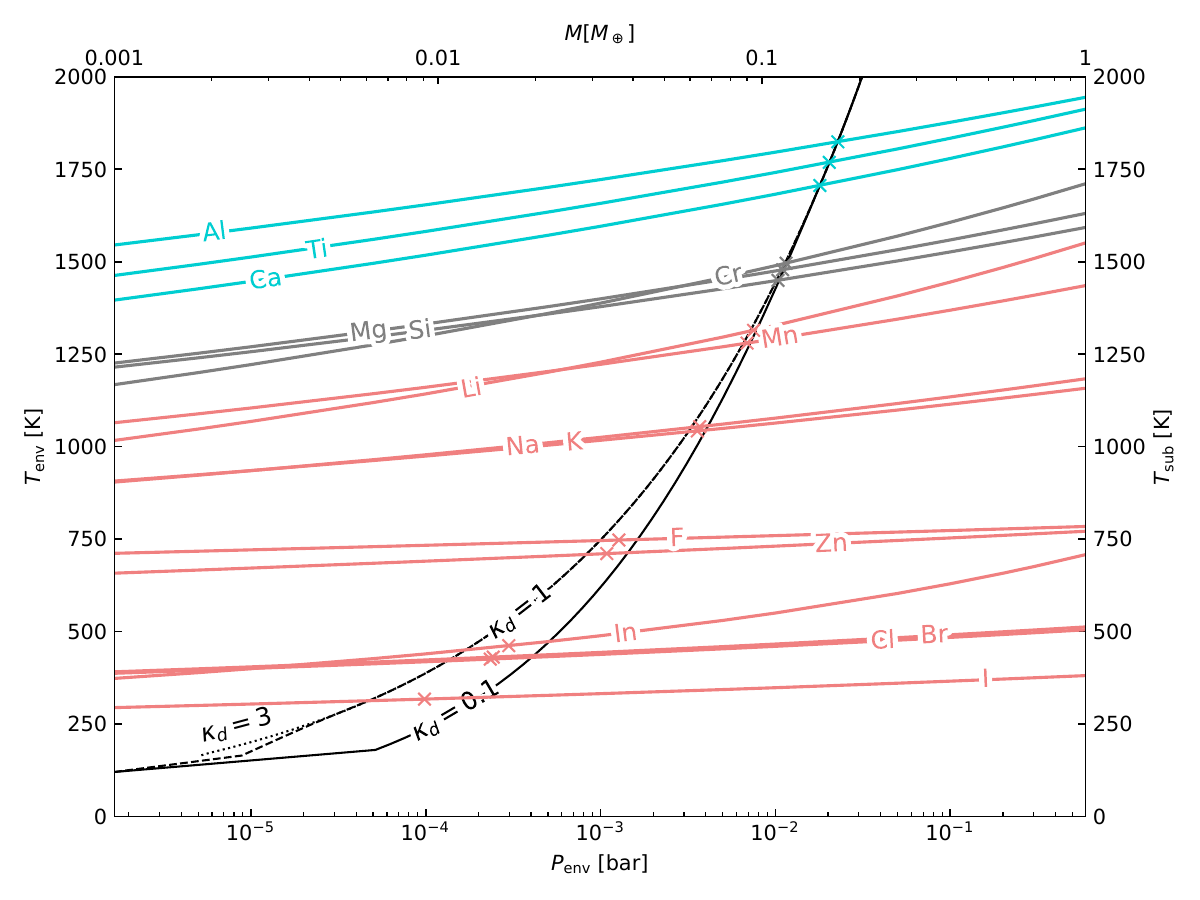} 
\caption{\textbf{The pressure ($P_{\rm env}$)--temperature ($T_{\rm env}$) profiles at the bottom of the envelope of the pebble-accreting body over its growth.} Those color lines indicate the pressure-dependent elemental sublimation
temperatures ($T_{\rm sub}$; Extended Data Table~1) for two different
opacity conditions in the surrounding protoplanetary disk
($\kappa = [0.1, 1] \mathrm{m^2\,kg^{-1}}$).
The adopted mean $T_{\rm sub}$ values are denoted by cross ($\times$) signs.
The dotted curve shows that a further increase of the opacity parameter
($\kappa$) by a factor of 3 over the unity value (resulting, for example,
from the release of dust and recondensed MVLEs from the pebbles) will not
change the bottom $T$ profile significantly, and thus the elemental
sublimation temperatures are robust to opacity changes.} 
\label{fig:PTenv}
\end{figure}

\begin{table}[htbp!]
\centering
	\caption{\textbf{Elemental sublimation temperatures ($T_{\rm sub}$)}. These are calculated for different disk opacity conditions ($\kappa_{\rm d}=[0.1, \,1]\,{\rm m^2\,kg^{-1}}$), while their mean values ($\overline{T_{\rm sub}}$; with negligible error bars) are adopted for model exploration. $T_{\rm sub}$ values at specific pressures ($P_{\rm sub}$) for different disk opacity conditions are interpolated from the recommended pressure-dependent $T_{\rm c}^{50}$ (equivalent to $T_{\rm sub}$ here) model \cite{Lodders2025} ($10^4/T_{\rm c}^{50} = \alpha\log_{10}P{\rm (bar)} + \beta$, where our best-fit values of $\alpha$ and $\beta$) are specified in the last two columns for reference). Note that the $T_{\rm sub}$ values for light halogens Cl, Br and I are only used for plotting but not for model fitting. \rev{S is not included for model fitting either, but its $T_{\rm sub}$ is used to infer the nominal S composition from the best-solution models of hybrid accretion.}}
	\begin{tabular}{c cc cc c cc}	
	\hline\hline

\multirow{2}{*}{X} & \multicolumn{2}{c}{$\kappa_d=0.1\,{,\rm m^2\,kg^{-1}}$} & \multicolumn{2}{c}{$\kappa_d=1\,{\rm m^2\,kg^{-1}}$} & \multirow{2}{*}{$\overline{T_{\rm sub}}$, K} &\multicolumn{2}{c}{Best-fit parameters} \\
& $T_{\rm sub}$, K & $P_{\rm sub}$, bar & $T_{\rm sub}$, K & $P_{\rm sub}$, bar & & $\alpha$ & $\beta$\\
\hline
Al & 1824 & $2.3 \times 10^{-2}$ & 1824 & $2.3 \times 10^{-2}$ & 1824 & -0.2395  & 5.0883\\
Ti & 1769 & $2.0 \times 10^{-2}$ & 1769 & $2.0 \times 10^{-2}$ & 1769   &-0.2896 & 5.1634\\
Ca & 1706 & $1.8 \times 10^{-2}$ & 1706 & $1.8 \times 10^{-2}$ & 1706  &-0.3230 & 5.2976 \\
Mg & 1479 & $1.1 \times 10^{-2}$ & 1477 & $1.1 \times 10^{-2}$  & 1478  &-0.3650  & 6.0502 \\
Si & 1450 & $1.1 \times 10^{-2}$ & 1449 &$1.0 \times 10^{-2}$  & 1450  &-0.3519 & 6.1664 \\
Cr & 1496 & $1.2 \times 10^{-2}$ & 1495 &$1.1 \times 10^{-2}$ & 1496  &-0.4903 & 5.7343\\
Li & 1316 & $7.8 \times 10^{-3}$ & 1312 &$7.2 \times 10^{-3}$ & 1314  &-0.6108 & 6.3101 \\
Mn & 1281 & $7.2 \times 10^{-3}$ & 1278 &$6.6 \times 10^{-3}$ &  1280 &-0.4375 & 6.8678 \\
Na & 1056 & $4.1 \times 10^{-3}$ & 1051 &$3.3 \times 10^{-3}$ & 1054  &-0.4706 & 8.3443 \\
K & 1045 & $4.0 \times 10^{-3}$ & 1040 &$3.2 \times 10^{-3}$ & 1042  &-0.4308 & 8.5403\\
F & 749 & $1.6 \times 10^{-3}$ & 746 &$9.9 \times 10^{-4}$ & 747  &-0.2346 & 12.7031 \\
Zn & 713 & $1.4 \times 10^{-3}$ & 708 &$8.2 \times 10^{-4}$ & 711  &-0.4028 & 12.8792 \\
In & 473 & $5.1 \times 10^{-4}$ & 450 &$1.7 \times 10^{-4}$ & 461  &-2.2867 & 13.6120 \\
Br & 435 & $4.2 \times 10^{-4}$ & 426 & $1.4 \times 10^{-4}$& 431  &-1.0928 & 19.2740 \\
Cl & 430 & $4.0 \times 10^{-4}$ & 420 &$1.3 \times 10^{-4}$ & 425  &-1.0857 & 19.5766 \\
I & 321 & $2.0 \times 10^{-4}$ & 313 &$4.8 \times 10^{-5}$ & 317  &-1.3935 & 25.9576 \\
S & 661 & $1.2 \times 10^{-3}$ & 661 &$6.5 \times 10^{-4}$ & 661  &0.0 & 15.1286 \\
  \hline

\end{tabular}
\label{tab:T_sub}
\end{table}

\begin{figure}
\centering
\includegraphics[width=\textwidth]{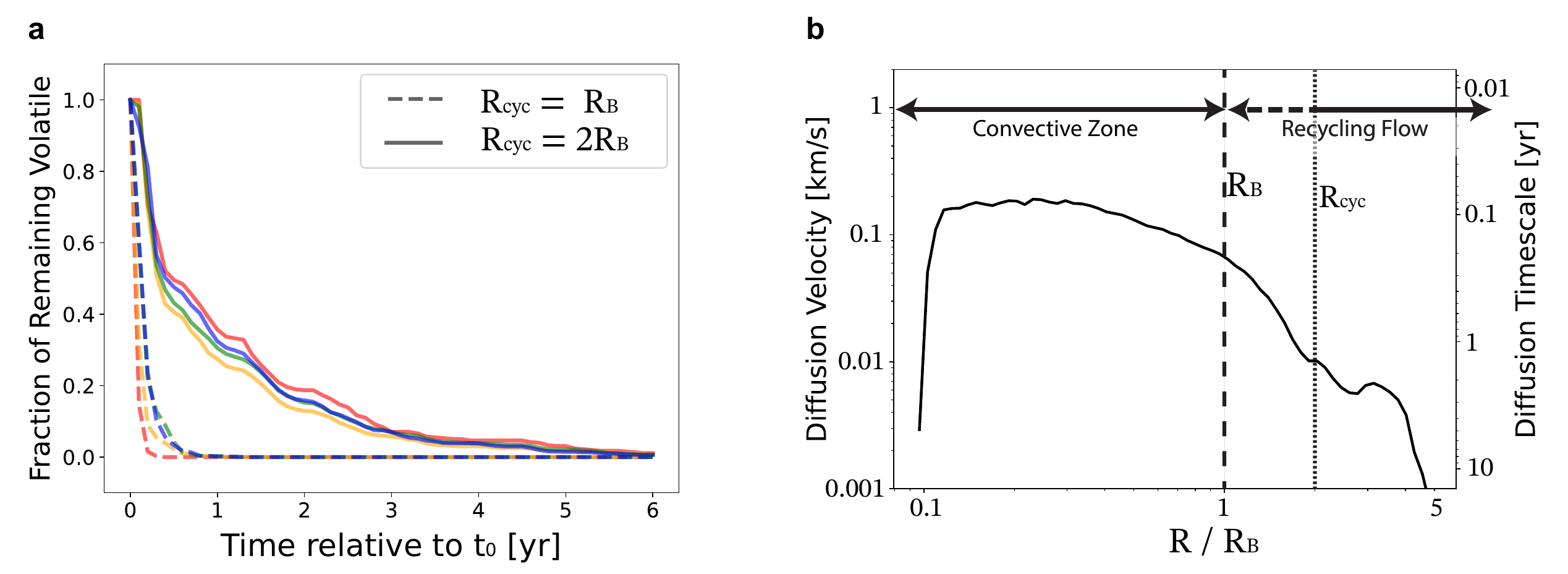} 
\caption{{\bf The effect of different assumptions of the recycling range $R_{\rm cyc}$ on diffusion timescale of particles.} (\textbf{a}) The surviving particle fraction as a function of time for two values of $R_{\rm cyc}$ and for different spatial intervals of particles initially placed in the envelope: 0.1-0.3 $R_{\rm B}$ (red), 0.3-0.6 $R_{\rm B}$ (yellow), 0.6-0.8 $R_{\rm B}$ (green), and 0.8-1 $R_{\rm B}$ (blue). We note that for our assumed two values of $R_{\rm cyc}$, the case of $R_{\rm cyc}\approx R_{\rm B}$ is the most realistic \cite{KurokawaTanigawa2018}, but loss of volatiles is rapid under both assumptions. (\textbf{b}) The measured diffusion velocity and diffusion time-scale as a function of distance from the outgassed atmosphere of the protoplanet. The reduction in the strength of the convection in the convective overshoot region beyond $R_{\rm B}$ is clearly visible.}
\label{fig:conv_suppl}
\end{figure}

\begin{figure}[htbp!]
\centering
\includegraphics[width=\textwidth]{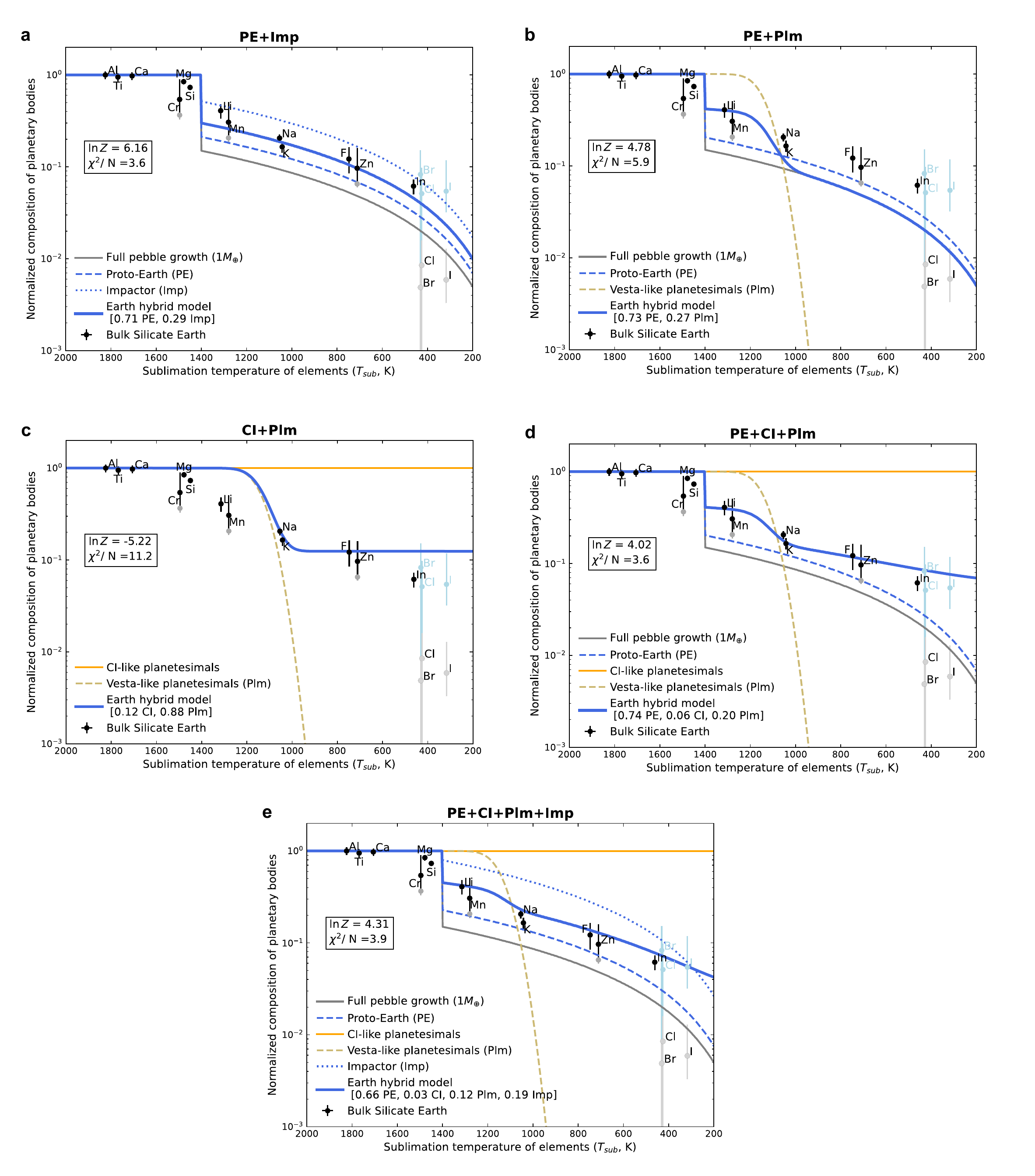}
\caption{\rev{\textbf{Tests of the essentiality of potential components ({\bf PE, Plm, Imp, CI}) for Earth's accretion.} Compositional data are presented as mean values $\pm 1\sigma$ SD. ({\bf a}) The {\bf PE+Imp} model. ({\bf b}) The {\bf PE+Plm} model. ({\bf c}) The {\bf CI+Plm} model. ({\bf d}) The {\bf PE+CI+Plm} model. ({\bf e}) The {\bf PE+CI+Plm+Imp} model. We refer to Supplementary Fig.~3 for corresponding posterior distributions of the Bayesian inferences.}}
\label{fig:earth_component_tests}
\end{figure}

\begin{figure}[htbp!]
\centering
\includegraphics[width=\textwidth]{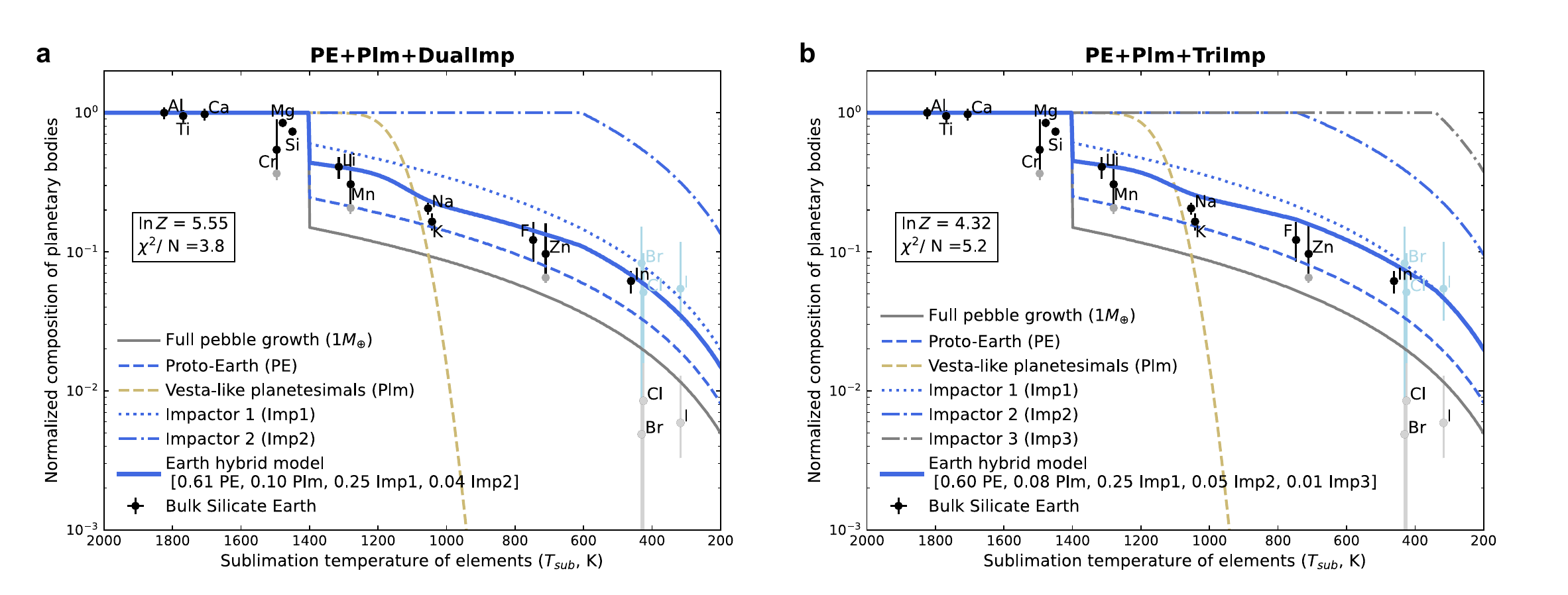}
\caption{\rev{\textbf{Tests of multi-impactor scenarios for Earth's accretion.} Compositional data are presented as mean values $\pm 1\sigma$ SD. ({\bf a}) The {\bf PE+Plm+DualImp} model. ({\bf b}) The {\bf PE+Plm+TriImp} model. We refer to Supplementary Fig.~4 for corresponding posterior distributions of the Bayesian inferences.}}
\label{fig:earth_multiimpactors}
\end{figure}

\begin{figure}[htbp!]
\centering
\includegraphics[width=\textwidth]{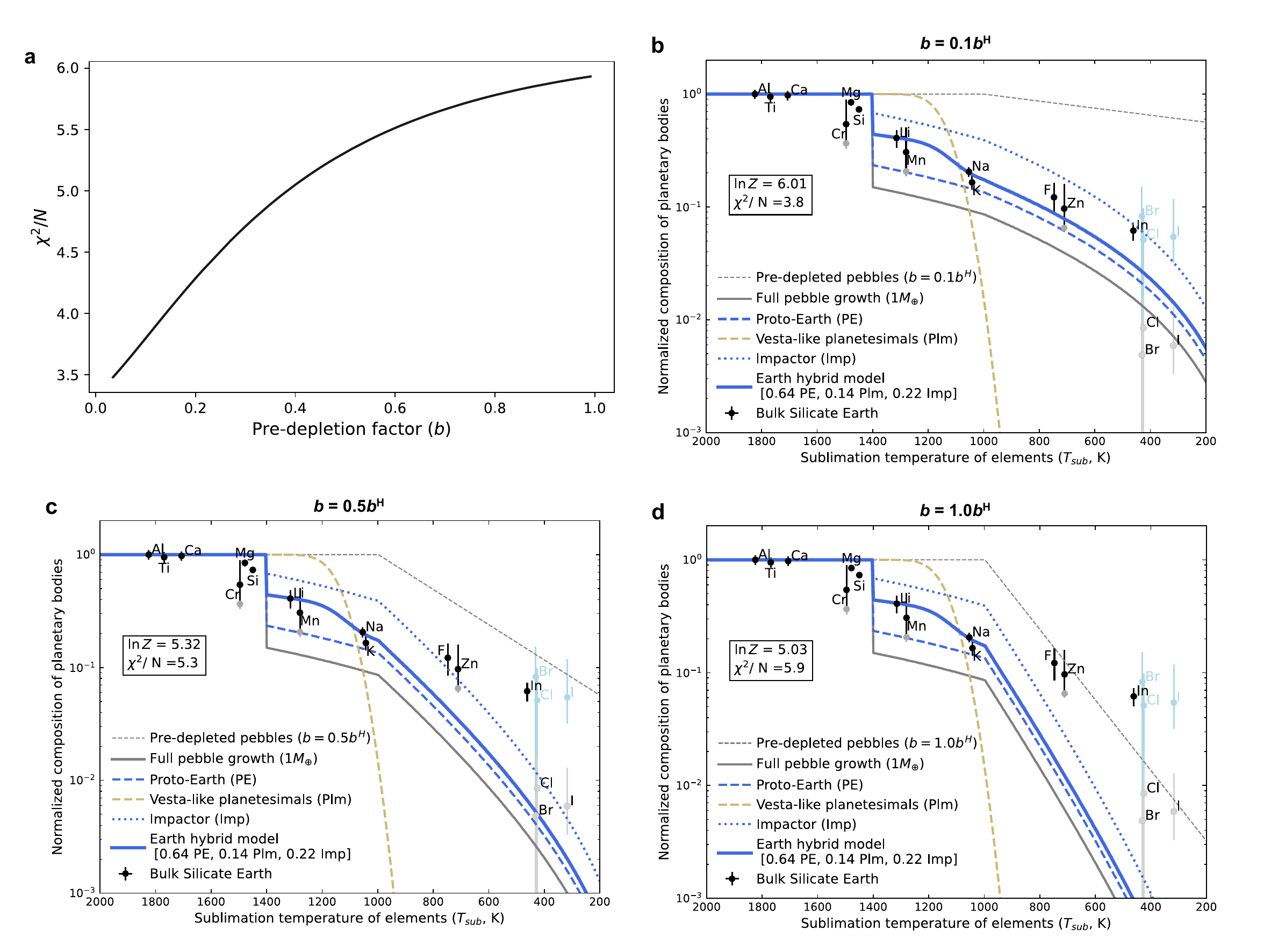}
\caption{\rev{\textbf{Tests of pre-depleted pebbles for Earth's accretion}. Compositional data are presented as mean values $\pm 1\sigma$ SD. ({\bf a}) The reduced-$\chi^2$ ($\chi^2/N$) as a function of pre-depletion factor ($b$), which is relative to the depletion slope of H-chondrite model ($b^{\rm H}=-3.12$; Methods). ({\bf b}) The case of $b=0.1b^{\rm H}$. ({\bf c}) The case of $b=0.5b^{\rm H}$. ({\bf d}) The case of $b=1.0b^{\rm H}$. }}
\label{fig:earth_pebble_predepeltion}
\end{figure}

\begin{figure}[htbp!]
\centering
\includegraphics[width=\textwidth]{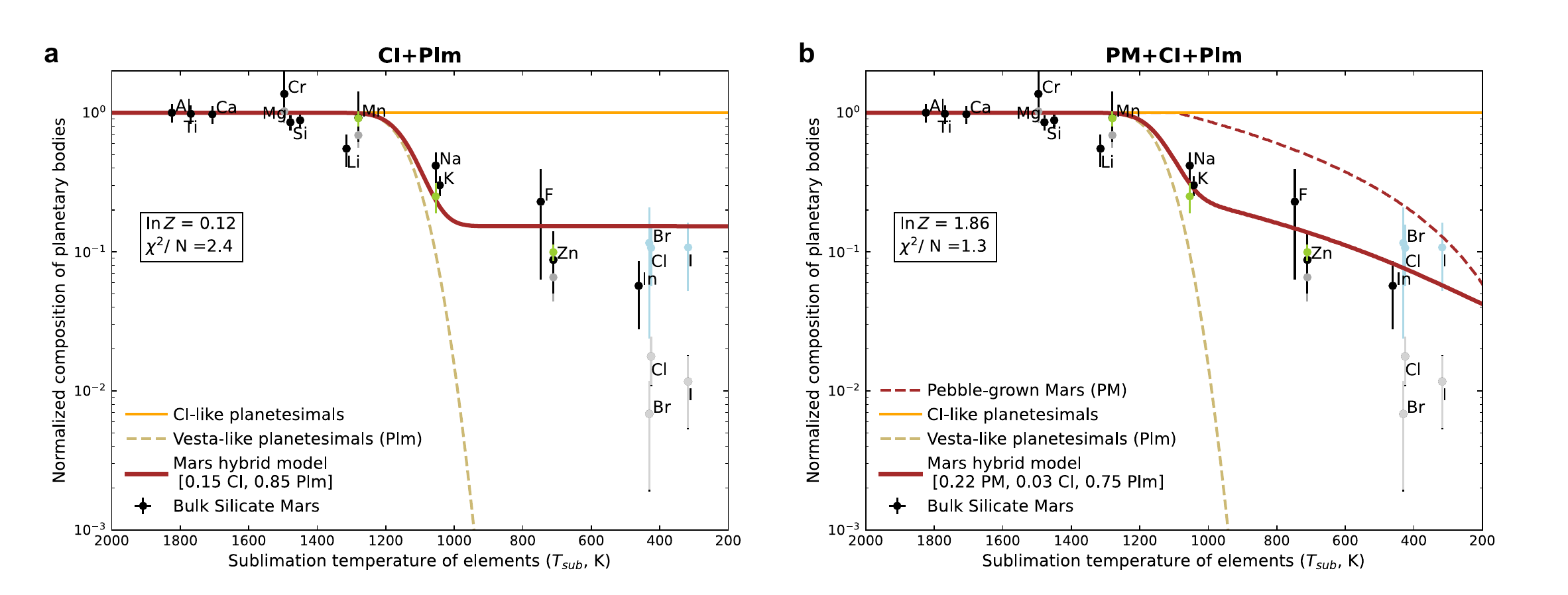}
\caption{\rev{\textbf{Tests of the essentiality of a volatile-rich CI-like component ({\bf CI}) for Mars's accretion.} Compositional data are presented as mean values $\pm 1\sigma$ SD. ({\bf a}) The {\bf CI+Plm} model. ({\bf b}) The {\bf PM+CI+Plm} model. We refer to Supplementary Fig.~5 for corresponding posterior distributions of the Bayesian inferences.}}
\label{fig:mars_CI_component}
\end{figure}

\begin{figure}[htbp!]
\centering
\includegraphics[width=\textwidth]{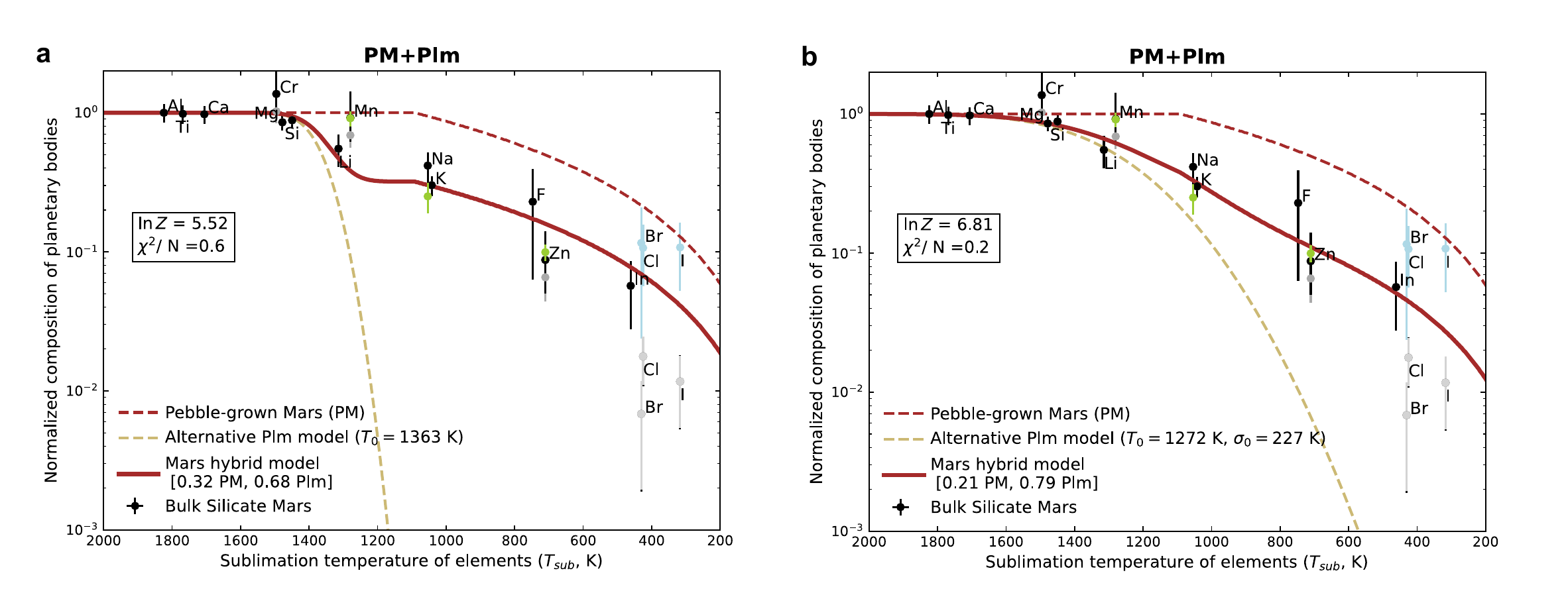}
\caption{\rev{\textbf{Tests of varied $T_0$ and/or $\sigma_0$ of the logistic model of volatile-depleted planetesimals (in the case of Mars)} Compositional data are presented as mean values $\pm 1\sigma$ SD. ({\bf a}) The {\bf PM+Plm (varied $T_0$)} model. ({\bf b}) The {\bf PM+Plm (varied $T_0\&\sigma_0$)} model. We refer to Supplementary Fig.~6 for corresponding posterior distributions of the Bayesian inferences.}}
\label{fig:mars_T0_sigma0}
\end{figure}

\begin{figure}[htbp!]
\centering
\includegraphics[width=\textwidth]{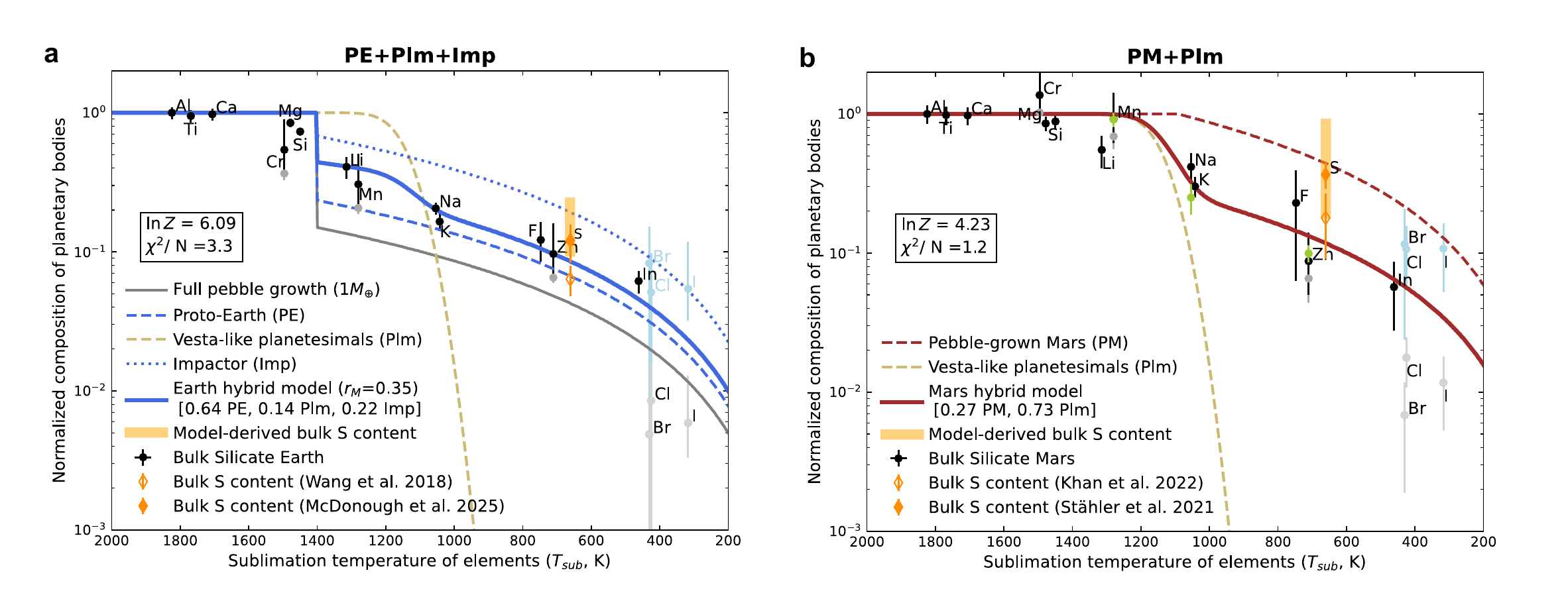}
\caption{\rev{\textbf{Model-derived sulfur (S) contents in bulk Earth and Mars}. The estimates (yellow bars) are based on the best-solutions of the reference hybrid-accretion models ({\bf PE+Plm+Imp} and {\bf PM+Plm}) of Earth and Mars. Compositional data are presented as mean values $\pm 1\sigma$ SD. ({\bf a}) The result for bulk Earth compared to two different estimates in the literature (yellow diamonds). ({\bf b}) The result for bulk Mars compared to two different estimates in the literature (yellow diamonds). For the Khan et al. estimate, we have adopted their calibrated uppermost and lowermost limits of core S content (3.1--9.1 wt\%), while using the same core mas fraction ($0.25\pm0.005$) as of St\"{a}hler et al. to convert to bulk S content in Mars.}}
\label{fig:earth_mars_S_content}
\end{figure}

\begin{table}[htbp!]
\centering
\caption{\textbf{A set of priors applied to perform the Bayesian inference.} The set of priors includes the fractions ($f$) of components, the Bayesian inference scatter ($\sigma_s$), and the parameters ($T_0$ and $\sigma_0$) of a planetesimal logistic model.} 
\begin{tabular}{p{0.2\linewidth}p{0.22\linewidth}p{0.55\linewidth}} 
\hline
\textbf{Parameter} & \textbf{Prior Condition} & \textbf{Description} \\
\hline
Fraction ($f$)           & $\mathcal{U}[0, 1]$ \& $\sum_i f_i=1$              & Following the Dirichlet prior, i.e., fraction of each component ($f_i$) is within 0--1 (uniform) and their sum is equal to 1. \\ 
Fraction ($f$)           & $f_{\rm Imp} \leq f_{\rm pp}$              & Where applicable, the fraction of an impactor ($f_{\rm Imp}$) is no greater than that of the protoplanet ($f_{\rm pp}$). \\
Fraction ($f$)           & $f_{\rm Imp}^{i+1} \leq f_{\rm Imp}^{i}$              & Where applicable, the fraction of a subsequent impactor ($f_{\rm Imp}^{i+1}$) is no greater than that of the preceding one ($f_{\rm Imp}^{i}$). \\
Scatter ($\sigma_s$)             & $\mathcal{U}[10^{-3}, 10^{1}]$ & A broad bound of the intrinsic scatter (uniform). \\
$T_0$ (K)             & $\mathcal{U}[500, 2000]$ & Where applicable, a broad bound of the inflection point ($T_0$) of a planetesimal logistic model (uniform). \\
$\sigma_0$ (K)             & $\mathcal{U}[10, 500]$ & Where applicable, a broad bound of the steepness ($\sigma_0$) of a planetesimal logistic model (uniform). \\
\hline
\end{tabular}
\label{tab:Bayesian_priors}
\end{table}

\clearpage

\subsection*{Supplementary Information for} 
\section*{Volatile depletion in rocky planets as a chemical fingerprint of hybrid accretion}
Haiyang S. Wang, Anders Johansen, Ziyan Xu, Marie-Luise Steinmeyer, Michiel Lambrechts, Elishevah van Kooten, Chao-Chin Yang, Zhaohuan Zhu, Dante S. Lauretta, Martin Bizzarro

\vspace{1cm}
\subsection*{Table of contents}
\vspace{2cm}
Supplementary Text
\begin{itemize}
\item Partition coefficients and sensitivity test for Cr, Mn and Zn that show slight siderophile tendency
\item Extended discussion of volatile pre-depletion of accreted pebbles
\item Posterior distributions of Bayesian inferences
\item Sensitivity tests for alternative models of volatile-depleted planetesimals in the case of Earth
\item Does the sublimation temperature correspond to the condensation temperature?
\end{itemize}
Supplementary Table 1\\
Supplementary Figures 1--9\\
References\\

\newpage


\subsection*{Partition coefficients and sensitivity test for Cr, Mn and Zn that show slight siderophile tendency} 

\begin{figure}[htbp]
\centering
\includegraphics[width=\textwidth]{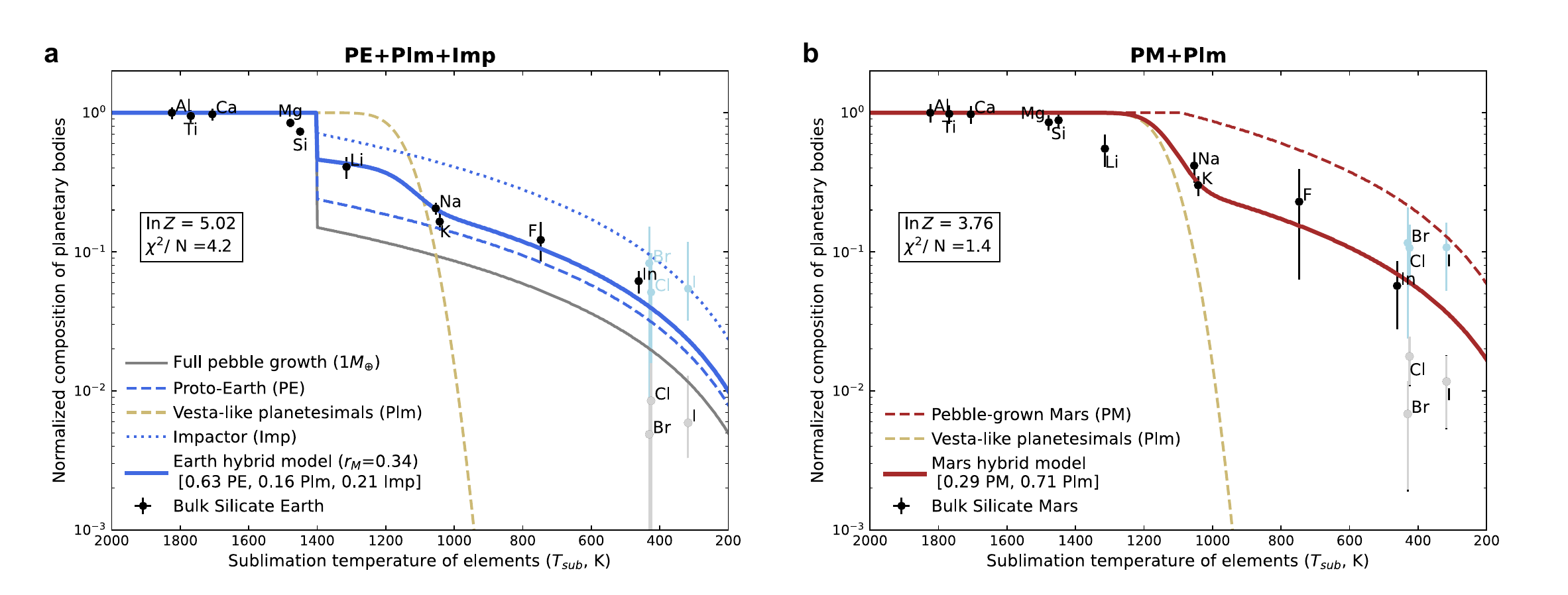}
\caption{\textbf{Sensitivity test by excluding Cr, Mn and Zn (lithophiles with slight siderophile tendency) for both Earth (\textbf{a}) and Mars (\textbf{b})}. The best solutions of a hybrid accretion scenario for both planets remain nearly identical to those found with those three elements considered, albeit with slightly higher post-calculated $\chi^2/N$ (due to the decreased number of fit elements, $N=10$ (instead of 13 otherwise)). The calculated Bayesian evidence ($\ln Z$) values for both Earth and Mars remain high (and equivalent to the cases with Cr, Mn and Zn considered, $\Delta\ln Z\lesssim1$).}
\label{fig:bestmodel_Earth_Mars_woCrMnZn}
\end{figure}

The partition coefficient ($D$, representing the equilibrium ratio of the concentration in liquid metal relative to the concentration in liquid silicate) 
of an element during core formation of a rocky planet is sensitive to both silicate melt composition and metal core composition (including particularly sulfur fraction), oxygen fugacity, as well as pressures and temperatures at an evolving core-mantle boundary \cite{Chabot2003S, Siebert2011S}. Experimentally determined values of $D$ for Cr, Mn and Zn at different conditions vary significantly, ranging from 0.5--4 for Cr \cite{Mann2009S, Fischer2015S, Huang2018S}, 0.1--2 for Mn \cite{Chabot2003S, Mann2009S, Siebert2011S}, and 0--7 for Zn \cite{Mann2009S, Wang2016S, Mahan2017S}. In pebble accretion, most of the moderately volatile elements including Mn and Zn are accreted already before the planet has grown {more} massive than Mars mass (Fig.~2a). In other words, the final masses of Earth and Mars do not necessarily discriminate much the partition coefficient of a moderately volatile element (given that all other conditions are the same). While the moderately refractory element Cr may be continuously accreted in the further growth of a proto-Earth, we have chosen to adopt the same partition coefficient, $D=1^{+2}_{-1}$, \revvv{where the value 1 is the nominal value of the parameter and the lower and upper error bars span our adopted range,} for Cr, Mn and Zn for both Earth and Mars, taking into account already that the higher concentration of sulfur in Martian core \cite{Yoshizaki2020S, Khan2022S} may elevate partitioning of chalcophile elements (even though other conditions in Mars may disfavor the process; \cite{Chabot2003S, Mann2009S}). \rev{The values for Cr, Mn and Zn before and after partition coefficient correction are all reported in Supplementary Table \ref{tab:comp_normalized}, along with the values for other adopted major lithophile elements.} 
Due to the abovementioned uncertainties (including those unquantified) in the partition coefficients for Cr, Mn, and Zn (those with a slight siderophile tendency), we have also made a test by removing those three elements ({Supplementary} Fig. \ref{fig:bestmodel_Earth_Mars_woCrMnZn}). We find that for both Earth and Mars (although the $\chi^2/N$ values slightly increase) the best model solutions, as well as the allowable model parameter spaces, constrained by the remaining set of elements, are equivalent to what are found with the full set of selected elements. Due to the fact that those three elements spread across a wide range of volatility, they are preferably included in our data selection of major rock-forming lithophiles. {We do not correct for partitioning of other lithophile elements with weaker siderophile tendency (e.g., Si and In, \cite{Wade2005S, Yi2000S}), but our sensitivity tests suggest that this non-correction does not alter our best solutions.}

\subsection*{Extended discussion of volatile pre-depletion of accreted pebbles}
\begin{figure}[htbp]
\centering
\includegraphics[width=\textwidth]{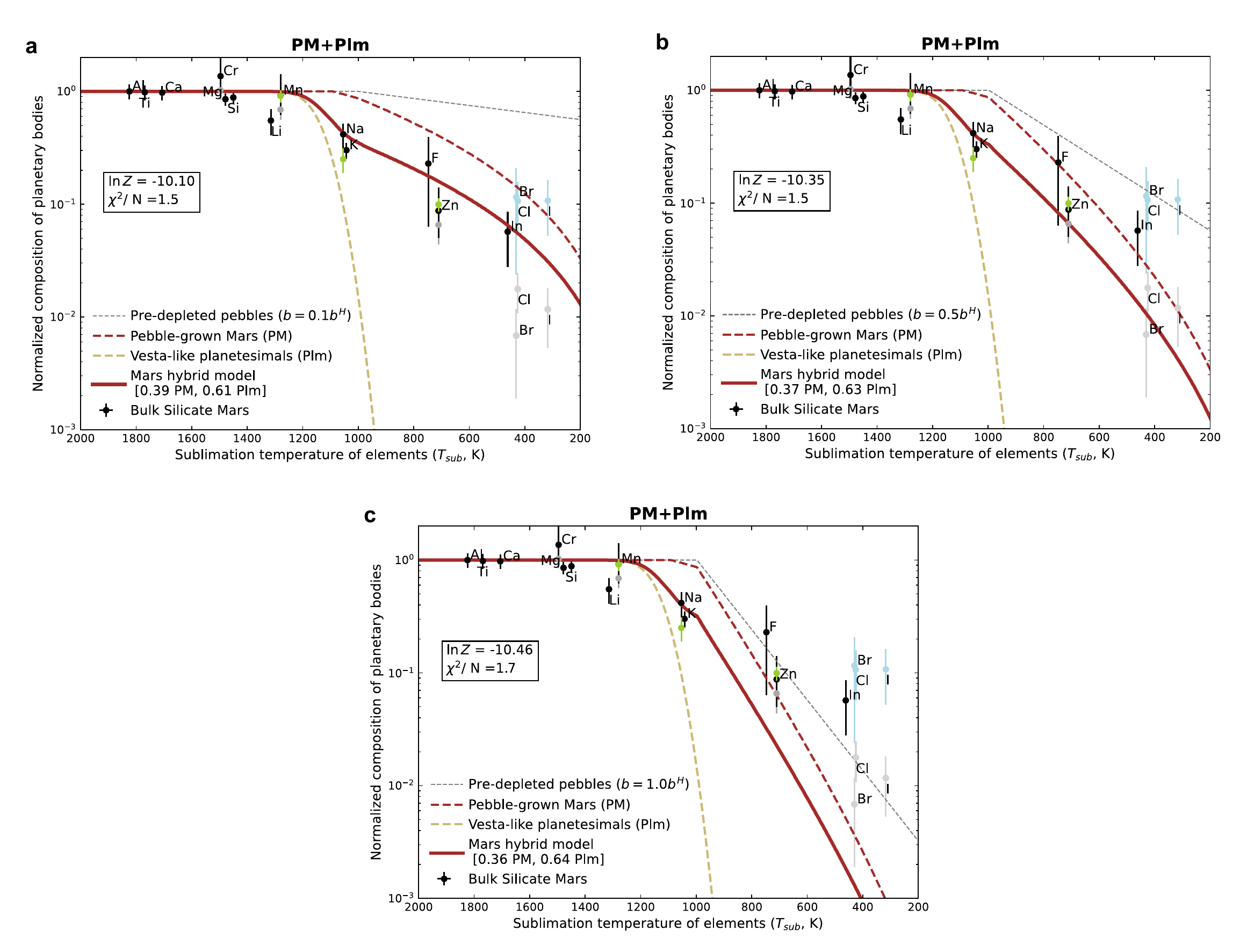}
\caption{\textbf{Tests of pre-depleted pebbles for Mars' accretion.} (a) The case of a pre-depletion factor $b=0.1b^H$; (b) The case of $b=0.5b^H$; (c) The case of $b=1.0b^H$.}
\label{fig:mars_pebble_predepleted}
\end{figure}

\revv{Similar to the case of Earth, we have also done tests for Mars in terms of accreting pre-depleted pebbles (Supplementary Fig. \ref{fig:mars_pebble_predepleted}). The relative contribution of the pebble-grown {\bf PM} increases slightly (by about 10\% relative to Mars' reference model) when pre-depleting the pebbles. However, the pre-depletion worsens the overall fits to the volatile depletion of Mars, in terms of both the reduced-$\chi^2$ and Bayesian evidence ($\ln Z$) due to the mismatch particularly for F, Zn, and In and to the (unnecessary) model complexity. Nonetheless, the change caused by including pre-depleted pebbles does not change our conclusion in terms of the hybrid-accretion nature of Mars, which remains dominated by planetesimal accretion.}

Extending the discussion of pre-depleted pebbles in the main text, it is noteworthy that chondrules present in the carbonaceous chondrites, whose parent bodies formed in the outer Solar System, are the most volatile-depleted \cite{Braukmuller2019S}. In contrast, chondrules from the ordinary chondrites that formed closer to the Sun maintain nearly solar composition for elements with volatility down to Na and K ($T_{\rm c}^{50}\sim$1000 K under a canonical pressure of $10^{-4}$ bar \cite{Lodders2025S}), but experienced extensive loss of more volatile elements \cite{Alexander2019aS,Lodders2025S}. Finally, the chondrules in enstatite chondrites (groups EH and EL) formed under very reducing (water-poor) conditions interior to the water ice line. The EH chondrites, which accreted an approximately solar level of both iron and sulfur, are the least volatile-depleted and maintain pristine levels of elements as volatile as Zn and In. The near-solar composition of the EH chondrules formed in the innermost regions of the disk, along with the broadly CI chondritic composition of the asteroids Bennu and Ryugu \cite{Nakamura2022S, Lauretta2024S}, is evidence that the solar protoplanetary disk did not experience significant bulk loss of volatiles during the inwards drift of dust-aggregate pebbles and chondrules \cite{Liu2022S,Colmenares2024S}.  The iron-poor compositions of many inner
Solar System meteorite groups (specifically, the groups L, LL and EL) are also evidence
that the sampled meteorite parent bodies underwent selective accretion of the various free-floating
components \cite{Alexander2022S,Garai2025S}, and hence that they are not necessarily representative of the mean composition of the
ambient material in the disk.

\subsection*{Posterior distributions of Bayesian inferences}
\begin{figure}[htbp!]
\centering
\includegraphics[width=\textwidth]{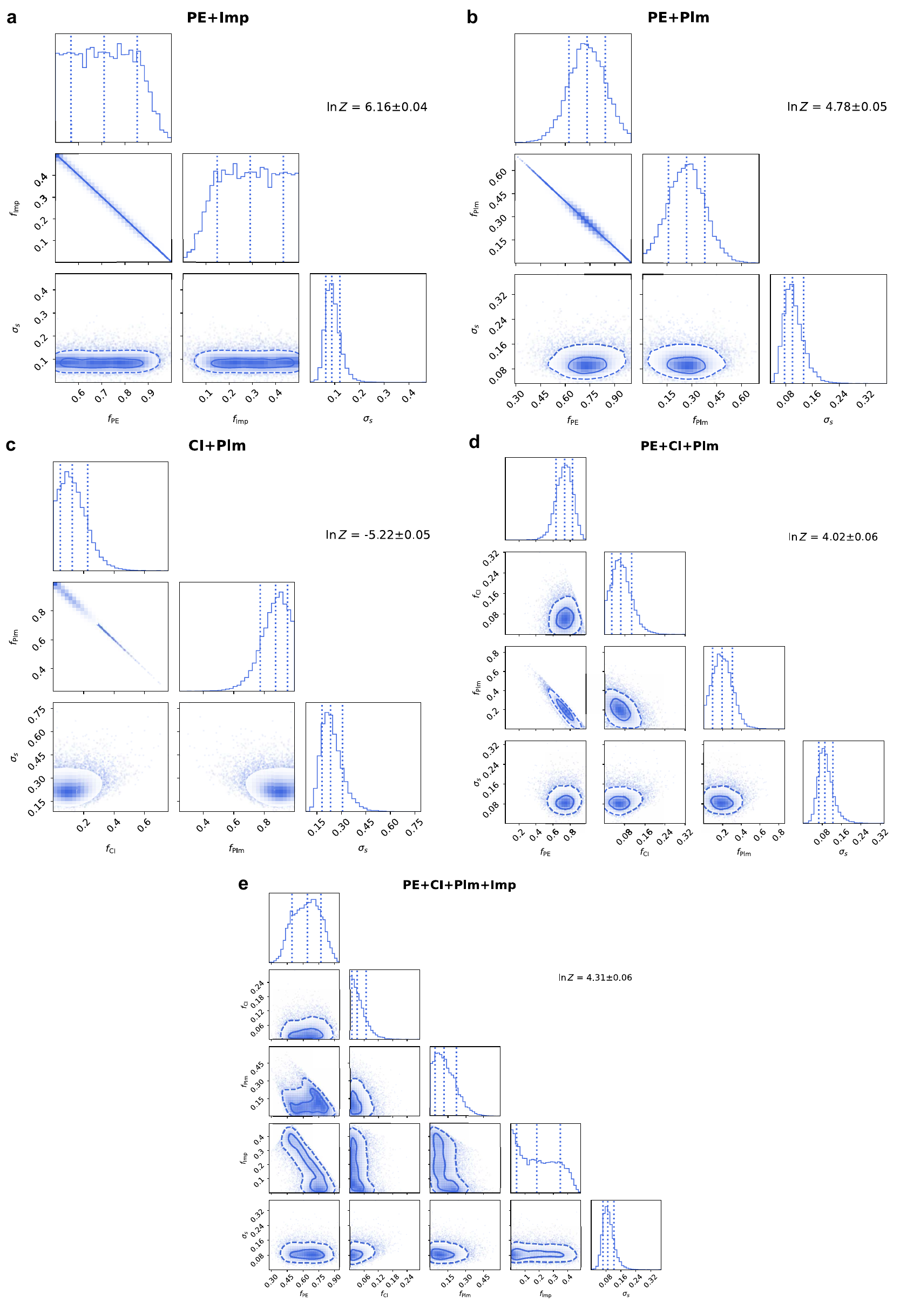}
\caption{\textbf{Posterior distributions corresponding to Extended Data Fig.~3.} ({\bf a}) The {\bf PE+Imp} model. ({\bf b}) The {\bf PE+Plm} model. ({\bf c}) The {\bf CI+Plm} model. ({\bf d}) The {\bf PE+CI+Plm} model. ({\bf e}) The {\bf PE+CI+Plm+Imp} model. }
\label{fig:earth_component_tests_posteriors}
\end{figure}

\begin{figure}[htbp!]
\centering
\includegraphics[width=\textwidth]{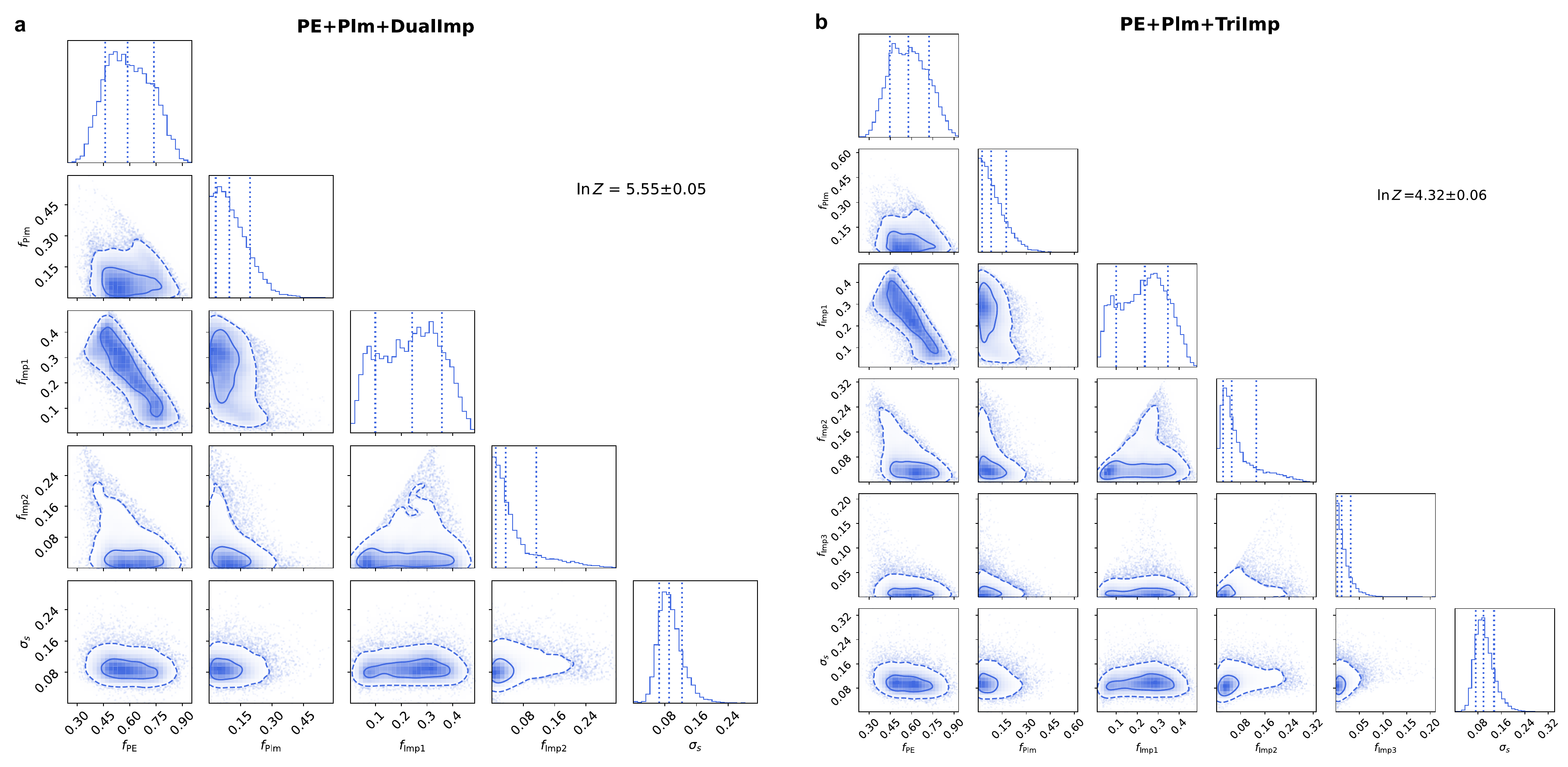}
\caption{\textbf{Posterior distributions corresponding to Extended Data Fig.~4.} ({\bf a}) The {\bf PE+Plm+DualImp} model. ({\bf b}) The {\bf PE+Plm+TriImp} model.}
\label{fig:earth_multiimpactors_posteriors}
\end{figure}

\begin{figure}[htbp!]
\centering
\includegraphics[width=\textwidth]{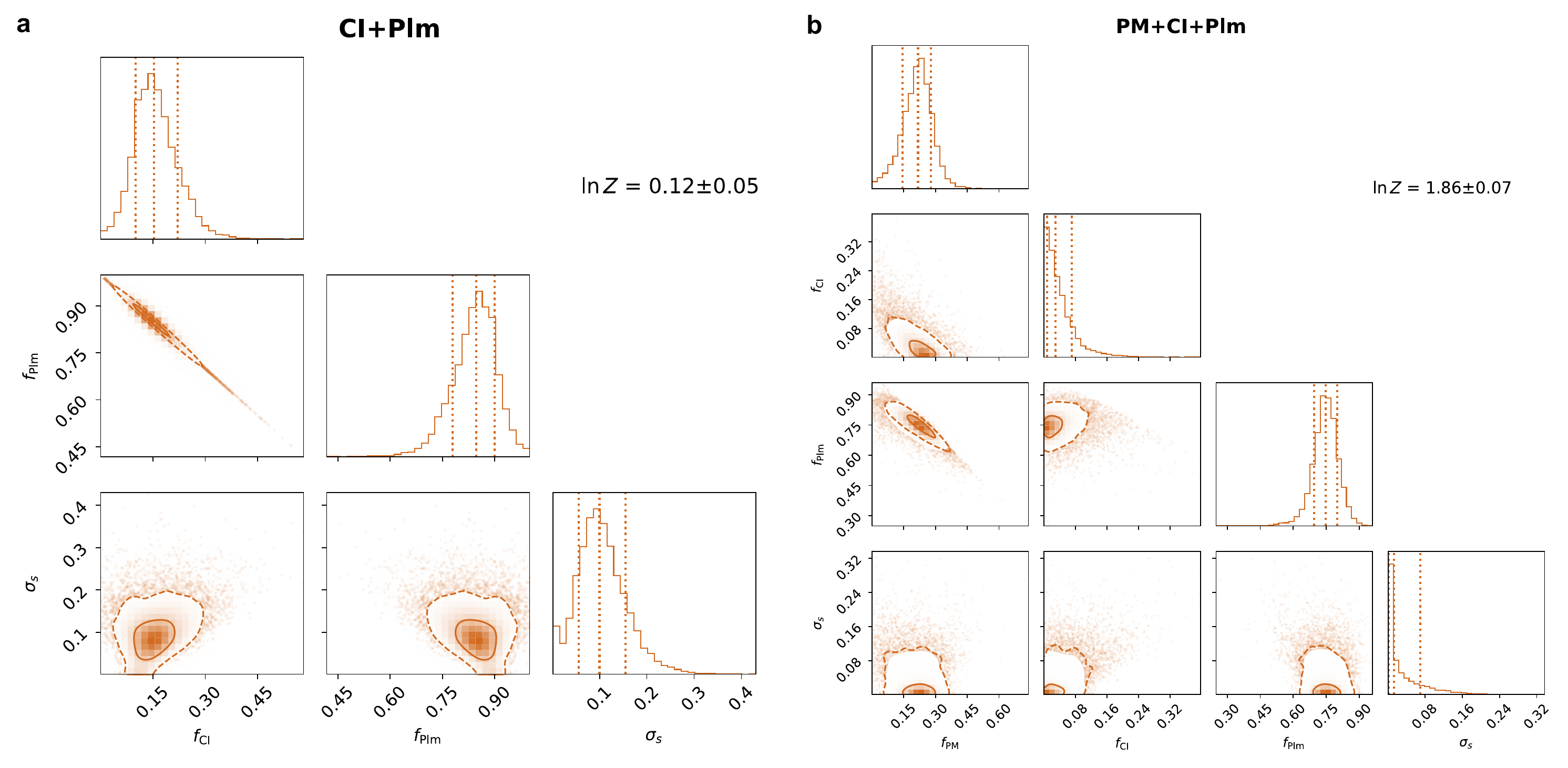}
\caption{\textbf{Posterior distributions corresponding to Extended Data Fig.~6.} ({\bf a}) The {\bf CI+Plm} model. ({\bf b}) The {\bf PM+CI+Plm} model.}
\label{fig:mars_CI_component_posteriors}
\end{figure}

\begin{figure}[htbp!]
\centering
\includegraphics[width=\textwidth]{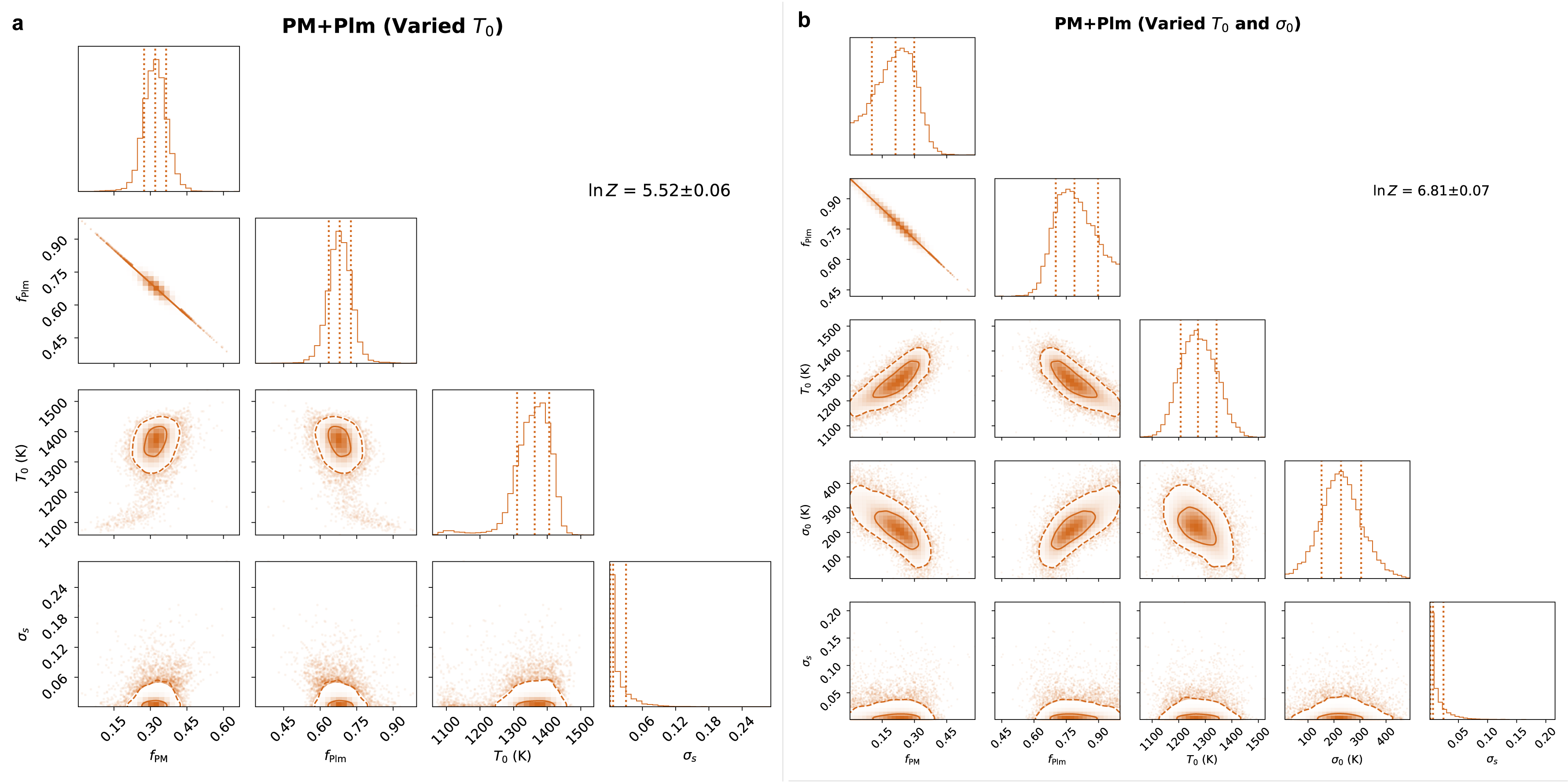}
\caption{\textbf{Posterior distributions corresponding to Extended Data Fig.~7.} ({\bf a}) The {\bf PM+Plm (varied $T_0$)} model. ({\bf b}) The {\bf PM+Plm (varied $T_0\&\sigma_0$)} model.}
\label{fig:mars_T0_sigma0_posteriors}
\end{figure}

\rev{Except for our reference models of Earth's and Mars' accretion, we have only presented the best-fit solutions for a set of alternative models examined in Table 1. Here, we present the posterior distributions (Supplementary Figs. \ref{fig:earth_component_tests_posteriors}--\ref{fig:mars_T0_sigma0_posteriors}) corresponding to those alternative best-fit solutions to better illustrate the model uncertainties.}

\subsection*{Sensitivity tests for alternative models of volatile-depleted planetesimals \rev{in the case of Earth}}
\rev{Similar to Mars' case (Extended Data Fig. 7), we have tested varying $T_0$ and $\sigma_0$ of the planetesimal logistic model for Earth's accretion. We show the result in Supplementary Fig.~\ref{fig:earth_T0_sigma0_posteriors}.} 
The Bayesian inference favors a planetesimal model with a higher $T_0$ ($1492^{+144}_{-77}$ K) and larger $\sigma_0$ ($198^{+107}_{-82}$ K) than those of the Vesta-like planetesimal model ($T_0=1135\pm35$ K and $\sigma_0=62\pm24$ K). This temperature increase can be mainly attributed to fitting the group of moderately refractory elements (Si, Mg and Cr), with a contribution of \revv{$40^{+15}_{-14}$\%} from planetesimals. \revvv{The values of those quantities are presented as medians (50th percentile) with the error bars spanning the range of 16th-84th percentiles.} The sublimation and loss of this group of elements are nevertheless not treated in our current pebble accretion model; loss of Si and Mg is physically plausible if the pebbles experience some degree of sublimative mass loss before reaching the protection of the SiO-dominated vapor atmosphere above the magma ocean \cite{SteinmeyerJohansen2024S, HabibPierrehumbert2024S}. Also, accretion of small amounts of ultra-refractory pebbles akin to calcium-aluminium rich inclusions (CAI) found in some meteorite classes can decrease the relative abundance of both Mg and Si in a rocky planet \cite{Alexander2022S,Garai2025S}. We leave the exact treatment of any potential loss of minor amounts of Si or extra accretion of ultra-refractory material during pebble accretion to a future work. We go on to show that by removing this group of elements (Si, Mg and Cr) from the fit, broader but consistent ranges of $T_0$ ($=1410^{+240}_{-220}$ K) and $\sigma_0$ ($=205^{+182}_{-124}$) are obtained for the volatile-depleted planetesimals ({Supplementary} Fig.~\ref{fig:earth_sensitivity_CrMgSi}a). 
The resultant contribution from such planetesimals to Earth's accretion in this case is $16^{+21}_{-12}$\%, which is still consistent (yet with larger uncertainties) with that ($12^{+7}_{-6}$\%) from adopting the Vesta-like logistic model for planetesimals in Earth's accretion (also without Si, Mg, and Cr). {Supplementary} Fig.~\ref{fig:earth_sensitivity_CrMgSi} compares the median (centra-estimate) solutions between these two cases. These solutions are nonetheless consistent with that ($14^{+10}_{-9}$\%) of our reference PE+Plm+Imp model of Earth's accretion (including Si, Mg, and Cr; Table 1). 

The fit of a logistic model to the composition of Vesta (Eq. 1) raises some concerns, given the clear deviations of Zn and In from the model (see Fig. 1), which may or may not be fully explained by partial condensation \cite{Fang2024S, Sossi2022S}. In this regard, we have also made tests with an alternative, exponential model for the Vesta composition as well as directly using the measured composition data points of Vesta. We show the results in  {Supplementary} Fig.~\ref{fig:earth_sensitivity_vesta_models}. The solutions found for both these alternative cases -- in terms of relative contributions of different components in the case the PM+Plm+Imp for Earth -- are equivalent to our main solutions (Fig. 3a). We also tested that a potentially larger partitioning of Zn and In into Vesta's core \cite{Steenstra2019S} does not change our solutions significantly, since Zn and In are so depleted in Vesta that any plausible core reservoir can not add significantly to the accreted volatile budget of Earth.

\subsection*{Does the sublimation temperature correspond to the condensation temperature?}

Our calculations of the {\it sublimation temperature} formally represent the {\it condensation temperature} derived under the assumption of cooling of a gas of solar composition under chemical equilibrium. However, in order to use this temperature to represent the sublimation temperature, we must assume that the moderately volatile species can sublimate from their host mineral phase on the relevant time-scale of their accretion. At 100 m/s convection speed (see {Extended Data} Fig. 2), pebbles descending with the convection flow take approximately 10 hours to fall over the size-scale of the planet (where the temperature and the sublimation rate are highest). If pebbles instead enter a circumplanetary pebble disk near the planet, as seen in hydrodynamical simulations \cite{JohansenLacerda2010S,Ormel2013S}, then the accretion time-scale through such a disk could be much longer than 10 hours. 

We discuss the mineral hosts here of some key elements that we use for fitting the formation models, in order of increasing volatility. {\it Li} is a trace element (Li/Si=$5.5\times10^{-5}$). Ref.~\cite{Maruyama2009S} reported that Li in the Allende CV chondrite was likely concentrated in the Na-rich mesostatis within chondrules prior to parent body heating (see discussion of sublimation of Na and K from the mesostatis below). {\it Mn} in CV chondrites is mostly concentrated in chondrule rims and in matrix grains \cite{vanKooten2019S}; a similar distribution holds for unequilibrated ordinary chondrites \cite{Alexander1995S}. {\it Na} and {\it K} are concentrated in the Na-rich mesostatis within chondrules. Ref.~\cite{Nagahara1989S} reported heating experiments on solid Na-rich plagioclase (a major component of the mesostatis), the most relevant experiment performed at 1573 K and 5.9 Pa H$_2$ ambient pressure. They found significant depletion of both Na and K within 200 microns of the surface of the mineral after 15.2 hours, which yields a loss speed of approximately 10 micrometers per hour. Loss rates at lower temperatures or higher H$_2$ pressures, which are conditions more relevant to the gas envelopes considered here, were not reported. {\it Zn} in the least thermally altered CV chondrites has highest concentration in the chondrule rims and in the interchondrule matrix, with very low concentrations within the chondrules themselves \cite{vanKooten2019S,Alexander2019bS}. Zn can also reside in FeS. Release of Zn from FeS will likely follow the sublimation of the major component S from the mineral. FeS sublimation displays sluggish kinetics, but the kinetics accelerate significantly at temperatures elevated above its approximately 700 K sublimation temperature \cite{Tachibana1998S,Steinmeyer2023S}, with sublimation speeds above 0.1 microns per hour reached above 1000 K. Zn$_2$SiO$_4$ is nevertheless the dominant Zn carrier in the chemical equilibrium calculations used to produce Fig. 2a. The fast rise of the equilibrium partial pressure of Zn with increasing ambient temperature implies efficient loss of Zn from Zn$_2$SiO$_4$ even at a temperature that is just slightly elevated above the equilibrium temperature of 750 K. {\it F} is mainly present in separate apatite grains within the matrix and as parts of chondrule rims \cite{Zhang2016S}. Finally, {\it In} is a trace element (In/Si$=1.8 \times 10^{-7}$) that in chemical equilibrium resides within FeS or as separate InS minerals \cite{Lodders2003S}. 

Planets can grow by accretion of both chondrules and dust-aggregate pebbles. As a general trend, the moderately volatile elements are concentrated in the matrix or in chondrule rims, and depleted within chondrules, which eases their release upon heating due to the small (micron-scale or smaller) particle size. Micron-sized dust coagulates easily to form millimeter-sized pebbles of high porosity; the high porosity ensures that sublimating agents such as H$_2$ can diffuse into the pebble and that sublimated species can rapidly diffuse out of the dust aggregate pebble. Na and K in chondrule plagioclase are  exceptions to the rule that chondrules interiors are depleted in moderately volatile elements, as they are concentrated in the chondrule mesostatis, but experiments (albeit performed at elevated temperatures) show favorable kinetics for Na-rich plagioclase sublimation \cite{Nagahara1989S}. It is possible that some  elements, such as Li and Zn, will be trapped within silicates and thus inherit the high sublimation temperatures  of enstatite (MgSiO$_3$) or forsterite (Mg$_2$SiO$_4$). However, condensation of these species from the gas phase (irrespective of whether this happened before the formation of the solar protoplanetary disk or within the disk) speaks against such trapping: the more volatile species should only begin to condense at temperatures significantly below the condensation temperature of enstatite and forsterite, favoring the presence of moderately volatile elements in chondrule rims (as observed for Mn in unequilibrated ordinary chondrites) \cite{Alexander1995S}, as rims on silicate matrix grains, or as separate minerals within the matrix where they are maximally exposed to the ambient gas.

\clearpage

\begin{table}[htbp!]
\centering
	\caption{\textbf{Protosolar- and Al-normalized abundances of major rock-forming lithophile elements in the bulk silicate Earth (BSE), bulk silicate Mars (BSM), Vesta (HED meteorites), Ryugu (sample return), Bennu (sample return), as well as EH and H condrites.} We refer to the main text and Fig.~1 for normalization and original data sources. Data are presented as mean values, with error bars, where applicable, being $1\sigma$ SD. A machine-readable table of this set of compositional data is also available \cite{Wang2026aS}.}
	\begin{tabular}{l ccc cc cc}	
	\toprule
Element & BSE & BSM & Vesta & Ryugu & Bennu & EH & H\\
\hline
Al & 1.000 & 1.000  & 1.000  & 1.000  & 1.000  & 1.000  & 1.000 \\
Ti & $0.948\pm0.100$ & $0.982\pm0.113$  & --  & 0.863  & 1.127  & 0.944  & 1.005 \\
Ca & $0.975\pm0.093$ & $0.976\pm0.108$  & $1.000\pm0.100$  & 1.0  & 1.0  & 1.0  & 1.0   \\
Mg & $0.846\pm0.059$ & $0.854\pm0.060$  & $1.000\pm0.100$   & 1.010  & 1.032  & 1.161  & 1.089   \\
Si & $0.732\pm0.048$ & $0.883\pm0.043$  & $1.000\pm0.100$   & --  & 1.036  & 1.626  & 1.209   \\
Cr & $0.366\pm0.038$ & $1.024\pm0.180$  & --  & 1.006  & 0.850  & 1.257  & 1.023   \\
Cr* & $0.542^{+0.357}_{-0.185}$ & $1.366^{+0.737}_{-0.439}$  &   &   &   &   &    \\
Li & $0.408\pm0.074$ & $0.552\pm0.133$  & $0.778\pm0.118$  & 1.007  & 0.912  & 1.222  & 0.892  \\
Mn & $0.207\pm0.020$ & $0.689\pm0.129$  & $1.020\pm0.046$  & 1.037  & 1.021  & 1.196  & 0.959   \\
Mn* & $0.306^{+0.201}_{-0.104}$ & $0.919^{+0.499}_{-0.301}$  &   &   &   &   &    \\
Na & $0.206\pm0.021$ & $0.417\pm0.096$  & $0.082\pm0.021$  & 1.223  & 1.110  & 1.563  & 0.994   \\
K & $0.165\pm0.023$ & $0.301\pm0.040$  & $0.073\pm0.010$  & 1.001  & 0.920  & 1.669  & 1.089   \\
F & $0.122\pm0.042$ & $0.229\pm0.164$  & --  & --  & --  & 4.482  & 0.680   \\
Zn & $0.065\pm0.006$ & $0.066\pm0.022$  & $0.002\pm0.000$  & 1.045  & 1.044  & 1.094  & 1.117   \\
Zn* & $0.097^{+0.063}_{-0.032}$ & $0.087^{+0.053}_{-0.037}$  &   &   &   &   &    \\
In & $0.062\pm0.011$ & $0.057\pm0.029$  & $0.003\pm0.001$  & 1.006  & 0.927  & 1.471  & 0.011   \\
  \bottomrule
\multicolumn{8}{p{13cm}}{* For Earth and Mars, these are the values (adopted for model fitting) after being corrected with a uniform partition coefficient (see Methods and 1st section of Supplementary Information).}  
\end{tabular}
\label{tab:comp_normalized}
\end{table}


\begin{figure}[htbp!]
\centering
\includegraphics[width=\textwidth]{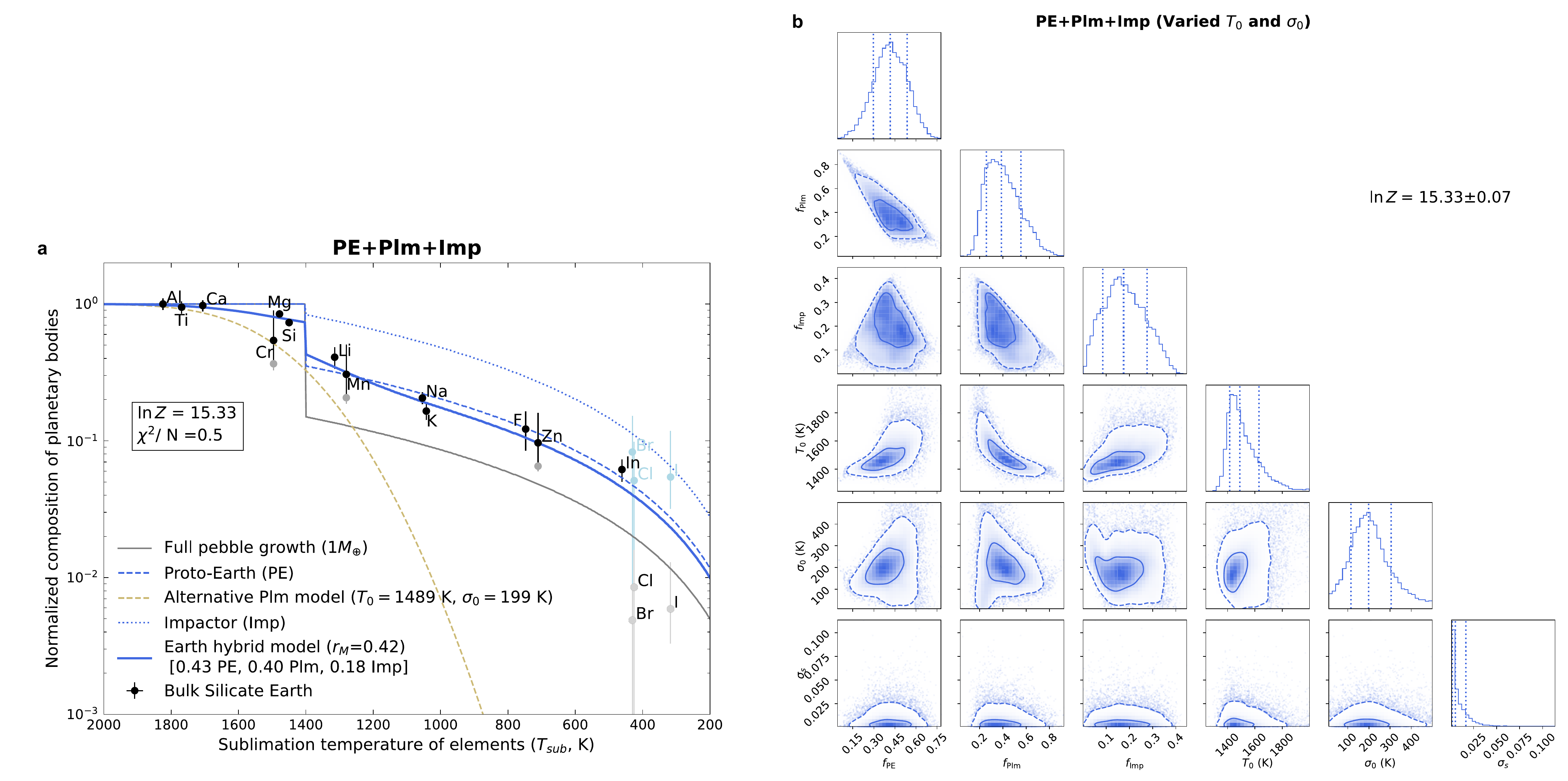}
\caption{\textbf{Test of varied $T_0$ and $\sigma_0$ of the logistic model of volatile-depleted planetesimals for Earth's accretion.} ({\bf a}) The best-solution model. ({\bf b}) The posterior distribution. We note that this test outcome is strongly affected by the group of moderately refractory elements (Si, Mg and Cr) -- see further tests in Supplementary Fig. \ref{fig:earth_sensitivity_CrMgSi}.}
\label{fig:earth_T0_sigma0_posteriors}
\end{figure}

\begin{figure}[htbp]
\centering
\includegraphics[width=\textwidth]{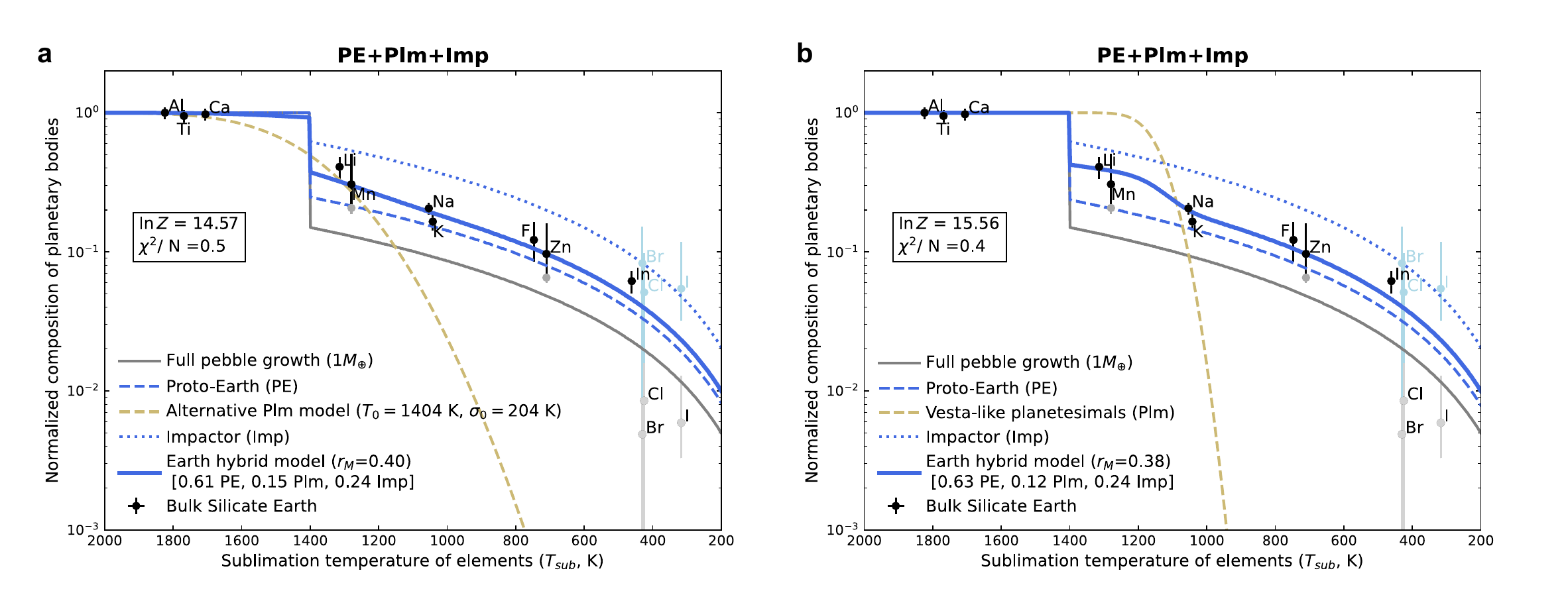}
\caption{\textbf{Sensitivity test of including the group of moderately refractory elements (Si, Mg and Cr) for Earth's accretion}. ({\bf a}) The case of a free planetesimal model that is fit without Si, Mg and Cr. Broader but consistent ranges of $T_0$ ($=1404^{+250}_{-227}$ K) and $\sigma_0$ ($=204^{+182}_{-124}$ K) are obtained compared to the case with Si, Mg and Cr included ($T_0=1489^{+141}_{-74}$ K and $\sigma_0=199^{+106}_{-83}$ K; \rev{Supplementary} Fig.~\ref{fig:earth_T0_sigma0_posteriors}). ({\bf b}) Similar to panel A (i.e., fit without Si, Mg and Cr) but with a specified Vesta-like planetesimal logistic model ($T_0=1135\pm35$ K and $\sigma_0=62\pm24$ K) as adopted in the reference model of Fig.~3a. \revvv{The values of $T_0$ and $\sigma_0$ are presented as medians (50th percentile) with the lower and upper error bars spanning the range of 16th-84 percentiles.} Despite the difference in shape between the logistic models of volatile-depleted planetesimals, the median solutions for the relative contributions of different components are equivalent between panels a and b; they are also equivalent to the reference model of  Fig.~3a. Given the large uncertainties in these free models, their full solutions are obviously consistent with each other.} 
\label{fig:earth_sensitivity_CrMgSi}
\end{figure}

\begin{figure}[htbp]
\centering
\includegraphics[width=\textwidth]{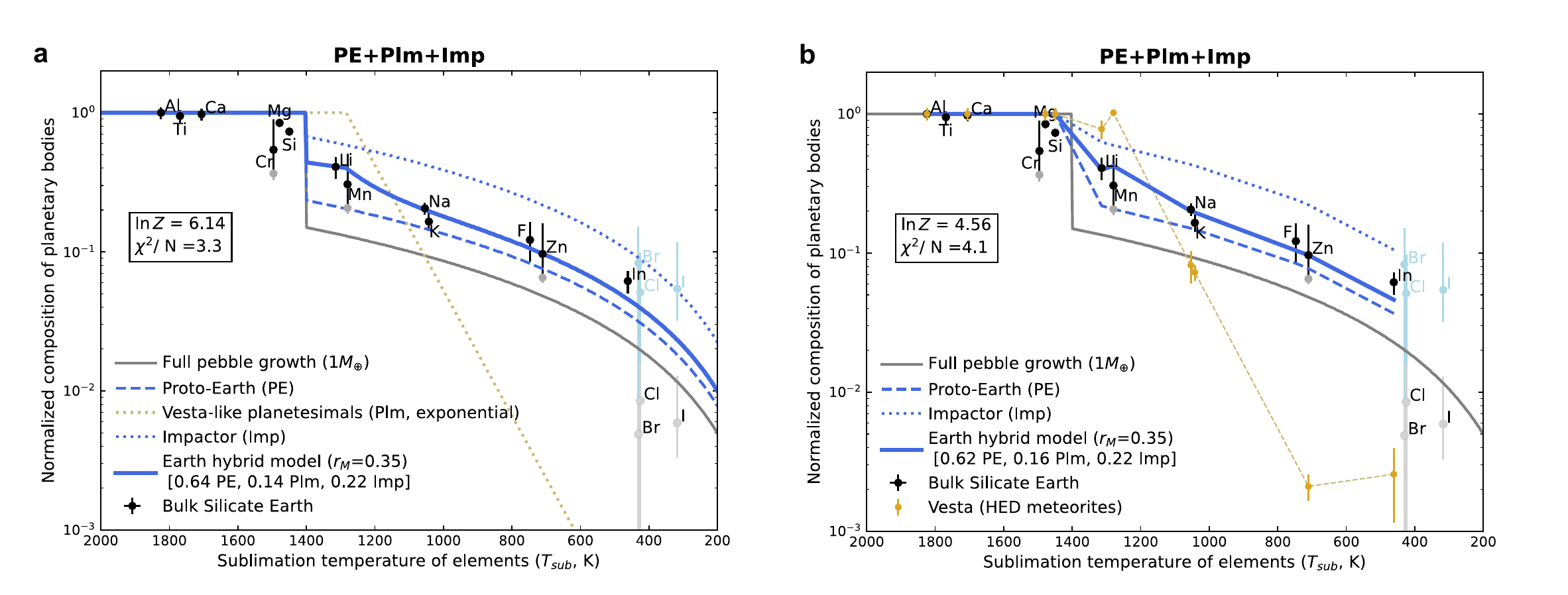}
\caption{\textbf{Using either an exponential model for Vesta-like planetesimals or Vesta's actual composition data to constrain Earth's accretion.} ({\bf a}) Result with an exponential model, which appears to be linear in the logarithmic-linear space, for the planetesimal component. ({\bf b}) Result  using the direct composition data of Vesta (based on HED meteorites) as input for the planetesimal component. These two alternatives do not significantly affect the median solution for the components accreted to Earth compared to the nominal logistic model presented in Fig.~3.} 
\label{fig:earth_sensitivity_vesta_models}
\end{figure}

\clearpage


\begin{thebibliography}{10}
\expandafter\ifx\csname url\endcsname\relax
  \def\url#1{\burl{#1}}\fi
\expandafter\ifx\csname urlprefix\endcsname\relax\def\urlprefix{URL }\fi
\providecommand{\bibinfo}[2]{#2}
\providecommand{\eprint}[2][]{\url{#2}}
\providecommand{\doi}[1]{\url{https://doi.org/#1}}
\bibcommenthead

\bibitem{Albarede2009}
\bibinfo{author}{Albarède, F.}
\newblock \bibinfo{title}{Volatile accretion history of the terrestrial planets and dynamic implications.}
\newblock \emph{\bibinfo{journal}{Nature}} \textbf{\bibinfo{volume}{461}}, \bibinfo{pages}{1227--1233} (\bibinfo{year}{2009}).

\bibitem{Wang2018}
\bibinfo{author}{Wang, H.~S.}, \bibinfo{author}{Lineweaver, C.~H.} \& \bibinfo{author}{Ireland, T.~R.}
\newblock \bibinfo{title}{{The elemental abundances (with uncertainties) of the most Earth-like planet}}.
\newblock \emph{\bibinfo{journal}{Icarus}} \textbf{\bibinfo{volume}{299}}, \bibinfo{pages}{460--474} (\bibinfo{year}{2018}).

\bibitem{Yoshizaki2020}
\bibinfo{author}{Yoshizaki, T.} \& \bibinfo{author}{McDonough, W.~F.}
\newblock \bibinfo{title}{{The composition of Mars}}.
\newblock \emph{\bibinfo{journal}{Geochim. Cosmochim. Acta}} \textbf{\bibinfo{volume}{273}}, \bibinfo{pages}{137--162} (\bibinfo{year}{2020}).

\bibitem{Wang2019}
\bibinfo{author}{Wang, H.~S.}, \bibinfo{author}{Lineweaver, C.~H.} \& \bibinfo{author}{Ireland, T.~R.}
\newblock \bibinfo{title}{{The volatility trend of protosolar and terrestrial elemental abundances}}.
\newblock \emph{\bibinfo{journal}{Icarus}} \bibinfo{pages}{287--305} (\bibinfo{year}{2019}).

\bibitem{Braukmuller2019}
\bibinfo{author}{Braukm{\"{u}}ller, N.}, \bibinfo{author}{Wombacher, F.}, \bibinfo{author}{Funk, C.} \& \bibinfo{author}{M{\"{u}}nker, C.}
\newblock \bibinfo{title}{{Earth's volatile element depletion pattern inherited from a carbonaceous chondrite-like source}}.
\newblock \emph{\bibinfo{journal}{Nat. Geosci.}} \textbf{\bibinfo{volume}{12}}, \bibinfo{pages}{564--568} (\bibinfo{year}{2019}).

\bibitem{Bond2010}
\bibinfo{author}{Bond, J.~C.}, \bibinfo{author}{O'Brien, D.~P.} \& \bibinfo{author}{Lauretta, D.~S.}
\newblock \bibinfo{title}{The compositional diversity of extrasolar terrestrial planets. i. in situ simulations}.
\newblock \emph{\bibinfo{journal}{Astrophys. J.}} \textbf{\bibinfo{volume}{715}}, \bibinfo{pages}{1050--1070} (\bibinfo{year}{2010}).

\bibitem{Sossi2022}
\bibinfo{author}{Sossi, P.~A.}, \bibinfo{author}{Stotz, I.~L.}, \bibinfo{author}{Jacobson, S.~A.}, \bibinfo{author}{Morbidelli, A.} \& \bibinfo{author}{O'Neill, H. S.~C.}
\newblock \bibinfo{title}{{Stochastic accretion of the Earth}}.
\newblock \emph{\bibinfo{journal}{Nat. Astron.}} \bibinfo{pages}{951--960} (\bibinfo{year}{2022}).

\bibitem{Lodders2025}
\bibinfo{author}{Lodders, K.}, \bibinfo{author}{Fegley, B.}, \bibinfo{author}{Mezger, K.} \& \bibinfo{author}{Ebel, D.}
\newblock \bibinfo{title}{Condensation and the volatility trend of the earth}.
\newblock \emph{\bibinfo{journal}{Space Sci. Rev.}} \textbf{\bibinfo{volume}{221}}, \bibinfo{pages}{54} (\bibinfo{year}{2025}).

\bibitem{Eatson2024}
\bibinfo{author}{Eatson, J.~W.}, \bibinfo{author}{Lichtenberg, T.}, \bibinfo{author}{Parker, R.~J.} \& \bibinfo{author}{Gerya, T.~V.}
\newblock \bibinfo{title}{Devolatilization of extrasolar planetesimals by $^{60}$Fe and $^{26}$Al heating}.
\newblock \emph{\bibinfo{journal}{Mon. Not. R. Astron. Soc.}} \textbf{\bibinfo{volume}{528}}, \bibinfo{pages}{6619--6630} (\bibinfo{year}{2024}).

\bibitem{Hin2017}
\bibinfo{author}{Hin, R.~C.} \emph{et~al.}
\newblock \bibinfo{title}{{Magnesium isotope evidence that accretional vapour loss shapes planetary compositions}}.
\newblock \emph{\bibinfo{journal}{Nature}} \textbf{\bibinfo{volume}{549}}, \bibinfo{pages}{511--515} (\bibinfo{year}{2017}).

\bibitem{Halliday2001}
\bibinfo{author}{Halliday, A.} \& \bibinfo{author}{Porcelli, D.}
\newblock \bibinfo{title}{In search of lost planets – the paleocosmochemistry of the inner solar system}.
\newblock \emph{\bibinfo{journal}{Earth Planet. Sci. Lett.}} \textbf{\bibinfo{volume}{192}}, \bibinfo{pages}{545--559} (\bibinfo{year}{2001}).

\bibitem{Norris2017}
\bibinfo{author}{Norris, C.~A.} \& \bibinfo{author}{Wood, B.~J.}
\newblock \bibinfo{title}{Earth’s volatile contents established by melting and vaporization}.
\newblock \emph{\bibinfo{journal}{Nature}} \textbf{\bibinfo{volume}{549}}, \bibinfo{pages}{507--510} (\bibinfo{year}{2017}).

\bibitem{Saurety2025}
\bibinfo{author}{Saurety, A.}, \bibinfo{author}{Caracas, R.} \& \bibinfo{author}{Raymond, S.~N.}
\newblock \bibinfo{title}{Impact-induced vaporization during accretion of planetary bodies}.
\newblock \emph{\bibinfo{journal}{Astrophys. J. Lett.}} \textbf{\bibinfo{volume}{981}}, \bibinfo{pages}{L13} (\bibinfo{year}{2025}).

\bibitem{Harrison2021}
\bibinfo{author}{Harrison, J.~H.}, \bibinfo{author}{Shorttle, O.} \& \bibinfo{author}{Bonsor, A.}
\newblock \bibinfo{title}{{Evidence for post-nebula volatilisation in an exo-planetary body}}.
\newblock \emph{\bibinfo{journal}{Earth Planet. Sci. Lett.}} \textbf{\bibinfo{volume}{554}}, \bibinfo{pages}{116694} (\bibinfo{year}{2021}).

\bibitem{Spina2021}
\bibinfo{author}{Spina, L.} \emph{et~al.}
\newblock \bibinfo{title}{Chemical evidence for planetary ingestion in a quarter of sun-like stars}.
\newblock \emph{\bibinfo{journal}{Nat. Astron.}} \textbf{\bibinfo{volume}{5}}, \bibinfo{pages}{1163–1169} (\bibinfo{year}{2021}).

\bibitem{Liu2024}
\bibinfo{author}{Liu, F.} \emph{et~al.}
\newblock \bibinfo{title}{{At least one in a dozen stars shows evidence of planetary ingestion}}.
\newblock \emph{\bibinfo{journal}{Nature}} \textbf{\bibinfo{volume}{627}}, \bibinfo{pages}{501--504} (\bibinfo{year}{2024}).

\bibitem{Morbidelli2025}
\bibinfo{author}{Morbidelli, A.}, \bibinfo{author}{Kleine, T.} \& \bibinfo{author}{Nimmo, F.}
\newblock \bibinfo{title}{{Did the terrestrial planets of the solar system form by pebble accretion?}}
\newblock \emph{\bibinfo{journal}{Earth Planet. Sci. Lett.}} \textbf{\bibinfo{volume}{650}}, \bibinfo{pages}{119120} (\bibinfo{year}{2025}).

\bibitem{Johansen2024}
\bibinfo{author}{Johansen, A.}, \bibinfo{author}{Olson, P.} \& \bibinfo{author}{Sharp, Z.}
\newblock \bibinfo{title}{{Comment on "Did the terrestrial planets of the Solar System form by pebble accretion?"}}.
\newblock \emph{\bibinfo{journal}{Preprint at arXiv:2411.17043}}  (\bibinfo{year}{2024}).

\bibitem{Johansen2021}
\bibinfo{author}{Johansen, A.} \emph{et~al.}
\newblock \bibinfo{title}{{A pebble accretion model for the formation of the terrestrial planets in the solar system}}.
\newblock \emph{\bibinfo{journal}{Sci. Adv.}} \textbf{\bibinfo{volume}{7}}, \bibinfo{pages}{1--14} (\bibinfo{year}{2021}).

\bibitem{Bizzarro2025}
\bibinfo{author}{Bizzarro, M.}, \bibinfo{author}{Johansen, A.} \& \bibinfo{author}{Dorn, C.}
\newblock \bibinfo{title}{The cosmochemistry of planetary systems}.
\newblock \emph{\bibinfo{journal}{Nat. Rev. Chem.}} \textbf{\bibinfo{volume}{9}}, \bibinfo{pages}{378--396} (\bibinfo{year}{2025}).

\bibitem{Burkhardt+etal2021}
\bibinfo{author}{{Burkhardt}, C.} \emph{et~al.}
\newblock \bibinfo{title}{{Terrestrial planet formation from lost inner solar system material}}.
\newblock \emph{\bibinfo{journal}{Sci. Adv.}} \textbf{\bibinfo{volume}{7}}, \bibinfo{pages}{eabj7601} (\bibinfo{year}{2021}).

\bibitem{Morbidelli2022}
\bibinfo{author}{{Morbidelli}, A.} \emph{et~al.}
\newblock \bibinfo{title}{{Contemporary formation of early Solar System planetesimals at two distinct radial locations}}.
\newblock \emph{\bibinfo{journal}{Nat. Astron.}} \textbf{\bibinfo{volume}{6}}, \bibinfo{pages}{72--79} (\bibinfo{year}{2022}).

\bibitem{Lambrechts2019}
\bibinfo{author}{{Lambrechts}, M.} \emph{et~al.}
\newblock \bibinfo{title}{{Formation of planetary systems by pebble accretion and migration. How the radial pebble flux determines a terrestrial-planet or super-Earth growth mode}}.
\newblock \emph{\bibinfo{journal}{Astron. Astrophys.}} \textbf{\bibinfo{volume}{627}}, \bibinfo{pages}{A83} (\bibinfo{year}{2019}).

\bibitem{Bethune2017}
\bibinfo{author}{{B{\'e}thune}, W.}, \bibinfo{author}{{Lesur}, G.} \& \bibinfo{author}{{Ferreira}, J.}
\newblock \bibinfo{title}{{Global simulations of protoplanetary disks with net magnetic flux. I. Non-ideal MHD case}}.
\newblock \emph{\bibinfo{journal}{Astron. Astrophys.}} \textbf{\bibinfo{volume}{600}}, \bibinfo{pages}{A75} (\bibinfo{year}{2017}).

\bibitem{Mori2021}
\bibinfo{author}{{Mori}, S.}, \bibinfo{author}{{Okuzumi}, S.}, \bibinfo{author}{{Kunitomo}, M.} \& \bibinfo{author}{{Bai}, X.-N.}
\newblock \bibinfo{title}{{Evolution of the Water Snow Line in Magnetically Accreting Protoplanetary Disks}}.
\newblock \emph{\bibinfo{journal}{Astrophys. J.}} \textbf{\bibinfo{volume}{916}}, \bibinfo{pages}{72} (\bibinfo{year}{2021}).

\bibitem{Grewal2021}
\bibinfo{author}{Grewal, D.~S.}, \bibinfo{author}{Dasgupta, R.} \& \bibinfo{author}{Marty, B.}
\newblock \bibinfo{title}{A very early origin of isotopically distinct nitrogen in inner solar system protoplanets}.
\newblock \emph{\bibinfo{journal}{Nat. Astron.}} \textbf{\bibinfo{volume}{5}}, \bibinfo{pages}{356--364} (\bibinfo{year}{2021}).

\bibitem{Grewal2022}
\bibinfo{author}{Grewal, D.~S.}, \bibinfo{author}{Seales, J.~D.} \& \bibinfo{author}{Dasgupta, R.}
\newblock \bibinfo{title}{Internal or external magma oceans in the earliest protoplanets – perspectives from nitrogen and carbon fractionation}.
\newblock \emph{\bibinfo{journal}{Earth Planet. Sci. Lett.}} \textbf{\bibinfo{volume}{598}}, \bibinfo{pages}{117847} (\bibinfo{year}{2022}).

\bibitem{Grewal2024}
\bibinfo{author}{{Grewal}, D.~S.}, \bibinfo{author}{{Nie}, N.~X.}, \bibinfo{author}{{Zhang}, B.}, \bibinfo{author}{{Izidoro}, A.} \& \bibinfo{author}{{Asimow}, P.~D.}
\newblock \bibinfo{title}{{Accretion of the earliest inner Solar System planetesimals beyond the water snowline}}.
\newblock \emph{\bibinfo{journal}{Nat. Astron.}} \textbf{\bibinfo{volume}{8}}, \bibinfo{pages}{290--297} (\bibinfo{year}{2024}).

\bibitem{Grewal2025}
\bibinfo{author}{{Grewal}, D.~S.}, \bibinfo{author}{{Bhattacharjee}, S.}, \bibinfo{author}{{Zhang}, B.}, \bibinfo{author}{{Nie}, N.~X.} \& \bibinfo{author}{{Miyazaki}, Y.}
\newblock \bibinfo{title}{{Enrichment of moderately volatile elements in first-generation planetesimals of the inner Solar System}}.
\newblock \emph{\bibinfo{journal}{Sci. Adv.}} \textbf{\bibinfo{volume}{11}}, \bibinfo{pages}{eadq7848} (\bibinfo{year}{2025}).

\bibitem{Nakamura2022}
\bibinfo{author}{Nakamura, E.} \emph{et~al.}
\newblock \bibinfo{title}{On the origin and evolution of the asteroid ryugu: A comprehensive geochemical perspective}.
\newblock \emph{\bibinfo{journal}{Proc. Jpn. Acad. Ser. B Phys. Biol. Sci.}} \textbf{\bibinfo{volume}{98}}, \bibinfo{pages}{227--282} (\bibinfo{year}{2022}).

\bibitem{Lauretta2024}
\bibinfo{author}{Lauretta, D.~S.} \emph{et~al.}
\newblock \bibinfo{title}{Asteroid (101955) bennu in the laboratory: Properties of the sample collected by OSIRIS-REx}.
\newblock \emph{\bibinfo{journal}{Meteorit. Planet. Sci.}} \textbf{\bibinfo{volume}{59}}, \bibinfo{pages}{2453--2486} (\bibinfo{year}{2024}).

\bibitem{Lichtenberg2016}
\bibinfo{author}{Lichtenberg, T.}, \bibinfo{author}{Golabek, G.~J.}, \bibinfo{author}{Gerya, T.~V.} \& \bibinfo{author}{Meyer, M.~R.}
\newblock \bibinfo{title}{The effects of short-lived radionuclides and porosity on the early thermo-mechanical evolution of planetesimals}.
\newblock \emph{\bibinfo{journal}{Icarus}} \textbf{\bibinfo{volume}{274}}, \bibinfo{pages}{350--365} (\bibinfo{year}{2016}).

\bibitem{Russell2012}
\bibinfo{author}{{Russell}, C.~T.} \emph{et~al.}
\newblock \bibinfo{title}{{Dawn at Vesta: Testing the Protoplanetary Paradigm}}.
\newblock \emph{\bibinfo{journal}{Science}} \textbf{\bibinfo{volume}{336}}, \bibinfo{pages}{684} (\bibinfo{year}{2012}).

\bibitem{Mittlefehldt2015}
\bibinfo{author}{Mittlefehldt, D.~W.}
\newblock \bibinfo{title}{Asteroid (4) vesta: I. the howardite-eucrite-diogenite (hed) clan of meteorites}.
\newblock \emph{\bibinfo{journal}{Chem. Erde.}} \textbf{\bibinfo{volume}{75}}, \bibinfo{pages}{155--183} (\bibinfo{year}{2015}).

\bibitem{Fang2024}
\bibinfo{author}{{Fang}, L.} \emph{et~al.}
\newblock \bibinfo{title}{{The origin of 4-Vesta's volatile depletion revealed by the zinc isotopic composition of diogenites}}.
\newblock \emph{\bibinfo{journal}{Sci. Adv.}} \textbf{\bibinfo{volume}{10}}, \bibinfo{pages}{eadl1007} (\bibinfo{year}{2024}).

\bibitem{Steenstra2018}
\bibinfo{author}{{Steenstra}, E.~S.} \emph{et~al.}
\newblock \bibinfo{title}{{Depletion of potassium and sodium in mantles of Mars, Moon and Vesta by core formation}}.
\newblock \emph{\bibinfo{journal}{Sci. Rep.}} \textbf{\bibinfo{volume}{8}}, \bibinfo{pages}{7053} (\bibinfo{year}{2018}).

\bibitem{Hu2022}
\bibinfo{author}{{Hu}, Y.}, \bibinfo{author}{{Moynier}, F.} \& \bibinfo{author}{{Bizzarro}, M.}
\newblock \bibinfo{title}{{Potassium isotope heterogeneity in the early Solar System controlled by extensive evaporation and partial recondensation}}.
\newblock \emph{\bibinfo{journal}{Nat. Commun.}} \textbf{\bibinfo{volume}{13}}, \bibinfo{pages}{7669} (\bibinfo{year}{2022}).

\bibitem{Steinmeyer2023}
\bibinfo{author}{Steinmeyer, M.-L.}, \bibinfo{author}{Woitke, P.} \& \bibinfo{author}{Johansen, A.}
\newblock \bibinfo{title}{{Sublimation of refractory minerals in the gas envelopes of accreting rocky planets}}.
\newblock \emph{\bibinfo{journal}{Astron. Astrophys.}} \textbf{\bibinfo{volume}{181}}, \bibinfo{pages}{1--16} (\bibinfo{year}{2023}).

\bibitem{KurokawaTanigawa2018}
\bibinfo{author}{{Kurokawa}, H.} \& \bibinfo{author}{{Tanigawa}, T.}
\newblock \bibinfo{title}{{Suppression of atmospheric recycling of planets embedded in a protoplanetary disc by buoyancy barrierc}}.
\newblock \emph{\bibinfo{journal}{Mon. Not. R. Astron. Soc.}} \textbf{\bibinfo{volume}{479}}, \bibinfo{pages}{635--648} (\bibinfo{year}{2018}).

\bibitem{Kuwahara2026}
\bibinfo{author}{Kuwahara, A.} \& \bibinfo{author}{Lambrechts, M.}
\newblock \bibinfo{title}{Interior dynamics of envelopes around disk-embedded planets}.
\newblock \emph{\bibinfo{journal}{Astron. Astrophys.}} \textbf{\bibinfo{volume}{707}}, \bibinfo{pages}{A26} (\bibinfo{year}{2026}).

\bibitem{Canup2001}
\bibinfo{author}{Canup, R.~M.} \& \bibinfo{author}{Asphaug, E.}
\newblock \bibinfo{title}{Origin of the moon in a giant impact near the end of the earth's formation}.
\newblock \emph{\bibinfo{journal}{Nature}} \textbf{\bibinfo{volume}{412}}, \bibinfo{pages}{708--712} (\bibinfo{year}{2001}).

\bibitem{Gabriel2023}
\bibinfo{author}{Gabriel, T.~S.} \& \bibinfo{author}{Cambioni, S.}
\newblock \bibinfo{title}{The role of giant impacts in planet formation}.
\newblock \emph{\bibinfo{journal}{Annu. Rev. Earth Planet. Sci.}} \textbf{\bibinfo{volume}{51}}, \bibinfo{pages}{671--695} (\bibinfo{year}{2023}).

\bibitem{Huang2025}
\bibinfo{author}{Huang, S.}, \bibinfo{author}{Ormel, C.~W.}, \bibinfo{author}{Zwart, S.~P.}, \bibinfo{author}{Kokubo, E.} \& \bibinfo{author}{Yi, T.}
\newblock \bibinfo{title}{A resonant beginning for the solar system’s terrestrial planets}.
\newblock \emph{\bibinfo{journal}{Astrophys. J.}} \textbf{\bibinfo{volume}{988}}, \bibinfo{pages}{137} (\bibinfo{year}{2025}).

\bibitem{Quintana2016}
\bibinfo{author}{{Quintana}, E.~V.}, \bibinfo{author}{{Barclay}, T.}, \bibinfo{author}{{Borucki}, W.~J.}, \bibinfo{author}{{Rowe}, J.~F.} \& \bibinfo{author}{{Chambers}, J.~E.}
\newblock \bibinfo{title}{{The frequency of giant impacts on Earth-like worlds}}.
\newblock \emph{\bibinfo{journal}{Astrophys. J.}} \textbf{\bibinfo{volume}{821}}, \bibinfo{pages}{126} (\bibinfo{year}{2016}).

\bibitem{Dauphas2011}
\bibinfo{author}{{Dauphas}, N.} \& \bibinfo{author}{{Pourmand}, A.}
\newblock \bibinfo{title}{{Hf-W-Th evidence for rapid growth of Mars and its status as a planetary embryo}}.
\newblock \emph{\bibinfo{journal}{Nature}} \textbf{\bibinfo{volume}{473}}, \bibinfo{pages}{489--492} (\bibinfo{year}{2011}).

\bibitem{Onyett2023}
\bibinfo{author}{Onyett, I.~J.} \emph{et~al.}
\newblock \bibinfo{title}{{Silicon isotope constraints on terrestrial planet accretion}}.
\newblock \emph{\bibinfo{journal}{Nature}} \textbf{\bibinfo{volume}{619}}, \bibinfo{pages}{539--544} (\bibinfo{year}{2023}).

\bibitem{Levison2015}
\bibinfo{author}{{Levison}, H.~F.}, \bibinfo{author}{{Kretke}, K.~A.} \& \bibinfo{author}{{Duncan}, M.~J.}
\newblock \bibinfo{title}{{Growing the gas-giant planets by the gradual accumulation of pebbles}}.
\newblock \emph{\bibinfo{journal}{Nature}} \textbf{\bibinfo{volume}{524}}, \bibinfo{pages}{322--324} (\bibinfo{year}{2015}).

\bibitem{Calogero2025}
\bibinfo{author}{Calogero, M.~A.}, \bibinfo{author}{Nimmo, F.} \& \bibinfo{author}{Hin, R.~C.}
\newblock \bibinfo{title}{Can impact-induced heating drive moderately volatile element loss?}
\newblock \emph{\bibinfo{journal}{Earth Planet. Sci. Lett.}} \textbf{\bibinfo{volume}{669}}, \bibinfo{pages}{119580} (\bibinfo{year}{2025}).

\bibitem{Kominami2002}
\bibinfo{author}{{Kominami}, J.} \& \bibinfo{author}{{Ida}, S.}
\newblock \bibinfo{title}{{The effect of tidal interaction with a gas disk on formation of terrestrial planets}}.
\newblock \emph{\bibinfo{journal}{Icarus}} \textbf{\bibinfo{volume}{157}}, \bibinfo{pages}{43--56} (\bibinfo{year}{2002}).

\bibitem{Fischer2025}
\bibinfo{author}{Fischer, R.~A.} \& \bibinfo{author}{McDonough, W.~F.}
\newblock \bibinfo{title}{Earth's core composition and core formation}.
\newblock \emph{\bibinfo{journal}{Treatise Geochem.}} \textbf{\bibinfo{volume}{1}}, \bibinfo{pages}{17--71} (\bibinfo{year}{2025}).

\bibitem{Stahler2021}
\bibinfo{author}{Stähler, S.~C.} \emph{et~al.}
\newblock \bibinfo{title}{Seismic detection of the martian core}.
\newblock \emph{\bibinfo{journal}{Science}} \textbf{\bibinfo{volume}{373}}, \bibinfo{pages}{443--448} (\bibinfo{year}{2021}).

\bibitem{Khan2022}
\bibinfo{author}{Khan, A.}, \bibinfo{author}{Sossi, P.~A.}, \bibinfo{author}{Liebske, C.}, \bibinfo{author}{Rivoldini, A.} \& \bibinfo{author}{Giardini, D.}
\newblock \bibinfo{title}{Geophysical and cosmochemical evidence for a volatile-rich mars}.
\newblock \emph{\bibinfo{journal}{Earth Planet. Sci. Lett.}} \textbf{\bibinfo{volume}{578}}, \bibinfo{pages}{117330} (\bibinfo{year}{2022}).

\bibitem{OlsonSharp2023}
\bibinfo{author}{{Olson}, P.~L.} \& \bibinfo{author}{{Sharp}, Z.~D.}
\newblock \bibinfo{title}{{Hafnium-tungsten evolution with pebble accretion during Earth formation}}.
\newblock \emph{\bibinfo{journal}{Earth Planet. Sci. Lett.}} \textbf{\bibinfo{volume}{622}}, \bibinfo{pages}{118418} (\bibinfo{year}{2023}).

\bibitem{Yu2011}
\bibinfo{author}{{Yu}, G.} \& \bibinfo{author}{{Jacobsen}, S.~B.}
\newblock \bibinfo{title}{{Fast accretion of the Earth with a late Moon-forming giant impact}}.
\newblock \emph{\bibinfo{journal}{Proc. Natl. Acad. Sci. U.S.A.}} \textbf{\bibinfo{volume}{108}}, \bibinfo{pages}{17604--17609} (\bibinfo{year}{2011}).

\bibitem{Bizzarro2025b}
\bibinfo{author}{Bizzarro, M.} \emph{et~al.}
\newblock \bibinfo{title}{Interstellar ices as carriers of supernova material to the early solar system}.
\newblock \emph{\bibinfo{journal}{Nat. Commun.}} \textbf{\bibinfo{volume}{16}}, \bibinfo{pages}{10657} (\bibinfo{year}{2025}).

\bibitem{Sossi2026}
\bibinfo{author}{Sossi, P.~A.} \& \bibinfo{author}{Bower, D.~J.}
\newblock \bibinfo{title}{Homogeneous accretion of the earth in the inner solar system}.
\newblock \emph{\bibinfo{journal}{Nat. Astron.}} \textbf{\bibinfo{volume}{10}}, \bibinfo{pages}{972–979} (\bibinfo{year}{2026}).

\bibitem{Houge2025}
\bibinfo{author}{Houge, A.} \emph{et~al.}
\newblock \bibinfo{title}{Burned to ashes: How the thermal decomposition of refractory organics in the inner protoplanetary disc impacts the gas-phase C/O ratio}.
\newblock \emph{\bibinfo{journal}{Astron. Astrophys.}} \textbf{\bibinfo{volume}{699}}, \bibinfo{pages}{A227} (\bibinfo{year}{2025}).

\bibitem{Raymond2009}
\bibinfo{author}{{Raymond}, S.~N.}, \bibinfo{author}{{O'Brien}, D.~P.}, \bibinfo{author}{{Morbidelli}, A.} \& \bibinfo{author}{{Kaib}, N.~A.}
\newblock \bibinfo{title}{{Building the terrestrial planets: Constrained accretion in the inner Solar System}}.
\newblock \emph{\bibinfo{journal}{Icarus}} \textbf{\bibinfo{volume}{203}}, \bibinfo{pages}{644--662} (\bibinfo{year}{2009}).

\bibitem{Lyubetskaya2007}
\bibinfo{author}{Lyubetskaya, T.} \& \bibinfo{author}{Korenaga, J.}
\newblock \bibinfo{title}{Chemical composition of earth's primitive mantle and its variance: 1. method and results}.
\newblock \emph{\bibinfo{journal}{J. Geophys. Res. Solid Earth}} \textbf{\bibinfo{volume}{112}}, \bibinfo{pages}{1--21} (\bibinfo{year}{2007}).

\bibitem{McDonough2008}
\bibinfo{author}{McDonough, W.~F.} \& \bibinfo{author}{Arevalo, R.}
\newblock \bibinfo{title}{Uncertainties in the composition of earth, its core and silicate sphere}.
\newblock \emph{\bibinfo{journal}{J. Phys.: Conf. Ser.}} \textbf{\bibinfo{volume}{136}}, \bibinfo{pages}{022006} (\bibinfo{year}{2008}).

\bibitem{Palme2014}
\bibinfo{author}{Palme, H.} \& \bibinfo{author}{O'Neill, H.}
\newblock \bibinfo{title}{{Cosmochemical estimates of mantle composition}}.
\newblock \emph{\bibinfo{journal}{Treatise Geochem.}} \bibinfo{pages}{1--39} (\bibinfo{year}{2014}).

\bibitem{McDonough2025}
\bibinfo{author}{McDonough, W.~F.}
\newblock \bibinfo{title}{Earth’s composition: Origin, energy budget, and insights from geoneutrinos}.
\newblock \emph{\bibinfo{journal}{Geochim. Cosmochim. Acta}} \bibinfo{pages}{9--34} (\bibinfo{year}{2025}).

\bibitem{Wang2026a}
\bibinfo{author}{Wang, H.~S.} \emph{et~al.}
\newblock \bibinfo{title}{Volatile depletion in rocky planets as a chemical fingerprint of hybrid accretion}.
\newblock \bibinfo{publisher}{figshare}, \bibinfo{type}{Dataset}
(\bibinfo{year}{2026}).
\newblock \doi{10.6084/m9.figshare.33085055}.

\bibitem{Clay2017}
\bibinfo{author}{{Clay}, P.~L.} \emph{et~al.}
\newblock \bibinfo{title}{{Halogens in chondritic meteorites and terrestrial accretion}}.
\newblock \emph{\bibinfo{journal}{Nature}} \textbf{\bibinfo{volume}{551}}, \bibinfo{pages}{614--618} (\bibinfo{year}{2017}).

\bibitem{Palme2021b}
\bibinfo{author}{Palme, H.} \& \bibinfo{author}{Zipfel, J.}
\newblock \bibinfo{title}{The composition of CI chondrites and their contents of chlorine and bromine: Results from instrumental neutron activation analysis}.
\newblock \emph{\bibinfo{journal}{Meteorit. Planet. Sci.}} \textbf{\bibinfo{volume}{17}}, \bibinfo{pages}{1--17} (\bibinfo{year}{2021}).

\bibitem{Lodders2023}
\bibinfo{author}{Lodders, K.} \& \bibinfo{author}{Fegley, B.}
\newblock \bibinfo{title}{Solar system abundances and condensation temperatures of the halogens fluorine, chlorine, bromine, and iodine}.
\newblock \emph{\bibinfo{journal}{Geochem.}} \textbf{\bibinfo{volume}{83}}, \bibinfo{pages}{125957} (\bibinfo{year}{2023}).

\bibitem{Campbell2012}
\bibinfo{author}{Campbell, I.~H.} \& \bibinfo{author}{O'Neill, H. S.~C.}
\newblock \bibinfo{title}{Evidence against a chondritic Earth}.
\newblock \emph{\bibinfo{journal}{Nature}} \textbf{\bibinfo{volume}{483}}, \bibinfo{pages}{553--558} (\bibinfo{year}{2012}).

\bibitem{PisoYoudin2014}
\bibinfo{author}{{Piso}, A.-M.~A.} \& \bibinfo{author}{{Youdin}, A.~N.}
\newblock \bibinfo{title}{{On the minimum core mass for giant planet formation at wide separations}}.
\newblock \emph{\bibinfo{journal}{Astrophys. J.}} \textbf{\bibinfo{volume}{786}}, \bibinfo{pages}{21} (\bibinfo{year}{2014}).

\bibitem{Ida2016}
\bibinfo{author}{{Ida}, S.}, \bibinfo{author}{{Guillot}, T.} \& \bibinfo{author}{{Morbidelli}, A.}
\newblock \bibinfo{title}{{The radial dependence of pebble accretion rates: A source of diversity in planetary systems. I. Analytical formulation}}.
\newblock \emph{\bibinfo{journal}{Astron. Astrophys.}} \textbf{\bibinfo{volume}{591}}, \bibinfo{pages}{A72} (\bibinfo{year}{2016}).

\bibitem{Brouwers2020}
\bibinfo{author}{Brouwers, M.~G.} \& \bibinfo{author}{Ormel, C.~W.}
\newblock \bibinfo{title}{How planets grow by pebble accretion}.
\newblock \emph{\bibinfo{journal}{Astron. Astrophys.}} \textbf{\bibinfo{volume}{634}}, \bibinfo{pages}{A15} (\bibinfo{year}{2020}).

\bibitem{Timmermann2023}
\bibinfo{author}{Timmermann, A.}, \bibinfo{author}{Shan, Y.}, \bibinfo{author}{Reiners, A.} \& \bibinfo{author}{Pack, A.}
\newblock \bibinfo{title}{{Revisiting equilibrium condensation and rocky planet compositions. Introducing the ECCOplanets code}}.
\newblock \emph{\bibinfo{journal}{Astron. Astrophys.}} \textbf{\bibinfo{volume}{52}}, \bibinfo{pages}{A52} (\bibinfo{year}{2023}).

\bibitem{Zaveri2026}
\bibinfo{author}{Zaveri, U.}, \bibinfo{author}{Wang, H.~S.} \& \bibinfo{author}{Sossi, P.~A.}
\newblock \bibinfo{title}{A chemical perspective on planet formation in reduced systems}.
\newblock \emph{\bibinfo{journal}{Astron. Astrophys.}} \textbf{\bibinfo{volume}{709}}, \bibinfo{pages}{A223} (\bibinfo{year}{2026}).

\bibitem{HabibPierrehumbert2024}
\bibinfo{author}{{Habib}, N.} \& \bibinfo{author}{{Pierrehumbert}, R.~T.}
\newblock \bibinfo{title}{{Modeling noncondensing compositional convection for applications to super-Earth and sub-Neptune Atmospheres}}.
\newblock \emph{\bibinfo{journal}{Astrophys. J.}} \textbf{\bibinfo{volume}{961}}, \bibinfo{pages}{35} (\bibinfo{year}{2024}).

\bibitem{Popovas2018}
\bibinfo{author}{{Popovas}, A.}, \bibinfo{author}{{Nordlund}, {\r{A}}.}, \bibinfo{author}{{Ramsey}, J.~P.} \& \bibinfo{author}{{Ormel}, C.~W.}
\newblock \bibinfo{title}{{Pebble dynamics and accretion on to rocky planets - I. Adiabatic and convective models}}.
\newblock \emph{\bibinfo{journal}{Mon. Not. R. Astron. Soc.}} \textbf{\bibinfo{volume}{479}}, \bibinfo{pages}{5136--5156} (\bibinfo{year}{2018}).

\bibitem{Wang2023}
\bibinfo{author}{Wang, Y.}, \bibinfo{author}{Ormel, C.~W.}, \bibinfo{author}{Huang, P.} \& \bibinfo{author}{Kuiper, R.}
\newblock \bibinfo{title}{Atmospheric recycling of volatiles by pebble-accreting planets}.
\newblock \emph{\bibinfo{journal}{Mon. Not. R. Astron. Soc.}} \textbf{\bibinfo{volume}{523}}, \bibinfo{pages}{6186--6207} (\bibinfo{year}{2023}).

\bibitem{Wang2019b}
\bibinfo{author}{Wang, H.~S.} \emph{et~al.}
\newblock \bibinfo{title}{Enhanced constraints on the interior composition and structure of terrestrial exoplanets}.
\newblock \emph{\bibinfo{journal}{Mon. Not. R. Astron. Soc.}} \textbf{\bibinfo{volume}{482}}, \bibinfo{pages}{2222--2233} (\bibinfo{year}{2019}).

\bibitem{Aguilera-Gomez2024}
\bibinfo{author}{{Aguilera-G{\'o}mez}, C.} \emph{et~al.}
\newblock \bibinfo{title}{{Host star and exoplanet composition: Polluted white dwarf reveals depletion of moderately refractory elements in planetary material}}.
\newblock \emph{\bibinfo{journal}{Astron. Astrophys.}} \textbf{\bibinfo{volume}{693}}, \bibinfo{pages}{A64} (\bibinfo{year}{2025}).

\bibitem{Stone2020}
\bibinfo{author}{{Stone}, J.~M.}, \bibinfo{author}{{Tomida}, K.}, \bibinfo{author}{{White}, C.~J.} \& \bibinfo{author}{{Felker}, K.~G.}
\newblock \bibinfo{title}{{The Athena++ Adaptive Mesh Refinement Framework: Design and Magnetohydrodynamic Solvers}}.
\newblock \emph{\bibinfo{journal}{Astrophys. J. Suppl. Ser.}} \textbf{\bibinfo{volume}{249}}, \bibinfo{pages}{4} (\bibinfo{year}{2020}).

\bibitem{HuLi2022}
\bibinfo{author}{Hu, X.}, \bibinfo{author}{Li, Z.-Y.}, \bibinfo{author}{Zhu, Z.} \& \bibinfo{author}{Yang, C.-C.}
\newblock \bibinfo{title}{Formation of dust rings and gaps in non-ideal mhd discs through meridional gas flows}.
\newblock \emph{\bibinfo{journal}{Mon. Not. R. Astron. Soc.}} \textbf{\bibinfo{volume}{516}}, \bibinfo{pages}{2006--2022} (\bibinfo{year}{2022}).

\bibitem{Baronett2024}
\bibinfo{author}{Baronett, S.~A.}, \bibinfo{author}{Yang, C.-C.} \& \bibinfo{author}{Zhu, Z.}
\newblock \bibinfo{title}{Dust–gas dynamics driven by the streaming instability with various pressure gradients}.
\newblock \emph{\bibinfo{journal}{Mon. Not. R. Astron. Soc.}} \textbf{\bibinfo{volume}{529}}, \bibinfo{pages}{275--295} (\bibinfo{year}{2024}).

\bibitem{Zhu2021}
\bibinfo{author}{{Zhu}, Z.} \emph{et~al.}
\newblock \bibinfo{title}{{Global 3D radiation hydrodynamic simulations of proto-Jupiter's convective envelope}}.
\newblock \emph{\bibinfo{journal}{Mon. Not. R. Astron. Soc.}} \textbf{\bibinfo{volume}{508}}, \bibinfo{pages}{453--474} (\bibinfo{year}{2021}).

\bibitem{Johansen2020}
\bibinfo{author}{Johansen, A.} \& \bibinfo{author}{Åke Nordlund}.
\newblock \bibinfo{title}{Transport, destruction, and growth of pebbles in the gas envelope of a protoplanet}.
\newblock \emph{\bibinfo{journal}{Astrophys. J.}} \textbf{\bibinfo{volume}{903}}, \bibinfo{pages}{102} (\bibinfo{year}{2020}).

\bibitem{Lodders2021}
\bibinfo{author}{Lodders, K.}
\newblock \bibinfo{title}{Relative atomic solar system abundances, mass fractions, and atomic masses of the elements and their isotopes, composition of the solar photosphere, and compositions of the major chondritic meteorite groups}.
\newblock \emph{\bibinfo{journal}{Space Sci. Rev.}} \textbf{\bibinfo{volume}{217}}, \bibinfo{pages}{44} (\bibinfo{year}{2021}).

\bibitem{Wang2026b}
\bibinfo{author}{Wang, H.~S.} \emph{et~al.}
\newblock \bibinfo{title}{Volatile depletion in rocky planets as a chemical fingerprint of hybrid accretion}.
\newblock \bibinfo{publisher}{Code Ocean}, \bibinfo{type}{Source code}
(\bibinfo{year}{2026}).
\newblock \doi{10.24433/CO.8499849.v1}.

\bibitem{Skilling2006}
\bibinfo{author}{Skilling, J.}
\newblock \bibinfo{title}{Nested sampling for general bayesian computation}.
\newblock \emph{\bibinfo{journal}{Bayesian Anal.}} \textbf{\bibinfo{volume}{1}}, \bibinfo{pages}{833--860} (\bibinfo{year}{2006}).

\bibitem{jeffreys1998theory}
\bibinfo{author}{Jeffreys, H.}
\newblock \emph{\bibinfo{title}{Theory of Probability}} Oxford Classic Texts in the Physical Sciences (\bibinfo{publisher}{Oxford University Press}, \bibinfo{address}{Oxford}, \bibinfo{year}{1998}).
\newblock \bibinfo{note}{3rd Edition (First published 1939)}.

\bibitem{Trotta2008}
\bibinfo{author}{Trotta, R.}
\newblock \bibinfo{title}{Bayes in the sky: Bayesian inference and model selection in cosmology}.
\newblock \emph{\bibinfo{journal}{Contemp. Phys.}} \textbf{\bibinfo{volume}{49}}, \bibinfo{pages}{71--104} (\bibinfo{year}{2008}).

\end{thebibliography}

\begin{thebibliography}{10}
\expandafter\ifx\csname url\endcsname\relax
  \def\url#1{\burl{#1}}\fi
\expandafter\ifx\csname urlprefix\endcsname\relax\def\urlprefix{URL }\fi
\providecommand{\bibinfo}[2]{#2}
\providecommand{\eprint}[2][]{\url{#2}}
\providecommand{\doi}[1]{\url{https://doi.org/#1}}
\bibcommenthead

\bibitem{Chabot2003S}
\bibinfo{author}{Chabot, N.~L.} \& \bibinfo{author}{Agee, C.~B.}
\newblock \bibinfo{title}{Core formation in the earth and moon: New experimental constraints from v, cr, and mn}.
\newblock \emph{\bibinfo{journal}{Geochim. Cosmochim. Acta}} \textbf{\bibinfo{volume}{67}}, \bibinfo{pages}{2077--2091} (\bibinfo{year}{2003}).

\bibitem{Siebert2011S}
\bibinfo{author}{Siebert, J.}, \bibinfo{author}{Corgne, A.} \& \bibinfo{author}{Ryerson, F.~J.}
\newblock \bibinfo{title}{Systematics of metal-silicate partitioning for many siderophile elements applied to earth's core formation}.
\newblock \emph{\bibinfo{journal}{Geochim. Cosmochim. Acta}} \textbf{\bibinfo{volume}{75}}, \bibinfo{pages}{1451--1489} (\bibinfo{year}{2011}).

\bibitem{Mann2009S}
\bibinfo{author}{Mann, U.}, \bibinfo{author}{Frost, D.~J.} \& \bibinfo{author}{Rubie, D.~C.}
\newblock \bibinfo{title}{Evidence for high-pressure core-mantle differentiation from the metal-silicate partitioning of lithophile and weakly-siderophile elements}.
\newblock \emph{\bibinfo{journal}{Geochim. Cosmochim. Acta}} \textbf{\bibinfo{volume}{73}}, \bibinfo{pages}{7360--7386} (\bibinfo{year}{2009}).

\bibitem{Fischer2015S}
\bibinfo{author}{Fischer, R.~A.} \emph{et~al.}
\newblock \bibinfo{title}{High pressure metal-silicate partitioning of ni, co, v, cr, si, and o}.
\newblock \emph{\bibinfo{journal}{Geochim. Cosmochim. Acta}} \textbf{\bibinfo{volume}{167}}, \bibinfo{pages}{177--194} (\bibinfo{year}{2015}).

\bibitem{Huang2018S}
\bibinfo{author}{Huang, D.} \& \bibinfo{author}{Badro, J.}
\newblock \bibinfo{title}{Fe-ni ideality during core formation on earth}.
\newblock \emph{\bibinfo{journal}{American Mineralogist}} \textbf{\bibinfo{volume}{103}}, \bibinfo{pages}{1707--1710} (\bibinfo{year}{2018}).

\bibitem{Wang2016S}
\bibinfo{author}{Wang, Z.}, \bibinfo{author}{Laurenz, V.}, \bibinfo{author}{Petitgirard, S.} \& \bibinfo{author}{Becker, H.}
\newblock \bibinfo{title}{Earth's moderately volatile element composition may not be chondritic: Evidence from in, cd and zn}.
\newblock \emph{\bibinfo{journal}{Earth Planet. Sci. Lett.}} \textbf{\bibinfo{volume}{435}}, \bibinfo{pages}{136--146} (\bibinfo{year}{2016}).

\bibitem{Mahan2017S}
\bibinfo{author}{Mahan, B.}, \bibinfo{author}{Siebert, J.}, \bibinfo{author}{Pringle, E.~A.} \& \bibinfo{author}{Moynier, F.}
\newblock \bibinfo{title}{Elemental partitioning and isotopic fractionation of zn between metal and silicate and geochemical estimation of the s content of the earth's core}.
\newblock \emph{\bibinfo{journal}{Geochim. Cosmochim. Acta}} \textbf{\bibinfo{volume}{196}}, \bibinfo{pages}{252--270} (\bibinfo{year}{2017}).

\bibitem{Yoshizaki2020S}
\bibinfo{author}{Yoshizaki, T.} \& \bibinfo{author}{McDonough, W.~F.}
\newblock \bibinfo{title}{{The composition of Mars}}.
\newblock \emph{\bibinfo{journal}{Geochim. Cosmochim. Acta}} \textbf{\bibinfo{volume}{273}}, \bibinfo{pages}{137--162} (\bibinfo{year}{2020}).

\bibitem{Khan2022S}
\bibinfo{author}{Khan, A.}, \bibinfo{author}{Sossi, P.~A.}, \bibinfo{author}{Liebske, C.}, \bibinfo{author}{Rivoldini, A.} \& \bibinfo{author}{Giardini, D.}
\newblock \bibinfo{title}{Geophysical and cosmochemical evidence for a volatile-rich mars}.
\newblock \emph{\bibinfo{journal}{Earth Planet. Sci. Lett.}} \textbf{\bibinfo{volume}{578}}, \bibinfo{pages}{117330} (\bibinfo{year}{2022}).

\bibitem{Wade2005S}
\bibinfo{author}{Wade, J.} \& \bibinfo{author}{Wood, B.~J.}
\newblock \bibinfo{title}{Core formation and the oxidation state of the earth}.
\newblock \emph{\bibinfo{journal}{Earth Planet. Sci. Lett.}} \textbf{\bibinfo{volume}{236}}, \bibinfo{pages}{78--95} (\bibinfo{year}{2005}).

\bibitem{Yi2000S}
\bibinfo{author}{Yi, W.} \emph{et~al.}
\newblock \bibinfo{title}{Cadmium, indium, tin, tellurium, and sulfur in oceanic basalts: Implications for chalcophile element fractionation in the earth}.
\newblock \emph{\bibinfo{journal}{J. Geophys. Res. Solid Earth}} \textbf{\bibinfo{volume}{105}}, \bibinfo{pages}{18927--18948} (\bibinfo{year}{2000}).

\bibitem{Braukmuller2019S}
\bibinfo{author}{Braukm{\"{u}}ller, N.}, \bibinfo{author}{Wombacher, F.}, \bibinfo{author}{Funk, C.} \& \bibinfo{author}{M{\"{u}}nker, C.}
\newblock \bibinfo{title}{{Earth's volatile element depletion pattern inherited from a carbonaceous chondrite-like source}}.
\newblock \emph{\bibinfo{journal}{Nat. Geosci.}} \textbf{\bibinfo{volume}{12}}, \bibinfo{pages}{564--568} (\bibinfo{year}{2019}).

\bibitem{Lodders2025S}
\bibinfo{author}{Lodders, K.}, \bibinfo{author}{Fegley, B.}, \bibinfo{author}{Mezger, K.} \& \bibinfo{author}{Ebel, D.}
\newblock \bibinfo{title}{Condensation and the volatility trend of the earth}.
\newblock \emph{\bibinfo{journal}{Space Sci. Rev.}} \textbf{\bibinfo{volume}{221}}, \bibinfo{pages}{54} (\bibinfo{year}{2025}).

\bibitem{Alexander2019aS}
\bibinfo{author}{{Alexander}, C. M.~O.}
\newblock \bibinfo{title}{{Quantitative models for the elemental and isotopic fractionations in the chondrites: The non-carbonaceous chondrites}}.
\newblock \emph{\bibinfo{journal}{Geochim. Cosmochim. Acta}} \textbf{\bibinfo{volume}{254}}, \bibinfo{pages}{246--276} (\bibinfo{year}{2019}).

\bibitem{Nakamura2022S}
\bibinfo{author}{Nakamura, E.} \emph{et~al.}
\newblock \bibinfo{title}{On the origin and evolution of the asteroid ryugu: A comprehensive geochemical perspective}.
\newblock \emph{\bibinfo{journal}{Proc. Jpn. Acad. Ser. B Phys. Biol. Sci.}} \textbf{\bibinfo{volume}{98}}, \bibinfo{pages}{227--282} (\bibinfo{year}{2022}).

\bibitem{Lauretta2024S}
\bibinfo{author}{Lauretta, D.~S.} \emph{et~al.}
\newblock \bibinfo{title}{Asteroid (101955) bennu in the laboratory: Properties of the sample collected by osiris-rex}.
\newblock \emph{\bibinfo{journal}{Meteorit. Planet. Sci.}} \textbf{\bibinfo{volume}{59}}, \bibinfo{pages}{2453--2486} (\bibinfo{year}{2024}).

\bibitem{Liu2022S}
\bibinfo{author}{{Liu}, B.}, \bibinfo{author}{{Johansen}, A.}, \bibinfo{author}{{Lambrechts}, M.}, \bibinfo{author}{{Bizzarro}, M.} \& \bibinfo{author}{{Haugb{\o}lle}, T.}
\newblock \bibinfo{title}{{Natural separation of two primordial planetary reservoirs in an expanding solar protoplanetary disk}}.
\newblock \emph{\bibinfo{journal}{Sci. Adv.}} \textbf{\bibinfo{volume}{8}}, \bibinfo{pages}{eabm3045} (\bibinfo{year}{2022}).

\bibitem{Colmenares2024S}
\bibinfo{author}{{Colmenares}, M.~J.}, \bibinfo{author}{{Lambrechts}, M.}, \bibinfo{author}{{van Kooten}, E.} \& \bibinfo{author}{{Johansen}, A.}
\newblock \bibinfo{title}{{Thermal processing of primordial pebbles in evolving protoplanetary disks}}.
\newblock \emph{\bibinfo{journal}{Astron. Astrophys.}} \textbf{\bibinfo{volume}{685}}, \bibinfo{pages}{A114} (\bibinfo{year}{2024}).

\bibitem{Alexander2022S}
\bibinfo{author}{{Alexander}, C. M.~O.}
\newblock \bibinfo{title}{{An exploration of whether Earth can be built from chondritic components, not bulk chondrites}}.
\newblock \emph{\bibinfo{journal}{Geochim. Cosmochim. Acta}} \textbf{\bibinfo{volume}{318}}, \bibinfo{pages}{428--451} (\bibinfo{year}{2022}).

\bibitem{Garai2025S}
\bibinfo{author}{Garai, S.}, \bibinfo{author}{Olson, P.~L.} \& \bibinfo{author}{Sharp, Z.~D.}
\newblock \bibinfo{title}{Building earth with pebbles made of chondritic components}.
\newblock \emph{\bibinfo{journal}{Geochim. Cosmochim. Acta}} \textbf{\bibinfo{volume}{390}}, \bibinfo{pages}{86--104} (\bibinfo{year}{2025}).

\bibitem{SteinmeyerJohansen2024S}
\bibinfo{author}{{Steinmeyer}, M.-L.} \& \bibinfo{author}{{Johansen}, A.}
\newblock \bibinfo{title}{{Vapor equilibrium models of accreting rocky planets demonstrate direct core growth by pebble accretion}}.
\newblock \emph{\bibinfo{journal}{Astron. Astrophys.}} \textbf{\bibinfo{volume}{683}}, \bibinfo{pages}{A217} (\bibinfo{year}{2024}).

\bibitem{HabibPierrehumbert2024S}
\bibinfo{author}{{Habib}, N.} \& \bibinfo{author}{{Pierrehumbert}, R.~T.}
\newblock \bibinfo{title}{{Modeling Noncondensing Compositional Convection for Applications to Super-Earth and Sub-Neptune Atmospheres}}.
\newblock \emph{\bibinfo{journal}{Astrophys. J.}} \textbf{\bibinfo{volume}{961}}, \bibinfo{pages}{35} (\bibinfo{year}{2024}).

\bibitem{Fang2024S}
\bibinfo{author}{{Fang}, L.} \emph{et~al.}
\newblock \bibinfo{title}{{The origin of 4-Vesta's volatile depletion revealed by the zinc isotopic composition of diogenites}}.
\newblock \emph{\bibinfo{journal}{Sci. Adv.}} \textbf{\bibinfo{volume}{10}}, \bibinfo{pages}{eadl1007} (\bibinfo{year}{2024}).

\bibitem{Sossi2022S}
\bibinfo{author}{Sossi, P.~A.}, \bibinfo{author}{Stotz, I.~L.}, \bibinfo{author}{Jacobson, S.~A.}, \bibinfo{author}{Morbidelli, A.} \& \bibinfo{author}{O'Neill, H. S.~C.}
\newblock \bibinfo{title}{{Stochastic accretion of the Earth}}.
\newblock \emph{\bibinfo{journal}{Nat. Astron.}} \bibinfo{pages}{951--960} (\bibinfo{year}{2022}).

\bibitem{Steenstra2019S}
\bibinfo{author}{Steenstra, E.} \emph{et~al.}
\newblock \bibinfo{title}{Significant depletion of volatile elements in the mantle of asteroid vesta due to core formation}.
\newblock \emph{\bibinfo{journal}{Icarus}} \textbf{\bibinfo{volume}{317}}, \bibinfo{pages}{669--681} (\bibinfo{year}{2019}).

\bibitem{JohansenLacerda2010S}
\bibinfo{author}{{Johansen}, A.} \& \bibinfo{author}{{Lacerda}, P.}
\newblock \bibinfo{title}{{Prograde rotation of protoplanets by accretion of pebbles in a gaseous environment}}.
\newblock \emph{\bibinfo{journal}{Mon. Not. R. Astron. Soc.}} \textbf{\bibinfo{volume}{404}}, \bibinfo{pages}{475--485} (\bibinfo{year}{2010}).

\bibitem{Ormel2013S}
\bibinfo{author}{{Ormel}, C.~W.}
\newblock \bibinfo{title}{{The steady-state flow pattern past gravitating bodies}}.
\newblock \emph{\bibinfo{journal}{Mon. Not. R. Astron. Soc.}} \textbf{\bibinfo{volume}{428}}, \bibinfo{pages}{3526--3542} (\bibinfo{year}{2013}).

\bibitem{Maruyama2009S}
\bibinfo{author}{{Maruyama}, S.}, \bibinfo{author}{{Watanabe}, M.}, \bibinfo{author}{{Kunihiro}, T.} \& \bibinfo{author}{{Nakamura}, E.}
\newblock \bibinfo{title}{{Elemental and isotopic abundances of lithium in chondrule constituents in the Allende meteorite}}.
\newblock \emph{\bibinfo{journal}{Geochim. Cosmochim. Acta}} \textbf{\bibinfo{volume}{73}}, \bibinfo{pages}{778--793} (\bibinfo{year}{2009}).

\bibitem{vanKooten2019S}
\bibinfo{author}{{van Kooten}, E.} \& \bibinfo{author}{{Moynier}, F.}
\newblock \bibinfo{title}{{Zinc isotope analyses of singularly small samples (<5 ng Zn): Investigating chondrule-matrix complementarity in Leoville}}.
\newblock \emph{\bibinfo{journal}{Geochim. Cosmochim. Acta}} \textbf{\bibinfo{volume}{261}}, \bibinfo{pages}{248--268} (\bibinfo{year}{2019}).

\bibitem{Alexander1995S}
\bibinfo{author}{{Alexander}, C.~M.~O.}
\newblock \bibinfo{title}{{Rims, Matrix and the Bulk Compositions of Ordinary Chondrites}}.
\newblock \emph{\bibinfo{journal}{Meteoritics}} \textbf{\bibinfo{volume}{30}}, \bibinfo{pages}{479} (\bibinfo{year}{1995}).

\bibitem{Nagahara1989S}
\bibinfo{author}{{Nagahara}, H.} \& \bibinfo{author}{{Kushiro}, I.}
\newblock \bibinfo{title}{{Vaporization experiments in the system plagioclase-hydrogen}}.
\newblock \emph{\bibinfo{journal}{Antarctic Meteorite Research}} \textbf{\bibinfo{volume}{2}}, \bibinfo{pages}{235} (\bibinfo{year}{1989}).

\bibitem{Alexander2019bS}
\bibinfo{author}{{Alexander}, C. M.~O.}
\newblock \bibinfo{title}{{Quantitative models for the elemental and isotopic fractionations in chondrites: The carbonaceous chondrites}}.
\newblock \emph{\bibinfo{journal}{Geochim. Cosmochim. Acta}} \textbf{\bibinfo{volume}{254}}, \bibinfo{pages}{277--309} (\bibinfo{year}{2019}).

\bibitem{Tachibana1998S}
\bibinfo{author}{{Tachibana}, S.} \& \bibinfo{author}{{Tsuchiyama}, A.}
\newblock \bibinfo{title}{{Incongruent evaporation of troilite (FeS) in the primordial solar nebula: an experimental study}}.
\newblock \emph{\bibinfo{journal}{Geochim. Cosmochim. Acta}} \textbf{\bibinfo{volume}{62}}, \bibinfo{pages}{2005--2022} (\bibinfo{year}{1998}).

\bibitem{Steinmeyer2023S}
\bibinfo{author}{Steinmeyer, M.-L.}, \bibinfo{author}{Woitke, P.} \& \bibinfo{author}{Johansen, A.}
\newblock \bibinfo{title}{{Sublimation of refractory minerals in the gas envelopes of accreting rocky planets}}.
\newblock \emph{\bibinfo{journal}{Astron. Astrophys.}} \textbf{\bibinfo{volume}{181}}, \bibinfo{pages}{1--16} (\bibinfo{year}{2023}).

\bibitem{Zhang2016S}
\bibinfo{author}{{Zhang}, A.-C.} \emph{et~al.}
\newblock \bibinfo{title}{{P-O-rich sulfide phase in CM chondrites: Constraints on its origin on the CM parent body}}.
\newblock \emph{\bibinfo{journal}{Meteorit. Planet. Sci.}} \textbf{\bibinfo{volume}{51}}, \bibinfo{pages}{56--69} (\bibinfo{year}{2016}).

\bibitem{Lodders2003S}
\bibinfo{author}{Lodders, K.}
\newblock \bibinfo{title}{{Solar System abundances and condensation temperatures of the elements}}.
\newblock \emph{\bibinfo{journal}{Astrophys. J.}} \textbf{\bibinfo{volume}{591}}, \bibinfo{pages}{1220--1247} (\bibinfo{year}{2003}).

\bibitem{Wang2026aS}
\bibinfo{author}{Wang, H.~S.} \emph{et~al.}
\newblock \bibinfo{title}{Volatile depletion in rocky planets as a chemical fingerprint of hybrid accretion}.
\newblock \bibinfo{publisher}{figshare}, \bibinfo{type}{Dataset}
(\bibinfo{year}{2026}).
\newblock \doi{10.6084/m9.figshare.33085055}.

\end{thebibliography}

\end{document}